**UNIVERSIDADE FEDERAL DO RIO DE JANEIRO**
**OBSERVATÓRIO DO VALONGO**

**ANÁLISE DE OBSERVAÇÕES DO DIÂMETRO NO CONTEXTO DA ATIVIDADE SOLAR**

**Sérgio Calderari Boscardin**

> **Tese de Mestrado apresentada ao Programa de Pós-Graduação do Observatório do Valongo como parte dos requisitos necessários à obtenção do título de Mestre em Astronomia.**
>
> **Orientador: Dr. Alexandre Humberto Andrei**

**Rio de Janeiro**
**dezembro de 2005**

# RESUMO

## ANÁLISE DE OBSERVAÇÕES DO DIÂMETRO NO CONTEXTO DA ATIVIDADE SOLAR

**Sérgio Calderari Boscardin**

**Orientador: Dr. Alexandre Humberto Andrei**

De 1997 a 2003 mais de 20.000 observações do Semidiâmetro Solar foram feitas pelo Astrolábio Solar CCD do Observatório Nacional no Rio de Janeiro. Os dados experimentais de 1998 a 2001 foram obtidos em dois trabalhos anteriores independentes, pelas correções de efeitos sistemáticos aos resultados observacionais brutos. Neste trabalho foram determinadas correções semelhantes para os resultados observacionais de 2002 e 2003. Visando a robustez das correções, as medidas de cada ano foram consideradas em separado, bem como foram separadas as séries de observações feitas a leste do meridiano local e a oeste deste. Inicialmente, para cada série anual utilizamos uma abordagem estatística, justificada pela quantidade de dados. Os valores a este e a oeste foram corrigidos em função de seu desvio quadrático às tendências médias locais, sem alterar, portanto as variações medidas. Finalmente, os valores foram corrigidos dos erros sistemáticos, utilizando coeficientes obtidos das correlações entre os parâmetros e as medidas observacionais.

A série de 16.523 dados coerentes da variação do semidiâmetro solar, entre 1998 e 2003, permitiu a comparação detalhada com séries de indicadores da atividade solar: Índice de Flares, Irradiância, Campo Magnético Integrado, Contagem de Manchas e Fluxo Rádio em 10,7cm. A hipótese de variação do semidiâmetro ligada à atividade solar, sendo reciprocamente seu estimador, foi examinada através das correlações entre os diferentes pares de indicadores. As correlações foram calculadas para espaçamentos contínuos das séries de valores, desde a divisão anual até a divisão mensal de dados. Foram obtidas fortes correlações entre alguns pares, interpretadas como forte interação física entre eles. A seguir obtiveram-se as mesmas correlações considerando atrasos temporais de uma série em relação à outra. Diversos pares mostraram aumento nas correlações apontando para um valor máximo quando há alguma defasagem entre as curvas. Para o par Semidiâmetro Solar e Irradiância obtivemos a moda das defasagens nas correlações máximas dos diversos períodos para dois casos. Quer para a série completa de dados e quer deixando de fora os dados relativos aos dois picos máximos do ciclo de atividade solar. A comparação mostra que o Semidiâmetro responde imediatamente às variações de Irradiância nas condições em que se consideram os picos de atividades e precede às variações de Irradiância em pelo menos cem dias quando aqueles valores são desconsiderados, apontando para a existência de dois regimes diferentes.

Estudamos também como o semidiâmetro do Sol varia ao longo de suas latitudes, observando que cada faixa de latitude tem uma resposta diferente embora haja uma variação global em certa direção. Foram detectadas faixas de latitude solar que têm módulos de variações do semidiâmetro mais suaves e outras com módulos de variações mais fortes.

Palavras chave: Sol, diâmetro solar, atividade solar, Astrometria

# ABSTRACT

**DIAMETER OBSERVATIONS ANALYSIS IN THE SOLAR ACTIVITY CONTEXT**

**Sérgio Calderari Boscardin**

**Supervisor: Dr. Alexandre Humberto Andrei**

From 1997 to 2003 the CCD Solar Astrolabe of the Observatório Nacional in Rio de Janeiro made more than 20,000 observations of the Solar Semidiameter. The experimental data from 1998 to 2001 had been obtained in two independent previous works, by the corrections of the systematic effects on the raw observational results. In the present work similar corrections for the observational results of 2002 and 2003 were determined. Aiming at the robustness of the corrections, the measurements of each year have been considered separately, as well as the series of observational data as taken at east or west from the local meridian were separately considered. Initially, for each annual series a statistical approach was used, justified by the amount of data. The values to east and west have been corrected in function of their mean quadratic offset to the local trend averages, therefore without modifying the measured variations. Finally, the values have been corrected of the bias, using coefficients obtained from the correlation between the parameters and the observational measures.

The series of 16.523 coherent data of the variation of the Solar Semidiameter, between 1998 and 2003, allowed the detailed comparison with series of pointers of the solar activity: Flare Index, Total Irradiance, Integrated Magnetic Field, Sunspot Number and 10.7cm Radio Flux. The hypothesis of variation of the Semidiameter tied to the solar activity, otherwise being its estimator, was examined through the correlations between the different pairs of pointers. The correlations have been calculated at continuous spacing of the series of values, from the data annual division until the monthly division. Strong correlations between some pairs were obtained, and interpreted as strong physical interaction between them. Next, the same correlations were obtained now considering time delays of one series in relation to the other. Several pairs have shown an increase on the correlation, indicating the presence of a significant phase shift between them. For the pair Solar Semidiameter and Irradiance the mode of the phase for maximum correlation was calculated for two distinct cases: either for the complete series of data, or leaving off the data relative to the epochs of the two summits of the solar activity cycle. The comparison shows that the Solar Semidiameter responds closely to variations of Irradiance in the conditions where the peaks of activity are considered; inversely, it precedes the variations of Irradiance, by at least one hundred days, when the peak values are discarded, thus indicating the existence of two distinct regimes.

We also studied how the Solar Semidiameter varies throughout its latitudes. It was noted that each band of latitude has a different response, even so there is a global variation in time. It was also determined the solar latitude bands that exhibit small modules for the Solar Semidiameter variation and those bands where the variation modules are much stronger.

# SUMÁRIO

# INTRODUÇÃO

Pelo menos duas perguntas importantes do momento atual exigem um conhecimento mais profundo dos mecanismos físicos que se passam no Sol. Primeiramente a questão de se saber como o Sol pode influenciar o clima da Terra. Estamos vivendo em uma época em que a humanidade se pergunta qual a real influência dos gases expelidos na atmosfera pela indústria e pelos veículos que queimam derivados de petróleo no aquecimento global que se observa. Devemos desativar o parque industrial, ou, na verdade, a maior parte deste aquecimento é provocada pelo Sol? A outra pergunta é da Astrofísica: o conhecimento mais exato de certos processos que ocorrem no Sol responderá importantes questionamentos desta ciência cujas respostas darão subsídios a seus estudiosos para conhecerem melhor outras estrelas e em conseqüência o universo.

Neste contexto está a importância de conhecermos as variações do Semidiâmetro do Sol e de sabermos como estas variações estão conectadas a outras observações da atividade solar. Estudos comparativos das variações de temperatura global da Terra e de variações da Irradiância total solar mostram haver uma correlação entre os dois da ordem de $0,46^{o}C/Wm^{2}$ (Lefebvre et Rozelot, 2003). Outros estudos apontam para uma possível correlação entre as variações do Semidiâmetro do Sol e as variações da Irradiância.

Pergunta-se também se o Sol varia seu diâmetro globalmente ou se estes são fenômenos apenas ligados às latitudes solares. O volume total do astro pode sofrer pequenas alterações ou mesmo nenhuma enquanto suas diversas latitudes trocam massas. Responder a esta pergunta exige o estudo da forma do Sol e de como esta forma varia temporalmente. Neste sentido devemos analisar as variações do Semidiâmetro Solar considerando também em que heliolatitude foram feitas as observações.

As medidas do Semidiâmetro Solar são feitas durante o dia quando o Sol é aparente, e portanto, quando as temperaturas e suas variações são bem maiores que à noite. Em conseqüência a agitação atmosférica é grande e correntes de vento são comuns. O Sol, que é o objeto de estudo, incide no instrumento externamente e diretamente em suas partes internas causando diversas deformações por efeito de dilatação térmica. A soma de todos estes efeitos somados a pequenos desvios do instrumento causa uma gama de erros da medida. Por estes motivos as séries de dados observados são muito ruidosas. Devemos, então, antes de utilizar os dados observados, tratá-los adequadamente a fim de retirar-lhes a maior parte possível de erros. A nosso favor temos a grande quantidade de observações e a leitura de diversos outros parâmetros concomitantes com as observações solares que nos permitem obter, através de tratamentos estatísticos, séries de dados menos ruidosas.

# O SEMIDIÂMETRO SOLAR.

Inicialmente um esclarecimento de porque se fala em Semidiâmetro do Sol e não em raio. Historicamente os observadores referiam-se ao raio do Sol, pois, imaginavam-no perfeitamente circular, e calculavam-lhe o raio. Mais recentemente, quando se percebeu que sua figura não era circular e, portanto, o diâmetro dependia da latitude medida, passaram a medir-lhe os diâmetros e por comparação com valores anteriormente determinados, dividiam-no por dois para comparar com o raio e assim passou-se a atribuir valores ao que hoje designamos por Semidiâmetro do Sol.

O diâmetro do Sol tem sido medido ao longo de toda a história. Arquimedes (287-212 AC) teria calculado um valor entre 27 e 33 minutos de arco. Aristarco (310-230 AC) atribuiu o valor de 30 minutos e Ptolomeu (87-151 DC) o valor de 31'20". Ptolomeu acompanhou a medida por um ano buscando verificar as variações sem contudo percebê-las. Em 1591, Tycho Brahe (1546-1601) realizou onze medidas que segundo Johannes Kepler (1571-1630) mostram um valor mínimo do diâmetro solar de 30'30". Dados históricos apontam para um raio solar maior durante o período conhecido como Mínimo de Maunder. Em 1891 Auwers obteve um valor de 959",53 para o Semidiâmetro do Sol (Reis Neto, 2002).

Mais recentemente as variações do Semidiâmetro Solar vêm sendo observadas por diversos pesquisadores. No Brasil além das observações feitas no Observatório Nacional (Penna et al., 1998) há as observações de M.Emílio em São Paulo (Emilio et Leister, 2005). Na França, no Observatório da Cote d'Azur (Laclare et al., 1996), na Turquia pela equipe do Observatório Nacional de Tubitak (Golbasi et al., 2000) e no Chile – observações visuais feitas por F.Noël da Universidade de Santiago (Noël, 1998). Todas estas medições apontam para uma variação temporal do Semidiâmetro do Sol.

O diâmetro solar pode variar no tempo e também ao longo de sua forma. Além da equipe do Observatório Nacional (Reis Neto, 2002) há pelo menos mais um autor que afirma haver variações do Semidiâmetro ao longo das latitudes havendo um aumento do diâmetro por volta dos 45 graus que se estende por uns 20 graus e uma depressão em torno dos 70 graus (Rozelot et Lefebvre, 2003).

# O ASTROLÁBIO SOLAR.

Muitos têm se preocupado com a medida do diâmetro solar e com as suas possíveis variações e, na lista das instituições que efetivaram algum programa neste sentido está o Observatório Nacional – ON que modificou em 1997 um Astrolábio Danjon, na sua sede no Rio de Janeiro, equipando-o com um prisma refletor de ângulo variável e uma câmara CCD, o que permitiu o monitoramento do diâmetro solar (Penna et al., 1998). Desde aquela data um vasto programa de observação e medida do diâmetro solar vem sendo desenvolvido e um total de mais de 20.000 observações está arquivado. A Tabela I a seguir mostra o número de observações a cada ano, bem como o número de dias em que se observou em cada ano.

**Tabela I – Número de observações do Semidiâmetro do Sol e o número de dias em que se observou de 1997 a 2003. (Penna et al., xxxx)**

| ANO | 1997 | 1998 | 1999 | 2000 | 2001 | 2002 | 2003 | Total |
|---|---|---|---|---|---|---|---|---|
| Número de Observações | 2.706 | 3.927 | 3.949 | 3.268 | 1.890 | 2.905 | 2.127 | 20.774 |
| Número de Dias | 158 | 162 | 157 | 163 | 122 | 154 | 134 | 1.050 |

O astrolábio solar do Observatório Nacional consiste de um telescópio refrator na frente do qual é colocado um prisma refletor e uma bacia com mercúrio. O prisma tem duas faces refletoras que são simétricas em relação ao plano horizontal. Os raios incidentes são separados em dois feixes: um devido à reflexão na face superior do prisma e o outro obtido por reflexão na face inferior do prisma após uma primeira reflexão na superfície de mercúrio. Nestas condições obtêm-se duas imagens do objeto observado. Quando o objeto se move sua variação azimutal provoca deslocamentos das imagens para o mesmo lado e a variação de sua distância zenital provoca deslocamentos opostos das imagens. Assim, o instrumento detecta a coincidência de duas imagens de um objeto pontual, quando este ponto cruza uma linha de distância zenital determinada pelo ângulo do prisma. Fazendo-se variar o ângulo com o plano horizontal, simultaneamente, das duas faces do prisma, obtêm-se imagens de objetos em diferentes distâncias zenitais. O instrumento instalado no ON permite a observação de objetos entre 25º e 55º de distância zenital. A Figura 1 mostra um esquema do Astrolábio Solar.

O Astrolábio fornece a uma câmara CCD as duas imagens do Sol. A cada observação de Semidiâmetro Solar escolhe-se determinada distância zenital, por onde deve passar o Sol. Um círculo de distância zenital constante é conhecido como almicantarado. À medida em que o Sol se aproxima do almicantarado escolhido, as duas imagens se aproximam e, quando o primeiro bordo do Sol cruza o almicantarado, as duas imagens se tocam. Quando o segundo bordo do Sol cruza esta linha, as imagens se separam ao se tocarem os bordos das duas imagens. Quando os bordos do Sol estão para cruzar o almicantarado, são feitas 46 imagens. O instante em que cada imagem é obtida é fornecido pelo relógio atômico do ON. A análise destas imagens fornece o instante em que as duas imagens, a direta e a refletida, se tocaram ou se separaram.

Conhecendo-se a marcha do Sol a cada dia do ano e a cada posição que ocupa na esfera celeste, pode-se calcular, a partir do tempo que o Sol levou para cruzar totalmente o almicantarado, o seu tamanho angular. Este tamanho é então reduzido para a distância média do Sol, isto é, para uma UA, obtendo-se a medida angular do seu diâmetro.

As imagens obtidas são dirigidas ao CCD que tem 512 linhas e 512 colunas de pixels. Por conta do entrelaçamento das imagens, tomam-se apenas 256 linhas. Um pixel corresponde a 0",56 e desta forma apenas uma parte do bordo solar é tomada na imagem. Em cada imagem são identificados 256 pontos do bordo direto e 256 pontos do bordo refletido do Sol, um para cada uma das linhas da imagem. As linhas apresentam uma curva de intensidade de luz. Esta curva tem a propriedade de apresentar um extremo de sua derivada ao longo do bordo solar. A segunda derivada da função de intensidade de luz é igual a zero no bordo solar. Assim, o bordo solar é o ponto em que a curva de luz tem seu ponto de inflexão. Na Figura 2 há uma esquema da determinação de um ponto do bordo solar. Na verdade a análise é um pouco mais complexa porque há dois bordos na mesma imagem, o direto e o refletido (Chollet et Golbasi, 2003).

O bordo solar é delineado pelo conjunto destes 256 pontos. Ocorre, porém, que estes pontos geralmente formam uma curva ruidosa em função da ação da atmosfera. É preciso passar por eles uma curva mais definida que indique o bordo solar. Ajusta-se a estes pontos uma parábola e não um arco de círculo, isto porque a parábola minimiza defeitos óticos, responde melhor à forma retangular dos pixels, e contempla o movimento do disco solar durante os 20ms de integração (Reis Neto, 2002). Há uma parábola para o bordo direto e uma parábola para o bordo refletido conforme é indicado no esquema apresentado na Figura 3.

Para cada imagem obtida existem então duas parábolas. Tomam-se as posições dos vértices destas parábolas e o instante de tempo em que a imagem foi obtida. Estas posições em função do tempo definem duas séries de pontos por onde se ajustam duas retas. Como uma das parábolas avança e a outra recua, estas retas têm inclinações opostas. O ponto de contacto destas retas define o instante de tempo em que os bordos direto e refletido do Sol se tocam. Este é o instante em que o bordo solar cruzou o almicantarado. Da mesma forma se encontra o instante em que o segundo bordo solar cruza o almicantarado. A Figura 4 mostra um esquema da determinação do instante de passagem do bordo solar pelo almicantarado.

Um conjunto de programas foi desenvolvido para calcular, a partir dos instantes de passagem dos dois bordos pelo mesmo almicantarado, o diâmetro vertical observado do Sol. Após a aplicação das correções necessárias o diâmetro é reduzido para 1 UA (Sinceac, 1998).

Aos bordos do Sol ajusta-se, não apenas uma, mas, na verdade, três parábolas com critérios diferentes de desvio padrão para remoção de pontos observados. O critério se baseia num teste de quartís, com o fator multiplicativo variando entre 1,5 (mínima rejeição) e 3,0 (máxima rejeição). Se o número de pontos utilizados for inferior a 50, então nenhuma parábola é ajustada. Em 1997 foram definidos três níveis de critérios: 1,7, 2,0 e 2,5. Se as condições forem boas, três soluções são obtidas, caso contrário apenas as soluções para 1,7 e 2,0 alcançam resultado ou, mais raramente, apenas a primeira (Reis Neto, 2002).

As imagens obtidas são analisadas pelos programas os quais calculam para cada observação os três valores do Semidiâmetro Solar e os respectivos erros de medida. Obtêm também a distância zenital e o azimute solar, a latitude solar na direção da qual o semidiâmetro é medido, o fator de Fried, os instantes de passagem dos bordos solares, os erros de medida destes instantes, a largura em pixels do bordo direto e do bordo refletido, o ajuste da parábola direta e o da refletida, o desvio padrão dos pontos da parábola direta e, da refletida, e a decalagem. Para cada seção de observações têm-se também os instantes inicial e final, e as temperaturas do ar e do mercúrio, a pressão atmosférica e a umidade do ar nestes instantes. O fator de Fried descreve a qualidade do 'seeing' da atmosfera, ele é definido como o comprimento de onda observado, dividido pela largura a meia altura de uma imagem pontual espalhada pela ação da atmosfera. Este fator é calculado a partir dos dados de observação (Lakhal et al., 1994). A decalagem é o desvio do centro do CCD ao ponto de contacto das duas parábolas.

Diante da câmera do CCD, há dois filtros de luz que definem uma banda de freqüências que podem ser detectadas. O intervalo onde 50% da luz é transmitida vai de 523,0 nm até 691,0 nm. Sendo o máximo em 563,5 nm com índice de transmissão de 75% (Jilinski et al., 1999).

Os dados finais processados são apresentados em um arquivo para cada sessão de trabalho. A cada dia, quando possível, são realizadas duas sessões: uma antes da passagem meridiana do Sol e outra após sua passagem. Os arquivos dispõem ao analista três conjuntos de dados para cada observação com os seguintes itens: ano, mês e dia da observação; lado (leste ou oeste); Semidiâmetro do Sol e erro do Semidiâmetro em segundos de arco; data Juliana com até quatro casas decimais; distância zenital, azimute e ângulo paralático do Sol; latitude solar do Semidiâmetro observado; fator de Fried; instante de passagem de cada bordo com hora, minuto e segundos até a quarta casa decimal; erro do instante de passagem; largura em pixel do bordo direto e do bordo refletido para os dois bordos; ajuste das parábolas da imagem direta e da imagem refletida para os dois bordos; desvio padrão destes ajustes; inclinação da observação a cada bordo; decalagem a cada bordo; inicio e final da sessão com hora e minuto; pressão barométrica ao início e ao final da sessão; temperatura do mercúrio ao início e ao final da sessão; temperatura do ar ao início e ao final da sessão; umidade do ar ao início e ao final da sessão.

**Figura 1 - Esquema do prisma e da bacia com mercúrio do Astrolábio, indicando o caminho que fazem os raios do Sol até chegarem ao detector.**

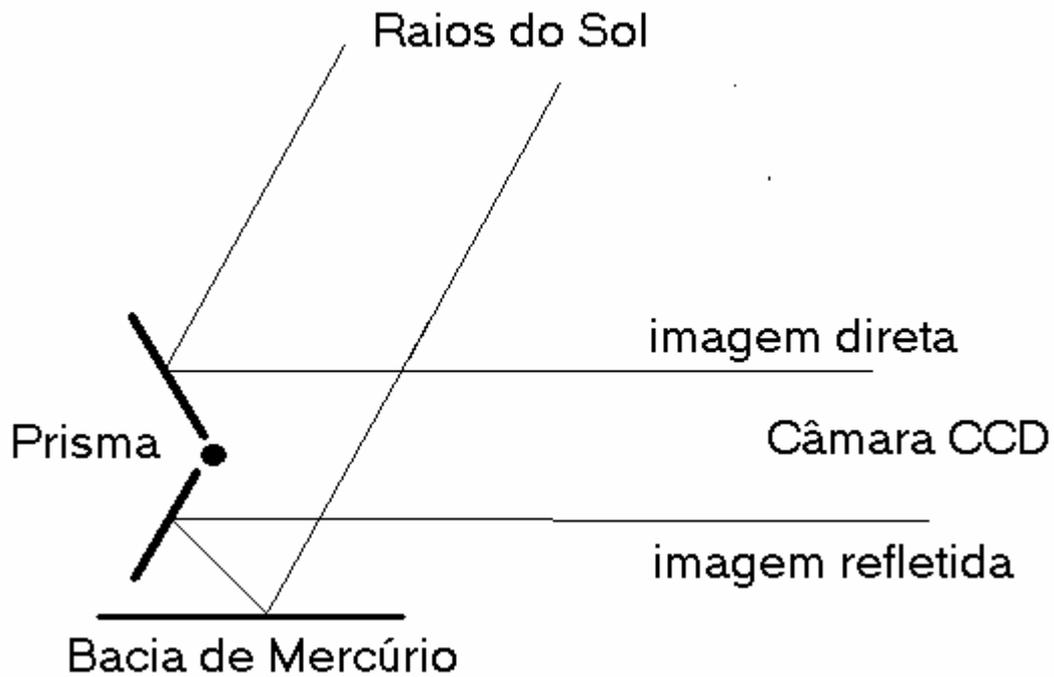

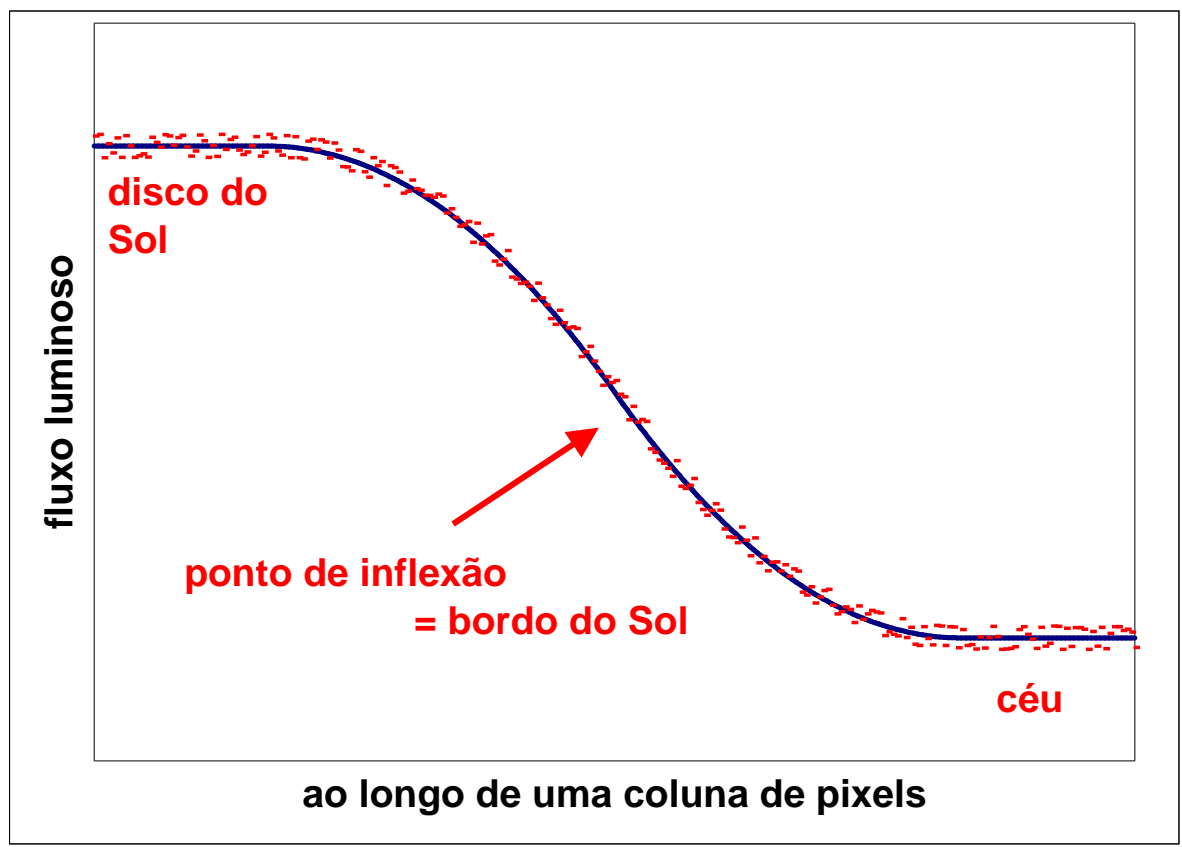

**Figura 2 - Determinação de um ponto do bordo solar. A curva de luz ao longo de uma coluna do CCD é ajustada e determina-se seu ponto de inflexão.**

**Figura 3 - Determinação do bordo solar. Ajusta-se uma parábola aos pontos do bordo solar anteriormente determinados.**

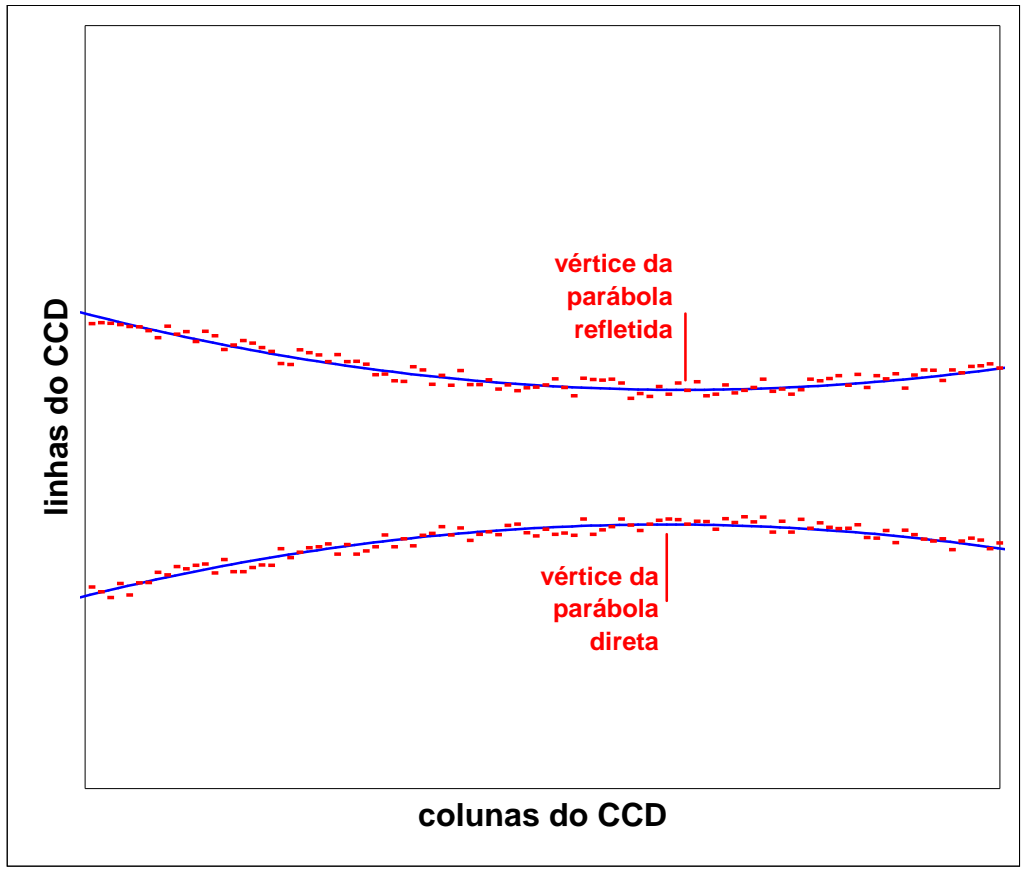

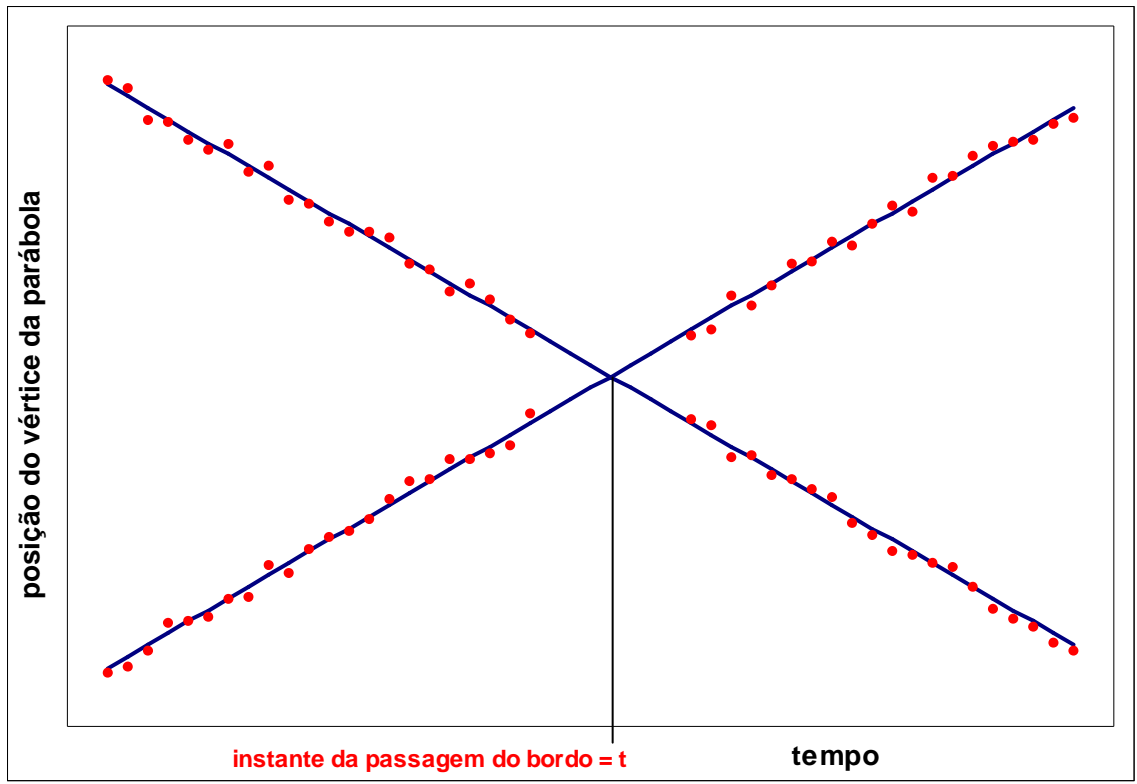

**Figura 4 - O instante de passagem do bordo solar é determinado pelo ponto de contacto das retas ajustadas às posições sucessivas dos vértices das parábolas que determinam o bordo solar.**

**DADOS UTILIZADOS.**

Este trabalho analisa dados obtidos com as observações realizadas pelo Astrolábio Solar do Observatório Nacional entre os anos de 1998 e 2003. Neste período foram feitas mais de dezoito mil observações do Semidiâmetro Solar em quase 900 dias de trabalho. Todas estas observações foram efetuadas com o mesmo instrumento, utilizando as mesmas técnicas de observação e de redução de imagens e foram obtidas em sua maior parte por apenas dois operadores. Estes dados foram posteriormente corrigidos para minimizar os efeitos causados pelo instrumento ou pela própria observação. Estas correções utilizaram os mesmos parâmetros sempre disponíveis embora tenham sido feitas em várias etapas de acordo com a disponibilidade dos dados.

O Sol foi observado em 1997, mas estes dados foram rejeitados porque a série se mostra muito ruidosa. Também porque, a partir de 1998 o Astrolábio recebeu filtros que estreitam a banda passante do espectro observado, diminuindo, a partir de então, o ruído da série.

Para todas as correções realizadas foram sempre retirados os valores de Semidiâmetro Solar superiores a 961,0 segundos de arco e inferiores a 957,0 segundos de arco que correspondem a pontos com desvio superior a 2,5 desvios padrão da média geral. Foram também retirados os pontos cujos erros de Semidiâmetro eram superiores a 1 segundo de arco. Foram ainda retirados os pontos em que o cálculo indica espuriamente o CCD como girado de mais de 2,5 graus em relação ao zero.

Em cada observação são ajustadas três parábolas aos bordos do Sol. Cada uma das parábolas é obtida por critérios diferentes de desvio dos pontos usados para seu cálculo. Em todas as análises foi sempre utilizado o caso em que o critério foi o mais relaxado. Verificou-se que as três soluções conduzem sempre a resultados muito semelhantes e optou-se por aquela que considera o maior número de pontos observados.

Os dados de 1998, de 1999 e de 2000 foram tratados no trabalho "Observações solares: Estudo das variações do diâmetro e suas correlações", tese de mestrado de Eugênio Reis Neto para o Observatório Nacional em abril de 2002 (Reis Neto, 2002). Estas observações foram corrigidas inicialmente do efeito causado pelas molas que mantém o ângulo do prisma objetivo do Astrolábio. O conjunto de molas não permanece estável durante o espaço de tempo em que ocorre a observação. Este erro é diferente para os lados leste e oeste porque no primeiro caso o Sol está diminuindo sua distância zenital e no outro caso o Sol está aumentando a distância zenital, enquanto que as molas atuam no sentido de fechar o ângulo entre as faces do prisma, fazendo com que a distância zenital observada seja respectivamente aumentada ou diminuída. Assim, para um dos lados haverá um acréscimo de tempo para o Sol cruzar a linha de almicantarado desejado enquanto que para o outro haverá um decréscimo. Esta diferença é de alguma forma proporcional ao tempo de observação.

Na segunda fase, os dados de 1998 a 2000 foram corrigidos dos efeitos causados por parâmetros de observação conhecidos. Dos parâmetros pesquisados foram selecionados para correção: as influências causadas pela temperatura do ar no momento da observação, pela diferença de temperatura durante a observação, pelo fator de Fried e pelo desvio padrão da parábola ajustada ao bordo direto observado do Sol.

As observações feitas no ON em 2001 se estendem do início de janeiro até quase ao final de setembro. A partir desta data o Astrolábio foi temporariamente desativado para sofrer uma manutenção, voltando a operar somente nos últimos dias do ano. As observações deste período consideradas úteis perfazem um total de 1890, sendo 976 de observações a leste do meridiano local e 914 de observações a oeste deste. Estes dados foram tratados e apresentados em maio de 2004 no trabalho final de graduação em Astronomia: "Análise de uma nova série de medidas de variação do Semidiâmetro do Sol" (Boscardin, 2004). Destes dados retiraram-se erros devidos ao instrumento e às condições de observação. A análise foi feita em três etapas. Na primeira foram retirados erros decorrentes da falta de estabilidade do prisma objetivo que causa um pequeno aumento dos valores de Semidiâmetro do Sol quando lidos a leste do meridiano local e uma pequena diminuição dos valores quando lidos a oeste. Na segunda etapa foram retirados erros devidos a parâmetros de observação dos quais foram selecionados como importantes causas de erros: o fator de Fried, a temperatura média do ar durante a observação e o desvio padrão do ajuste da parábola ao bordo refletido do Sol. Na última etapa foram corrigidos erros devidos a desvios de nivelamento do astrolábio que conduzem a erros em função do azimute de observação. A Tabela II mostra a evolução dos valores observados e corrigidos a cada etapa do processo de correção separando os valores a leste e a oeste. São mostrados a média e o desvio em cada etapa. Pode-se ver que as correções a leste são maiores que as correções a oeste e que os valores a leste são mais ruidosos que os outros.

**Tabela II –** Evolução da média e do desvio padrão ao longo do processo de correção dos valores observados de Semidiâmetro do Sol em 2001. (valores em segundos de arco)

| Etapas | LESTE | | OESTE | | TOTAL | |
|---|---|---|---|---|---|---|
| | Média | Desvio | Média | Desvio | Média | Desvio |
| brutos | 959,259 | 0,643 | 959,115 | 0,564 | 959,189 | 0,610 |
| 1ª etapa | 959,191 | 0,628 | 959,190 | 0,556 | 959,191 | 0,594 |
| 2ª etapa | 959,186 | 0,621 | 959,195 | 0,558 | 959,191 | 0,591 |
| 3ª etapa | 959,186 | 0,601 | 959,195 | 0,557 | 959,190 | 0,580 |

Os tratamentos dos dados de 2002 e de 2003 estão dentro do escopo deste trabalho e passamos a apresentá-los a seguir.

# CORREÇÃO DOS DADOS DE 2002

Os dados de 2002 se compõem de 3015 observações do Sol, sendo 1505 feitas a leste do meridiano local e 1510 a oeste deste. Estão incluídas nestes números, 30 observações a leste e 16 a oeste do meridiano que foram feitas nos últimos dias de 2001, quando o instrumento voltou a funcionar após três meses parado. A Figura 5 mostra a distribuição das observações ao longo dos meses do ano (dados de 2001 incluídos em janeiro), dividindo-as em duas séries: os valores observados a leste e os observados a oeste do meridiano local.

As duas séries de valores observados do Semidiâmetro Solar podem ser vistas no gráfico da Figura 6 que os mostra e suas médias corridas de 150 pontos em função da data juliana modificada cujo valor, neste gráfico e em todos os demais ao longo deste trabalho, é o da data Juliana diminuida de 2.450.000. Estas médias apresentam-se destacadamente desviadas. A média dos valores a leste apresenta-se sempre acima da dos valores a oeste. Foram traçadas duas retas que mostram a tendência linear das médias que têm um valor descendente semelhante: (-0,8) milisegundos de arco por dia a leste e (-0,7) milisegundos de arco por dia a oeste.

Os valores de Semidiâmetro Solar observados em 2002 estão espalhados em torno de duas retas de tendência, mesmo as médias corridas de 150 pontos também oscilam em torno destas retas de tendência. Podemos dizer que há uma tendência geral decrescente ao longo do tempo e que uma série de eventos e erros espalha os pontos em torno desta tendência. Um erro, provavelmente causado por um pequeno desvio do prisma do Astrolábio, causa a separação entre as séries de pontos observados a leste e a oeste do meridiano.

Erros de observação e instrumentais precisam ser considerados. O fato de a observação ser diurna, o que não é trivial em Astronomia, torna os erros maiores, uma vez que durante o dia, as temperaturas são maiores e suas variações também são maiores. Em conseqüência a turbulência atmosférica é maior. As diferenças de temperatura entre manhã e tarde bem como as diferenças de variação de temperatura durante a manhã e durante a tarde e as diferenças de umidade podem causar mudanças na turbulência e na refração atmosférica bem diferentes pela manhã e a tarde. A maior causa de erros é a temperatura que varia também ao longo do ano criando condições diferentes em épocas diferentes do ano. Na nas Figuras 7 e 8 pode ser vista a evolução da temperatura durante as observações ao longo do ano de 2002.

As condições de observação mudam ao longo do dia criando diferenças nas causas de erros, assim, os diferentes horários de observação podem influir nas causas de erros. Os horários de observação das duas séries são bem diferentes: as observações a leste são feitas entre 8 e 11 horas e as observações a oeste entre 13 e 16 horas, horário local. Os horários mudam também ao longo do ano aproximando-se as duas séries no solstício de inverno e afastando-se no solstício de verão. As horas em que o Semidiâmetro do Sol foi observado durante o ano de 2002 são mostradas na Figura 9.

Alguns erros têm uma função crescente com tempo de duração da observação. Observações mais longas estão sujeitas a um maior erro. O tempo entre as passagens dos dois bordos do Sol por um almicantarado varia bastante com os meses do ano, varia também ao longo do dia, muito pouco no verão e bastante no inverno. As observações vão de dois minutos e meio no verão até entre quatro e seis minutos e meio no solstício de inverno. As Figuras 10 e 11 ilustram estas durações para os valores de 2002.

Há erros causados pelo instrumento de observação. Uma destas causas de erros de observação é o apontamento do Astrolábio em ângulos diferentes com o zênite. A distância zenital do Sol varia bastante durante o ano. Ele é observado mais baixo próximo ao solstício de inverno e mais alto no outro solstício. Ocorrem outras variações, principalmente nas observações a leste, decorrentes da adaptação dos observadores, que, em certas épocas do ano, precisam mudar os horários das sessões de observação para se adequar aos limites de ângulos possíveis de observação. A variação da distância zenital de observação ao longo do ano de 2002 pode ser vista nas Figuras 12 e 13. Um pequeno desvio no nivelamento do aparelho causa erros quando ele gira no seu apontamento azimutal, erros que dependem do ângulo de azimute de observação. Este erro foi apontado e corrigido nos dados de 2001. As Figuras 14 e 15 mostram a variação sazonal do ângulo de azimute de observação do Sol ao longo do ano de 2002.

Além destas causas citadas porque são evidentes e podem ser medidas, há uma série de outras que podem levar a erros de observação, são causas que têm origem em mínimos desvios do instrumento ou em fenômenos associados ao ato de se observar. Assim, dividimos nossa investigação em duas partes: primeiramente a correção da componente de erro que não admitiu correlação com parâmetros conhecidos, usando-se uma abordagem puramente estatística. A seguir os erros, ainda presentes, e linearmente correlacionados a parâmetros conhecidos.

Na primeira parte da análise a metodologia utilizada foi a de dividir as séries em 38 grupos temporalmente ao longo do ano. Os grupos se interpenetram, isto é, os pontos podem pertencer a mais de um grupo. Isto foi feito para se observar como o Semidiâmetro Solar varia dentro de cada grupo. Cada grupo foi escolhido de modo a perfazer no mínimo 28 dias de observações, ter início pelo menos sete dias após o início do grupo anterior e conter observações a leste e a oeste nos primeiros e últimos dias de seu período, de modo a se ter um bom cálculo da variação do Semidiâmetro Solar dentro do grupo. Dos 38 grupos, 30 abrangem 28 dias, seis abrangem 29 dias, um abrange 30 e um abrange 34 dias. A escolha do tamanho e da defasagem dos períodos foi feita para que se obtivessem os períodos mais semelhantes em tamanho e defasagem em função da distribuição temporal das observações.

Verificamos que as variações de Semidiâmetro dentro dos grupos têm uma distribuição próxima da normal. As médias do Semidiâmetro Solar dentro de cada grupo têm também uma distribuição próxima da normal com respeito aos desvios para a tendência linear temporal de cada série. As Figuras 16 e 17 mostram as distribuições das variações do Semidiâmetro a leste e a oeste dos grupos comparando-os com uma distribuição normal. Todos os pontos se afastam da normal menos de 0,4 desvios padrão sendo o desvio padrão deste afastamento igual a 0,11 a leste e 0,16 a oeste. A Figura 18 mostra as médias dos grupos a leste e a oeste, bem como as retas de tendência de cada uma. Na abscissa está a data média de cada período. A figura mostra a abrangência de cada grupo de dados e a equação de tendência de cada uma das séries. As curvas indicam que a tendência geral dos dados durante o ano é de um decréscimo em torno do milésimo de segundo de arco por dia. As tendências das curvas a leste e a oeste são bastante semelhantes. As médias a leste e as médias a oeste afastam-se em cerca de três décimos de segundo de arco. As Figuras 19 e 20 mostram os desvios das médias dos grupos para as retas de tendência, comparando-as com uma normal. Todos os pontos se afastam da normal menos de 0,4 desvios padrão sendo o desvio padrão deste afastamento igual a 0,22 a leste e 0,18 a oeste. Os valores de semidiâmetro são medidos em segundos de arco.

São, portanto, pontos oscilando aproximadamente de maneira normal em relação às retas de tendência. Tal oscilação é devida ao efeito dos erros ou de pequenos eventos locais. Temos duas curvas, uma de médias de valores observados a leste e outra a oeste, cada uma apresenta pontos que se afastam de quantidades diferentes das retas de tendência. Levando em consideração estas constatações impusemos uma correção a estes dados admitindo os seguintes critérios: a média a leste mais oeste ponderada pelo inverso dos quadrados dos desvios de leste e oeste às retas de tendência linear deve ser preservada, a correção de cada grupo deve ser proporcional ao quadrado de seu desvio para a reta de tendência. Uma vez que cada valor observado pode estar contido dentro de mais de um grupo de pontos, para cada valor adota-se por correção a média das correções dos grupos de que faz parte.

Antes, porém, de se fazer as correções assim propostas, um exame dos dados mostra que a partir de outubro de 2002, ou seja, durante os meses de novembro e dezembro, fica evidente uma mudança da tendência decrescente da curvas de leste para uma tendência crescente. Esta constatação fica mais relevante ao se observar os valores observados em 2003 que têm, em ambas as séries, uma tendência crescente. Os valores a partir de outubro estão mais de acordo com a série de valores observados em 2003 e é bastante razoável estudá-los dentro daquele outro conjunto. Além disto, há uma pausa de 15 dias nas observações entre os dias 30 de outubro e 13 de novembro, incluídos os extremos. Não houve observações no Astrolábio nestas duas semanas, o que acrescenta motivos a se considerar estes últimos valores de 2002 anexados à série de valores do ano que lhe segue e analisá-los dentro daquele outro conjunto sem que haja qualquer perda para a análise dos dados. O principal intuito é, caso a tendência dos dados se revele real, não analisá-los num contexto com tendência oposta a sua, mas em outro contexto onde a tendência geral é coincidente com a sua.

Estudamos, então, a série de valores observados do Semidiâmetro Solar de 2002 dividindo-a em duas partes. Analisando inicialmente os valores que vão até o final de outubro e anexando os valores observados em novembro e dezembro aos valores observados em 2003 e estudando-os no conjunto daqueles outros dados.

As médias dos grupos de dados das séries reduzidas de observações do Semidiâmetro Solar de dezembro de 2001 a outubro de 2002 podem ser vistas na Figura 21. Nela são mostradas as retas de tendência linear que se ajustam aos valores médios dos 33 grupos restantes. A tendência é decrescente para ambas. Para leste a variação é de –1,092 milisegundos de arco por dia, para oeste é de –0,768 milisegundos de arco por dia.

Os valores de correção para os pontos dentro de cada grupo, de acordo com a proposta descrita acima, foram calculados com base nos desvios de cada média de grupo às retas de tendência: [y = -1,09220E-03x + 9,62022E+02], para leste e [y = -7,67784E-04x + 9,60951E+02], para oeste. Na Tabela III estão as datas de início e de final de cada grupo, os desvios de cada grupo para a reta de tendência e os valores de correção obtidos.

**Tabela III – Correções impostas aos grupos de dados de 2002 em função de seus desvios às retas de tendência.**

|    | Início do período | Final do período | Desvios |        | Correções |        |
|----|-------------------|------------------|---------|--------|-----------|--------|
|    | (data Juliana)    |                  | (segundos de arco) |        |           |        |
| 1  | 2263,7 | 2297,7 | -0,175 | 0,148  | -0,001 | -0,001 |
| 2  | 2278,7 | 2307,7 | -0,148 | 0,054  | -0,085 | -0,011 |
| 3  | 2294,7 | 2322,7 | -0,098 | 0,114  | 0,007  | 0,009  |
| 4  | 2304,7 | 2332,7 | -0,129 | 0,104  | -0,011 | -0,007 |
| 5  | 2311,7 | 2339,7 | -0,166 | -0,024 | -0,164 | -0,004 |
| 6  | 2319,7 | 2347,7 | -0,128 | -0,063 | -0,122 | -0,030 |
| 7  | 2328,7 | 2356,7 | -0,108 | -0,054 | -0,123 | -0,031 |
| 8  | 2338,7 | 2366,7 | 0,005  | -0,053 | 0,003  | 0,357  |
| 9  | 2346,7 | 2374,7 | 0,043  | -0,040 | -0,014 | -0,012 |
| 10 | 2352,7 | 2380,7 | 0,090  | -0,114 | 0,045  | 0,072  |
| 11 | 2361,7 | 2389,7 | 0,174  | -0,103 | -0,205 | -0,072 |
| 12 | 2368,7 | 2396,7 | 0,165  | -0,070 | -0,317 | -0,056 |
| 13 | 2375,7 | 2403,7 | 0,190  | -0,028 | -0,481 | -0,011 |
| 14 | 2382,7 | 2410,7 | 0,156  | 0,077  | -0,181 | -0,044 |
| 15 | 2394,7 | 2422,7 | 0,218  | 0,042  | -0,415 | -0,016 |
| 16 | 2402,7 | 2430,7 | 0,197  | 0,072  | -0,279 | -0,037 |
| 17 | 2410,7 | 2438,7 | 0,178  | 0,004  | -0,458 | 0,000  |
| 18 | 2419,7 | 2447,7 | 0,024  | 0,004  | -0,283 | -0,006 |
| 19 | 2429,7 | 2457,7 | -0,080 | -0,044 | -0,100 | -0,030 |
| 20 | 2444,7 | 2472,7 | -0,057 | -0,075 | 0,029  | 0,052  |
| 21 | 2452,7 | 2480,7 | 0,034  | -0,070 | 0,044  | 0,187  |
| 22 | 2458,7 | 2486,7 | 0,065  | -0,097 | 0,051  | 0,113  |
| 23 | 2469,7 | 2497,7 | 0,154  | -0,052 | -0,336 | -0,038 |
| 24 | 2479,7 | 2507,7 | 0,072  | -0,034 | -0,187 | -0,043 |
| 25 | 2486,7 | 2514,7 | 0,064  | 0,000  | -0,324 | 0,000  |
| 26 | 2495,7 | 2524,7 | -0,001 | 0,022  | 0,001  | 0,230  |
| 27 | 2502,7 | 2530,7 | -0,063 | 0,010  | -0,167 | -0,004 |
| 28 | 2508,7 | 2536,7 | -0,036 | -0,003 | -0,215 | -0,002 |
| 29 | 2515,7 | 2544,7 | -0,066 | 0,003  | -0,181 | 0,000  |
| 30 | 2523,7 | 2551,7 | -0,158 | 0,019  | -0,067 | -0,001 |
| 31 | 2534,7 | 2562,7 | -0,095 | 0,066  | -0,020 | -0,009 |
| 32 | 2542,7 | 2570,7 | -0,194 | 0,082  | 0,020  | 0,004  |
| 33 | 2549,7 | 2577,7 | -0,111 | 0,099  | -0,002 | -0,001 |

Lembrando que a correção de cada ponto considera a média das correções dos grupos a que pertence o ponto, as correções a leste foram quase todas negativas variando entre +0,02 e –0,38 segundos de arco com uma média de 0,11 segundos de arco. As correções a oeste foram menores variando entre +0,21 e –0,05 segundos de arco com uma média menor que 0,01 segundos de arco. As Figuras 22 e 23 mostram os histogramas das correções. Pode-se ver que as correções a leste foram bem maiores que as correções a oeste ficando a grande maioria destas entre –0,05 e 0,00 e a grande maioria daquelas entre –0,20 e 0,00 segundos de arco. As Figuras 24 e 25 mostram as correções distribuídas temporalmente. Pode-se ver que elas concentram-se em determinados períodos, onde as médias dos valores observados mais se afastaram das retas de tendência. A Figura 26 mostra o aspecto final da curva de Semidiâmetro Solar em função do tempo após esta correção onde se nota que as curvas de valores a leste e a oeste do meridiano se aproximaram embora a primeira ainda esteja acima da outra em quase todo o período.

.A segunda parte de nossa análise verifica a influência que parâmetros de observação medidos exercem sobre os dados observados. No próximo item tratamos dela.

**Figura 5 – Número de observações do Semidiâmetro do Sol feitas ao longo dos meses do ano de 2002. Valores acima do zero são observações a leste do meridiano. Valores abaixo do zero são observações a oeste deste.**

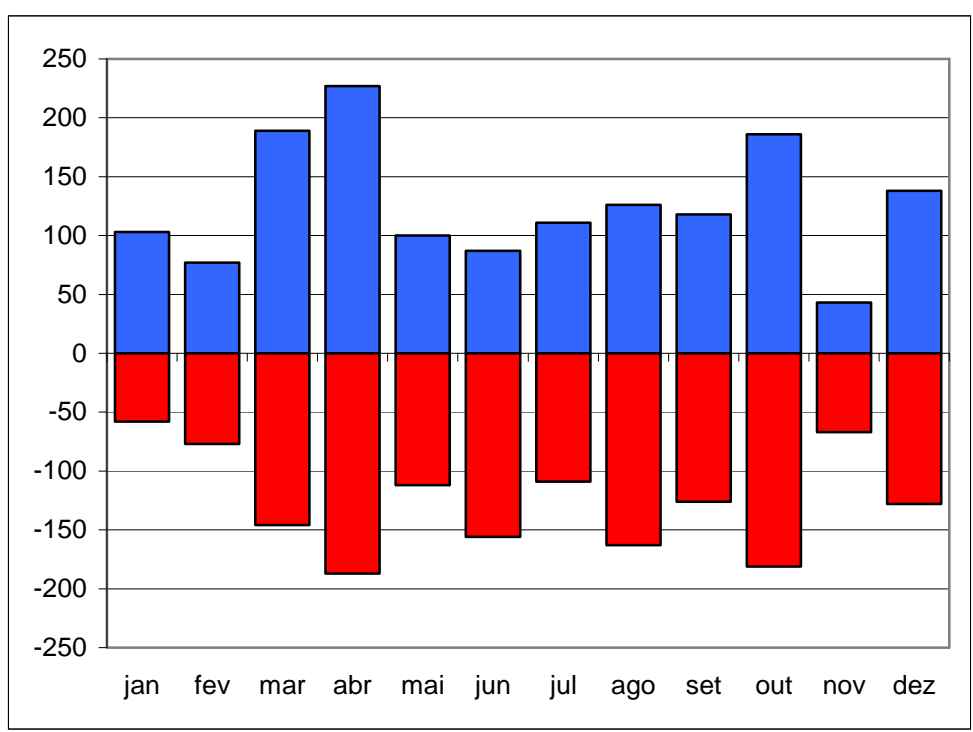

**Figura 6 – Os dados de Semidiâmetro Solar observados em 2002 e as retas que se ajustam às duas séries. Acima a da série a leste e abaixo a da série oeste.**

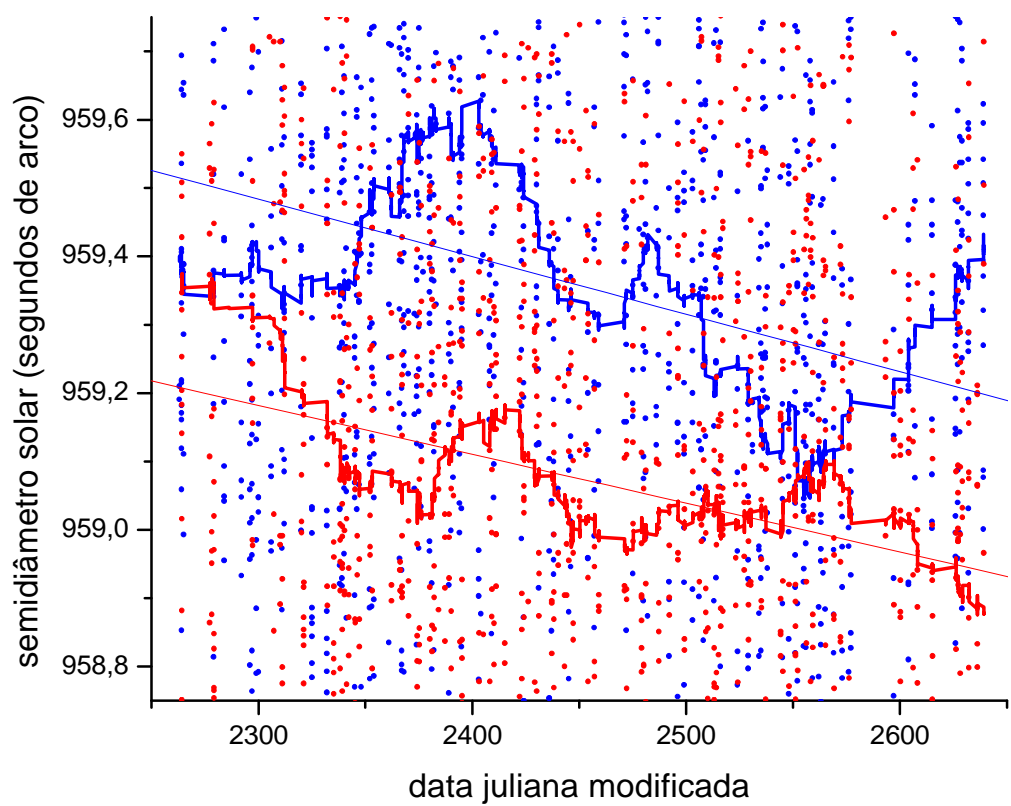

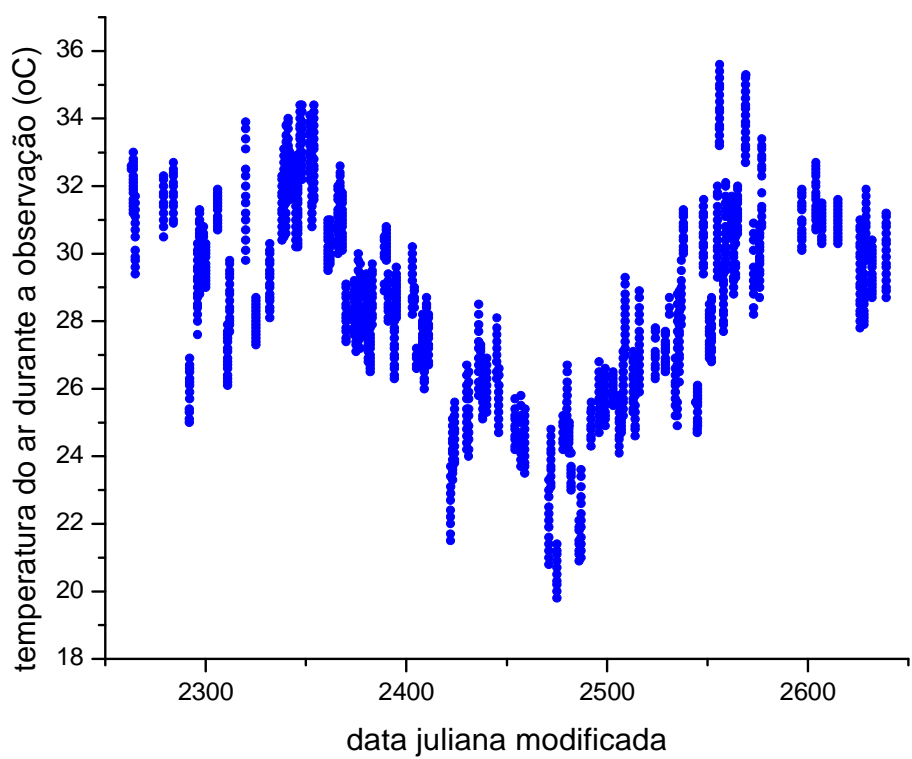

**Figura 7 – Variação anual da temperatura do ar durante as observações no ano de 2002 para os valores a leste do meridiano.**

**Figura 8 – Variação anual da temperatura do ar durante as observações no ano de 2002 para os valores a oeste do meridiano.**

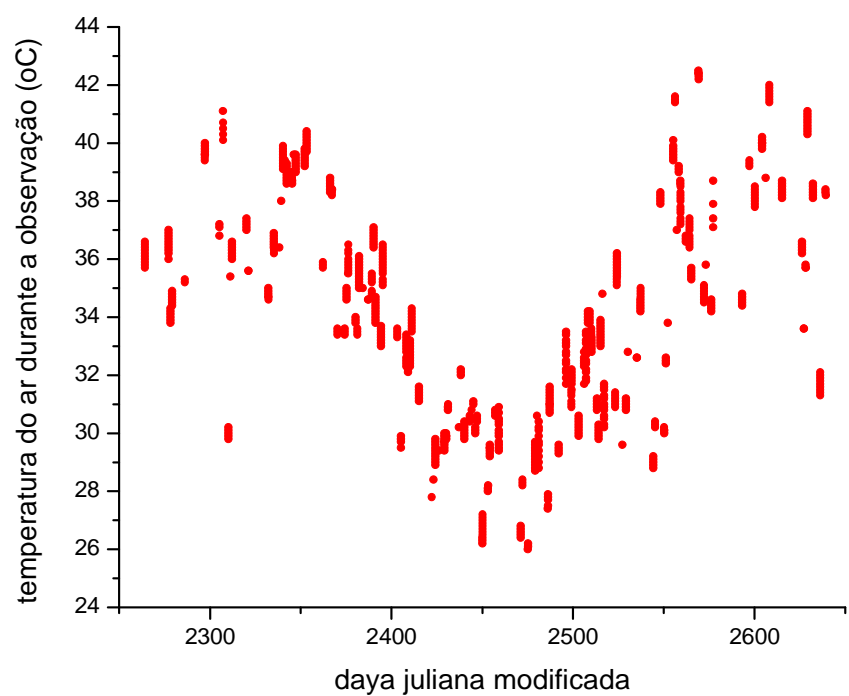

**Figura 9 – Horário das duas séries de observações: antes e depois da passagem meridiana.**

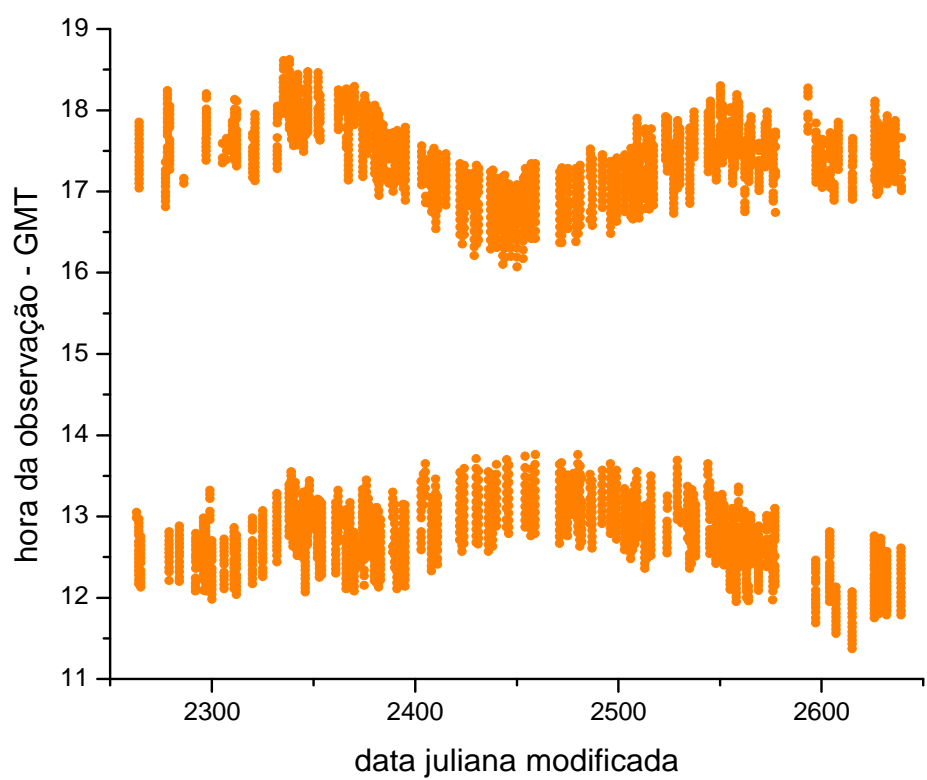

**Figura 10 – Duração das observações ao longo do ano de 2002 para os dados observados a leste do meridiano.**

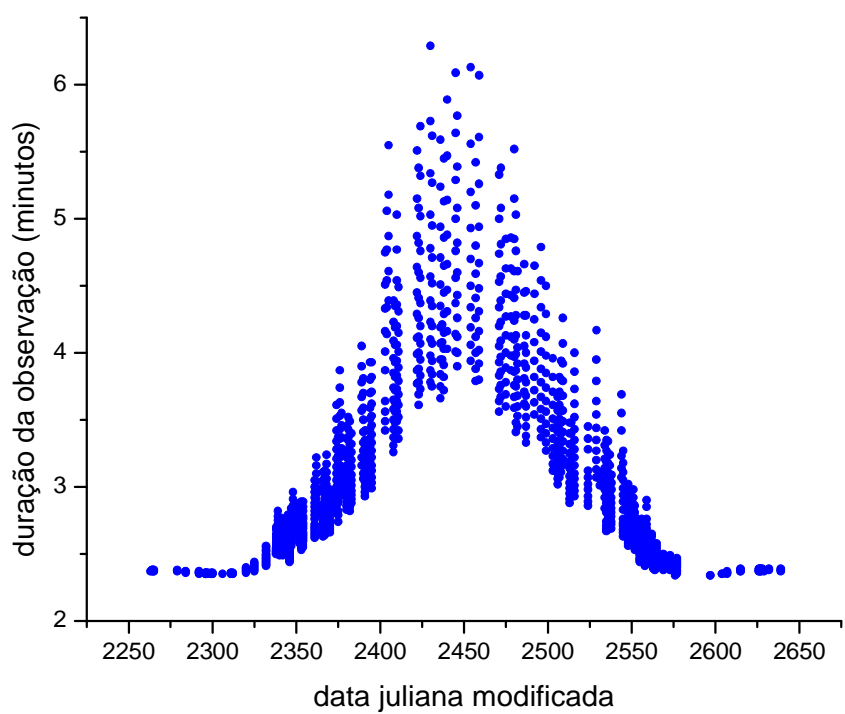

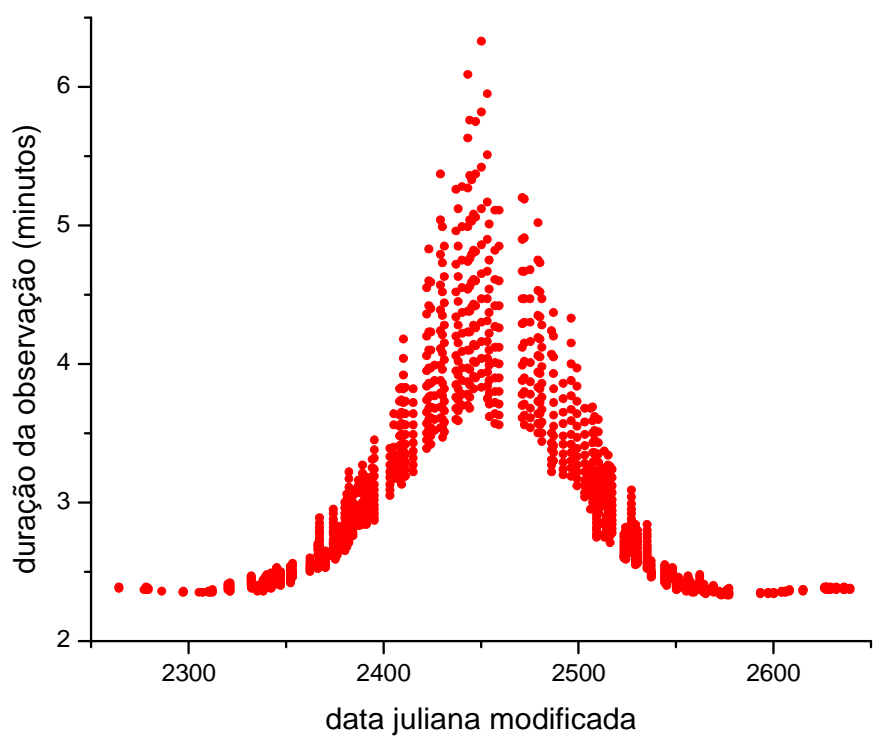

**Figura 11 – Duração das observações ao longo do ano de 2002 para os dados observados a oeste do meridiano.**

**Figura 12 – Variação anual da distância zenital observada durante o ano de 2002 para os valores observados a leste do meridiano.**

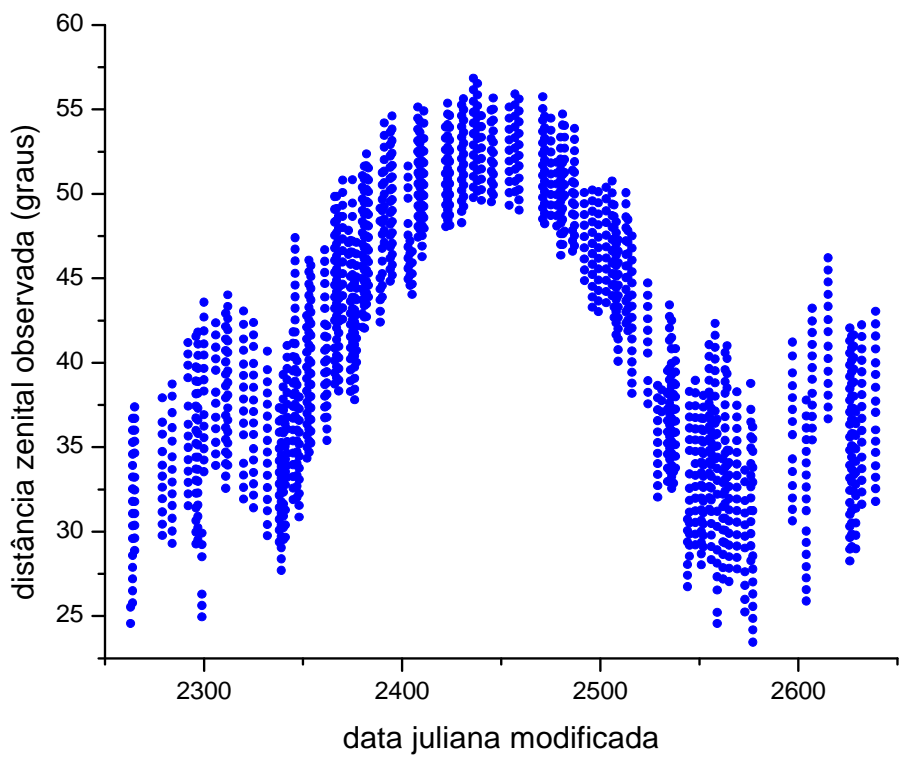

**Figura 13 – Variação anual da distância zenital observada durante o ano de 2002 para os valores observados a oeste do meridiano.**

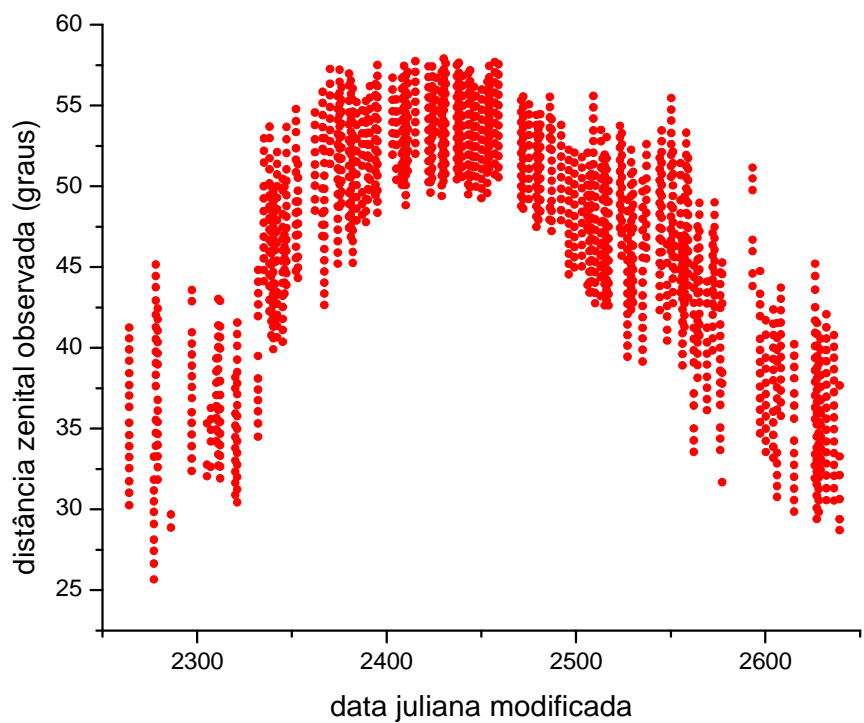

**Figura 14 – Variação anual do azimute observado durante o ano de 2002 para os valores a leste do meridiano. O valor do gráfico é o módulo da diferença entre a direção norte e o azimute observado.**

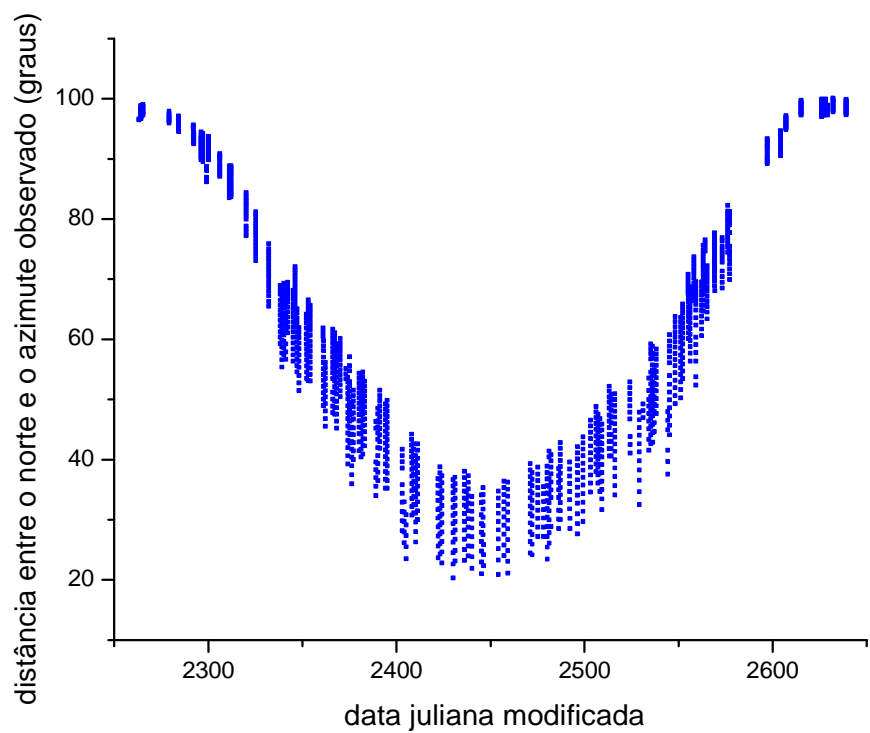

**Figura 15 – A variação anual do azimute observado durante o ano de 2002 para os valores a oeste do meridiano. O valor do gráfico é o módulo da diferença entre a direção norte e o azimute observado.**

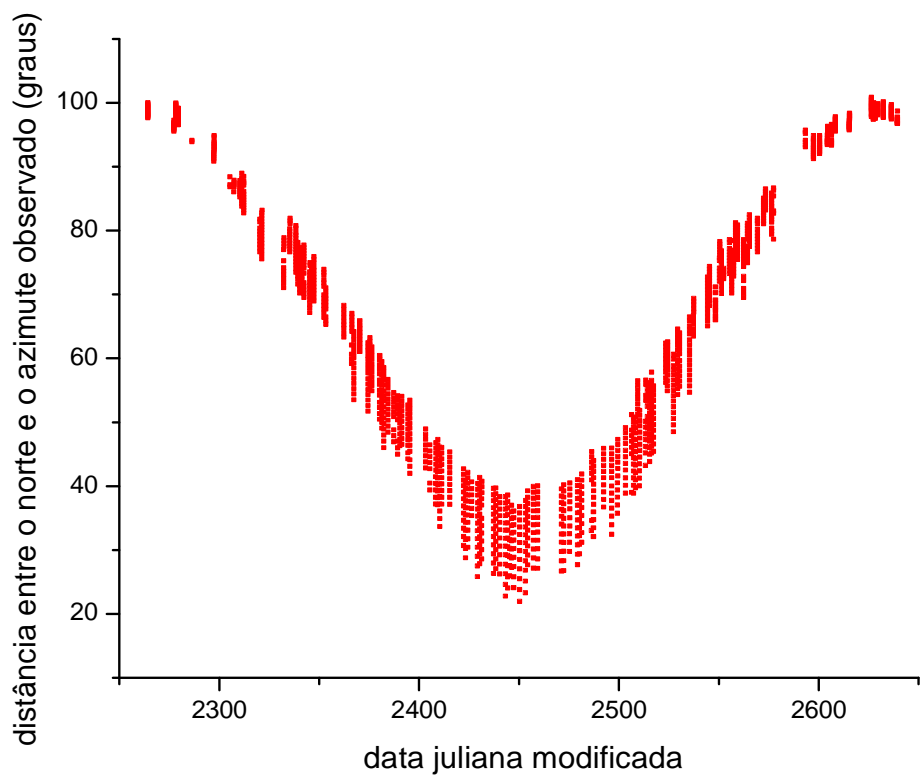

**Figura 16 - Distribuições das variações do Semidiâmetro a leste e comparação com a distribuição normal. À direita os desvios entre os pontos e a normal com a marca de um desvio padrão à esquerda e à direita da distribuição de desvios.**

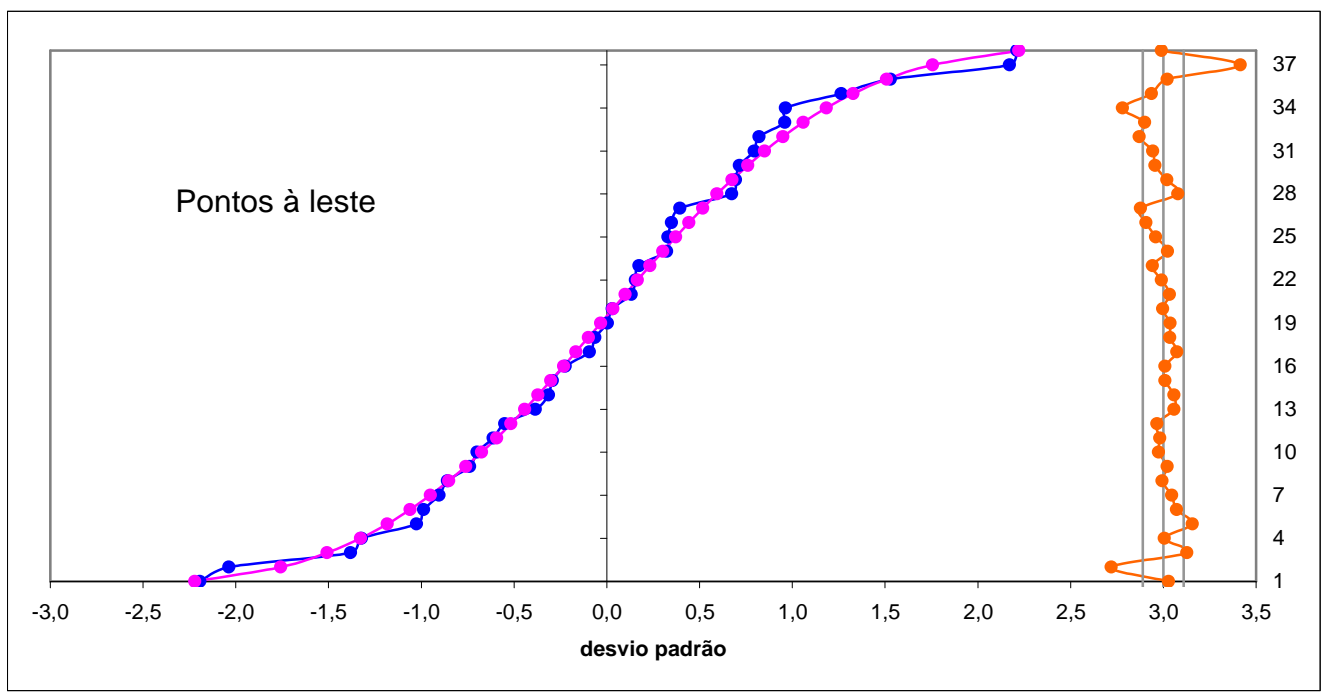

**Figura 17 - Distribuições das variações do Semidiâmetro a oeste e comparação com a distribuição normal. À direita os desvios entre os pontos e a normal com a marca de um desvio padrão à esquerda e à direita da distribuição de desvios.**

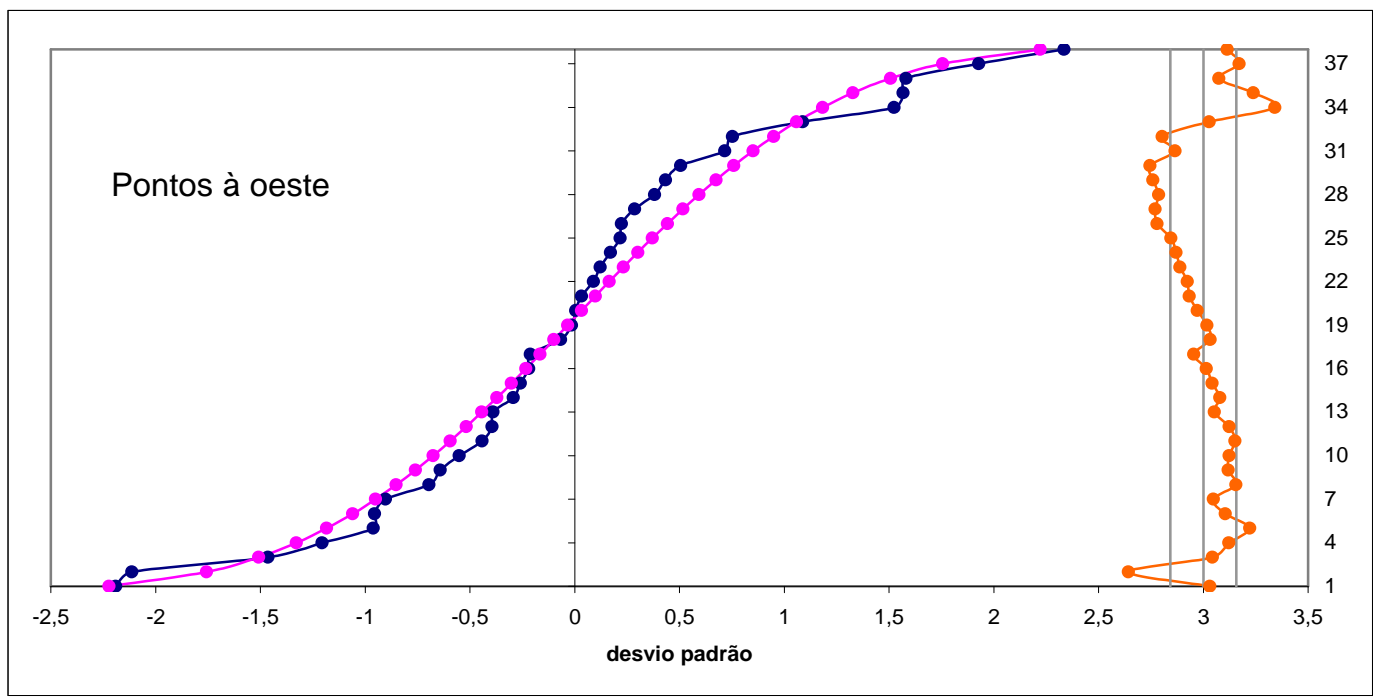

**Figura 18 – Média dos valores observados dentro de cada grupo. Mostra-se a abrangência temporal de cada grupo e uma linha de tendência para os grupos a leste (acima) e para os grupos a oeste (em baixo). As linhas de tendência têm sua equação exibida.**

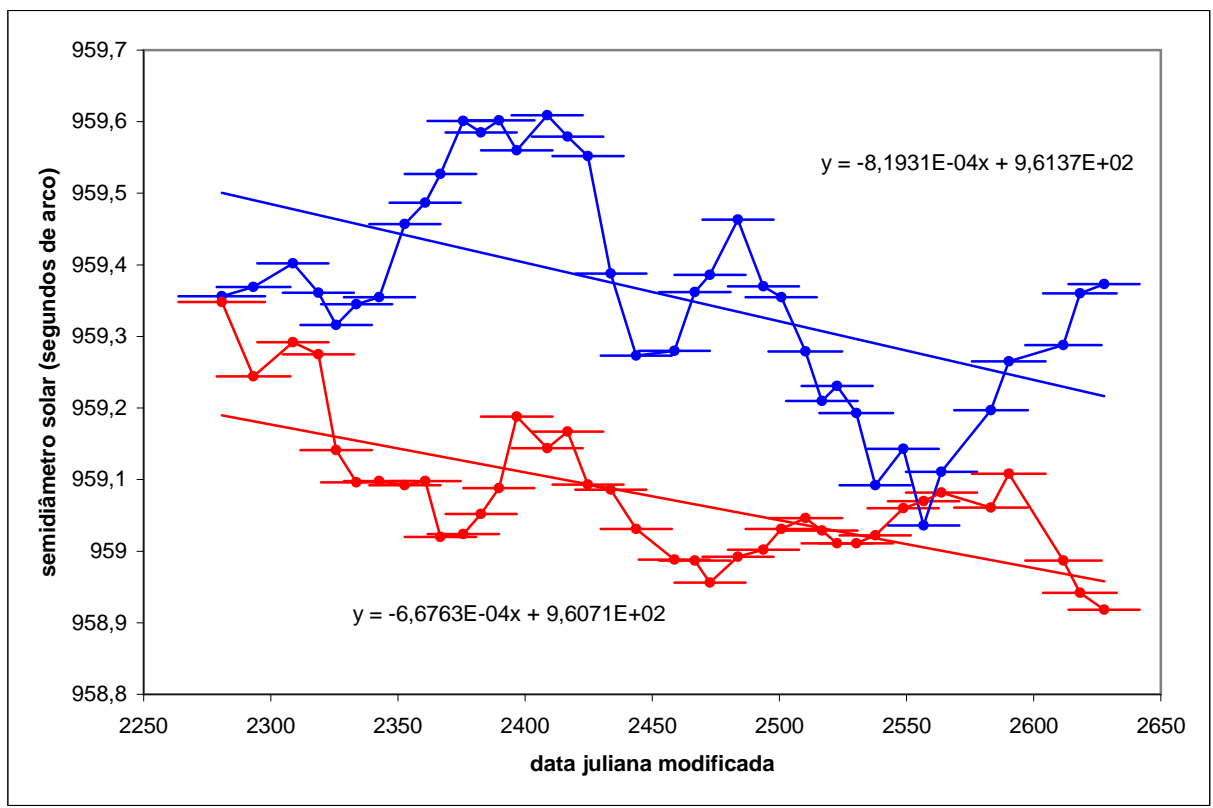

**Figura 19 - Distribuições das médias de Semidiâmetro dos grupos a leste e comparação com a distribuição normal. À direita os desvios entre os pontos e a normal marcando-se um desvio padrão à esquerda e à direita da distribuição de desvios.**

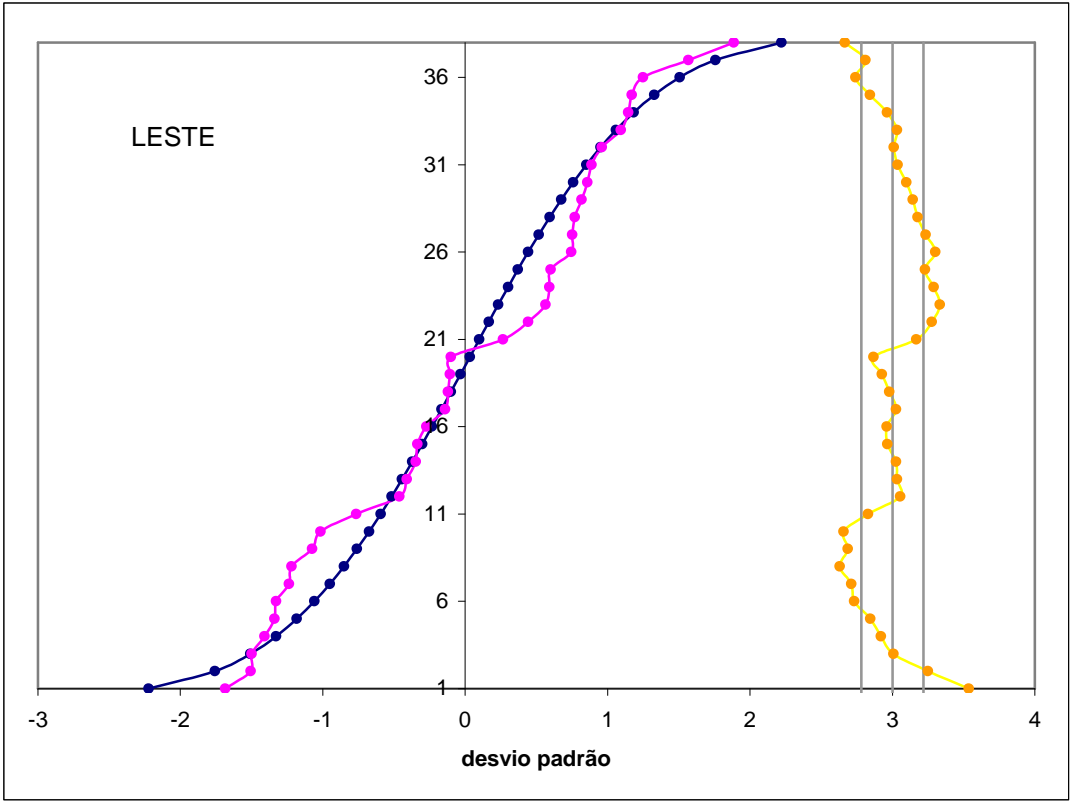

**Figura 20 - Distribuições das médias de Semidiâmetro dos grupos a oeste e comparação com a distribuição normal. À direita os desvios entre os pontos e a normal marcando-se um desvio padrão à esquerda e à direita da distribuição de desvios.**

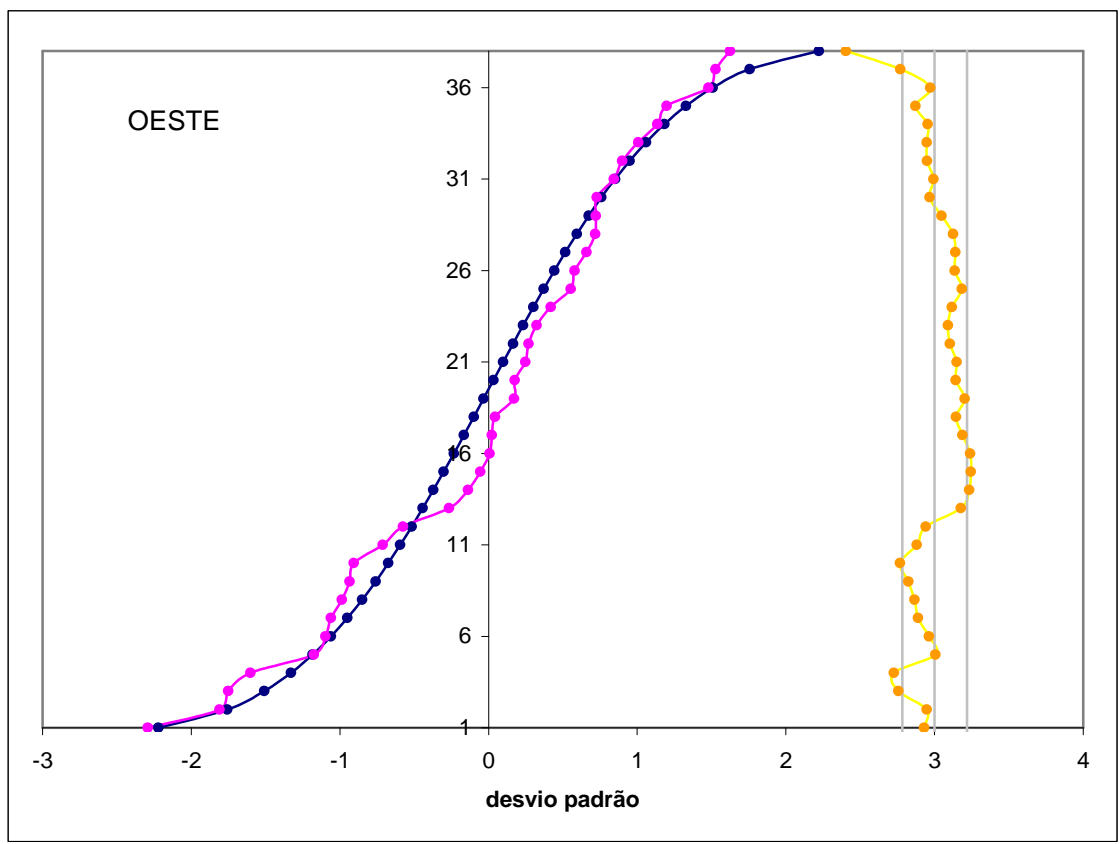

**Figura 21 - Médias dos grupos da série reduzida de 2003 e suas retas de tendência que têm a equação mostrada.**

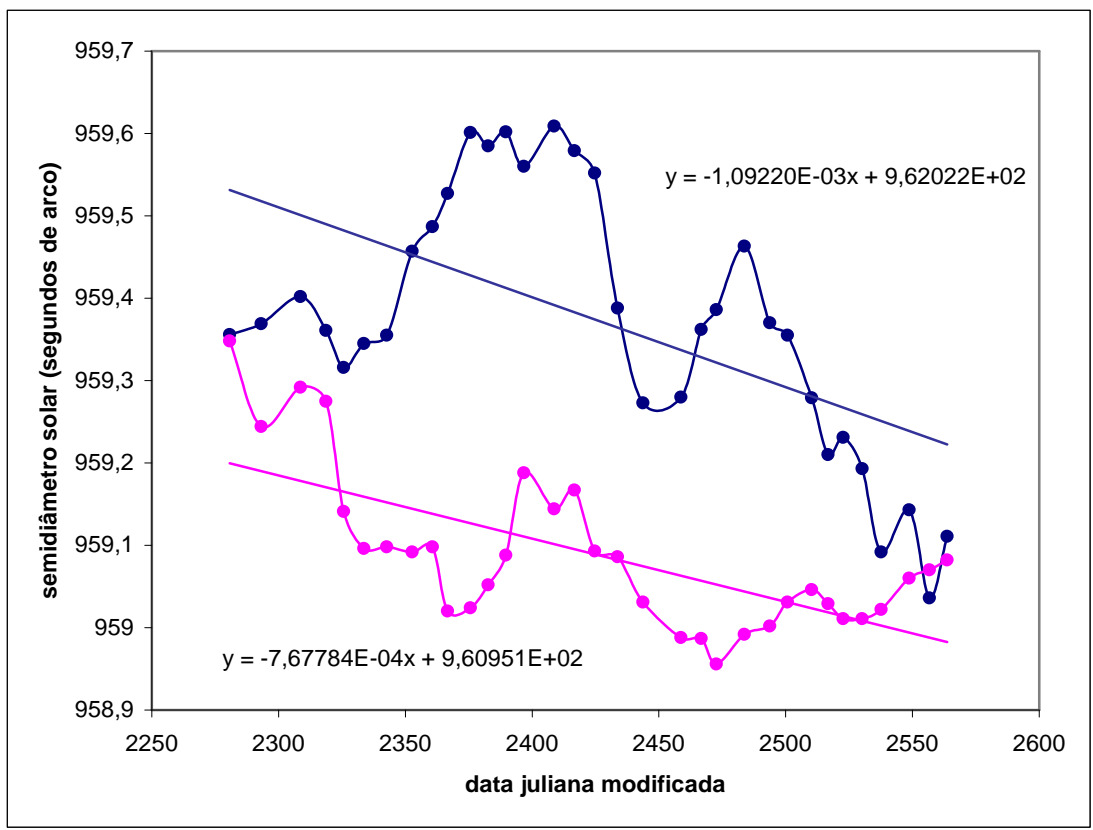

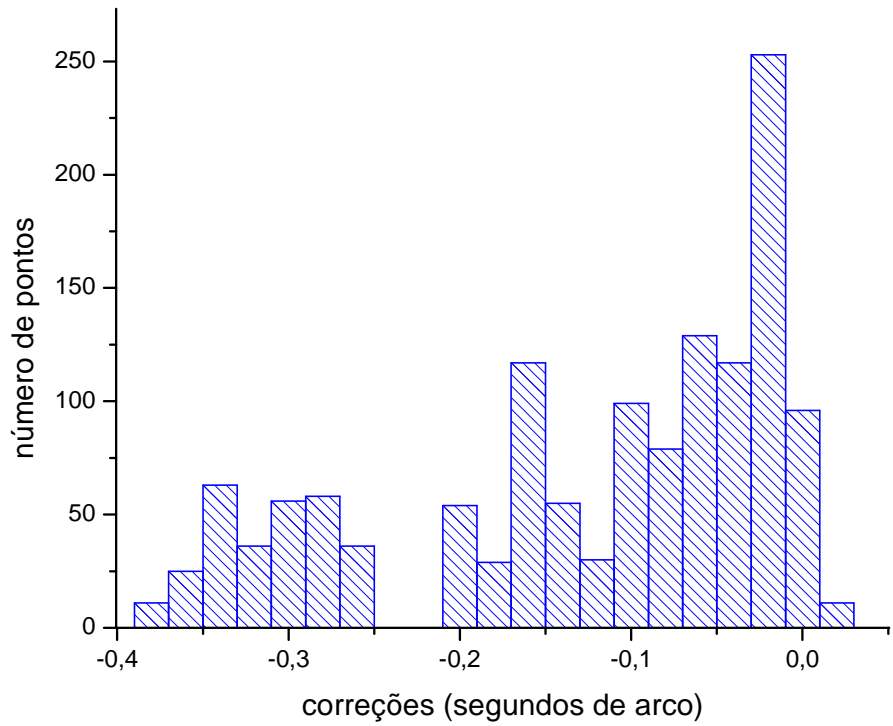

**Figura 22 - Histograma da primeira correção feita aos valores de 2002 observados a leste.**

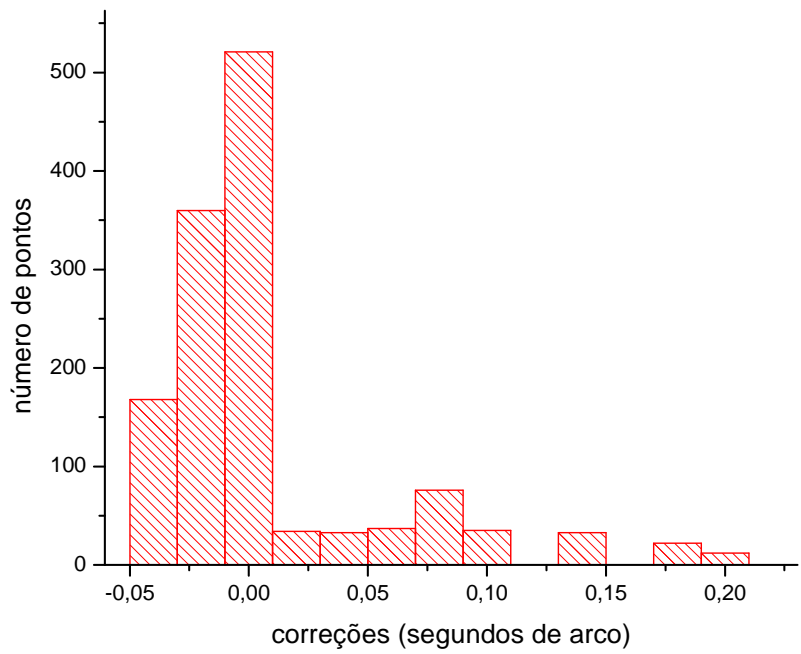

**Figura 23 - Histograma da primeira correção feita aos valores de 2002 observados a oeste.**

**Figura 24 – Primeira correção imposta aos valores de 2002 a leste em função do tempo.**

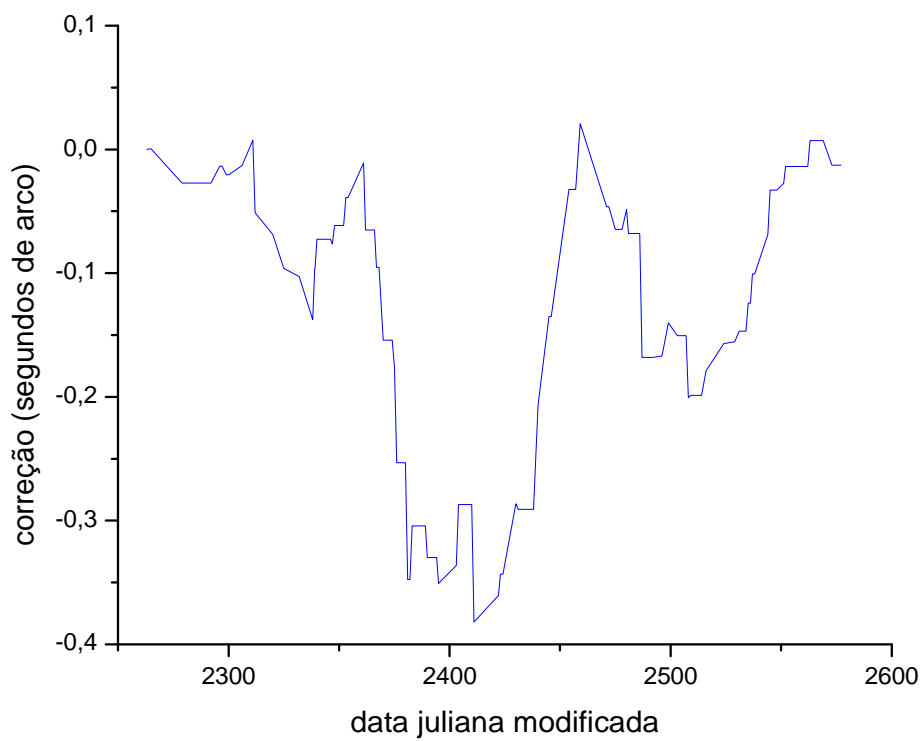

**Figura 25 – Primeira correção impostas aos valores de 2002 a oeste em função do tempo.**

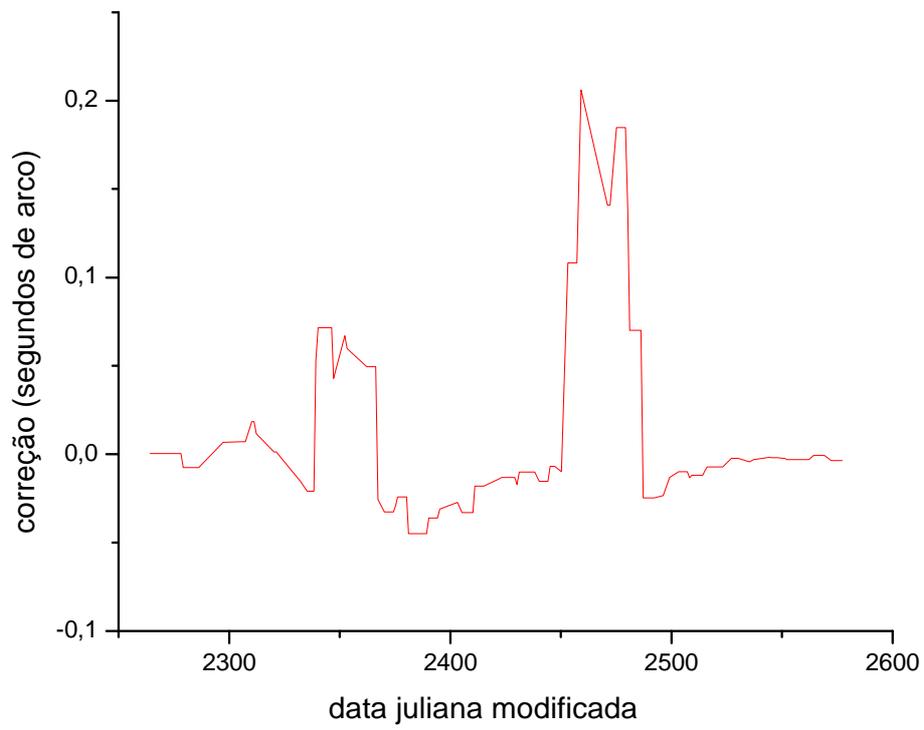

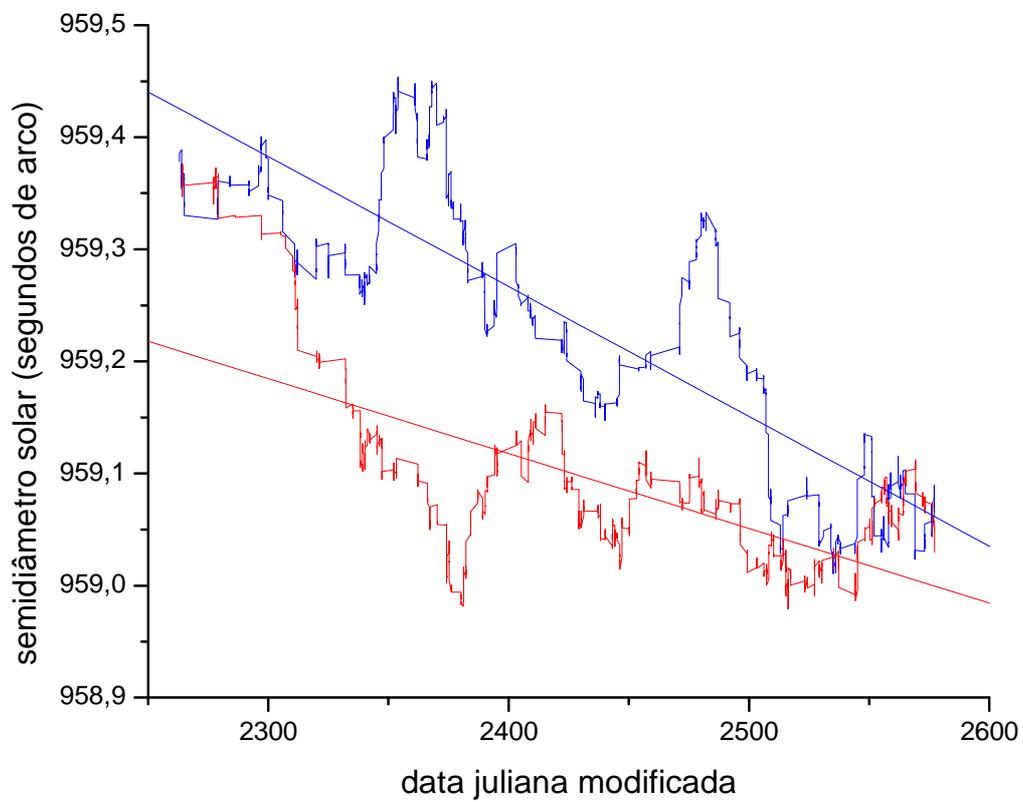

**Figura 26 – Semidiâmetro Solar em função do tempo após a primeira correção dos dados de 2002. A curva superior é dos valores a leste, a inferior dos valores a oeste.**

**CORREÇÃO DOS PARÂMETROS DAS OBSERVAÇÕES PARA 2002.**

Para cada observação do Semidiâmetro Solar temos disponível uma série de dados adicionais que são chamados de parâmetros da observação. Cada um destes parâmetros pode separadamente, ou em conjunto com outros, influir e alterar os valores medidos. Selecionamos vários destes parâmetros e verificamos separadamente e em conjunto, de que forma estes parâmetros poderiam estar influindo nos resultados. Para tal análise foram selecionados: a data Juliana, a distância zenital de observação do Sol, o azimute de observação do Sol, o parâmetro de Fried, o desvio padrão do ajuste da parábola ao bordo direto do Sol, o desvio padrão do ajuste da parábola ao bordo refletido do Sol, a temperatura do ar no instante médio da observação, a variação da temperatura durante a observação e a pressão atmosférica no instante médio da observação.

A possível influência destes parâmetros é verificada por meio do cálculo de correlações dos pontos observados com os parâmetros propostos verificando-se o valor do coeficiente angular da relação linear e o desvio padrão do ajuste. Os pontos são assim verificados, no seu conjunto total, nos subconjuntos a leste e a oeste e num conjunto total onde se tomam os pontos a leste e os pontos a oeste com sinais invertidos, procurando-se assim estabelecer a existência de assimetria nas influências de observações em lados opostos. O ajuste de retas aos dados é executado com a utilização da técnica de mínimos quadrados. Nestes cálculos é retirada sempre a tendência linear da evolução temporal do Semidiâmetro. Os dados de Semidiâmetro Solar utilizados nesta análise são os valores observados e já corrigidos pela análise anterior.

Para se fazer estes cálculos, todos os parâmetros são inicialmente normalizados, isto é, diminuídos de sua média e divididos por seu desvio padrão. Os cálculos assim realizados mostram que cinco dos parâmetros inicialmente selecionados, têm alguma influência nos dados de Semidiâmetro do Sol, uma vez que os coeficientes angulares das retas ajustadas têm valores relevantes e seus desvios padrão são, no mínimo, um terço menores que os valores.

Esta análise nos conduz a selecionar para correção, quatro parâmetros que são: o fator de Fried, o desvio padrão do ajuste da parábola ao bordo direto do Sol, a temperatura média do ar durante a observação e a diferença de temperatura durante a observação. O quinto parâmetro é a data Juliana. Sua influência nos valores observados não configura um erro, mas uma real variação temporal da série de dados. Como estes parâmetros influenciam de maneira diferente os dados observados a leste e os dados observados a oeste, calculamos valores diferentes para cada um dos lados. A Tabela IV mostra os resultados, nela estão os coeficientes angulares das retas que se ajustam aos dados do Semidiâmetro Solar tomados em função dos parâmetros analisados. Os valores estão em segundos de arco e correspondem ao desvio que cada um dos parâmetros causa ao Semidiâmetro Solar. Os valores se referem aos parâmetros normalizados, isto é, diminuídos de sua média e divididos por seu desvio padrão. Os desvios causados ao Semidiâmetro Solar são os produtos dos valores da tabela multiplicados pelo parâmetro normalizado.

**Tabela IV – Influência dos parâmetros relevantes no Semidiâmetro Solar. Valores em segundos de arco. Parâmetros normalizados.**

| PARÂMETRO | VALORES A LESTE | VALORES A OESTE |
|---|---|---|
| **Fator de Fried** | 0,046 | 0,027 |
| **Desvio do Bordo** | 0,113 | -0,002 |
| **Temperatura** | 0,005 | 0,014 |
| **Diferença de Temperatura** | 0,042 | 0,008 |

Os valores de Semidiâmetro corrigidos passam a ser os valores anteriores somados à correção, que é o desvio calculado acima com sinal invertido e multiplicado pelo valor normalizado do parâmetro considerado. Assim procedendo, estamos retirando das observações a influência introduzida a elas pelo parâmetro em questão. As quatro correções, assim calculadas, são somadas para a obtenção da correção total do Semidiâmetro.

As correções dos valores a leste variam entre –0,340 e 0,199 e têm a média de -0,068, dos valores a oeste variam entre -0,34 e 0,323 e têm média de 0,069. As Figuras 27 e 28 mostram as correções impostas aos valores de Semidiâmetro e sua média corrida de 150 pontos. As Figuras 29 e 30 mostram os histogramas destas correções. E a Figura 31 mostra como ficou finalmente a série de dados do Semidiâmetro Solar após as duas correções implementadas. As correções promoveram uma redução significativa dos erros que afastavam as duas curvas, os valores a leste e a oeste estão agora bem mais próximos se afastando um pouco em alguns períodos embora as tendências lineares já não sejam tão semelhantes, mas mantêm as tendências decrescentes; a leste de –0,911 milisegundos de arco por dia e a oeste de –0,461 milisegundos de arco por dia.

**Figura 27 – Correções da influência dos parâmetros nos valores de Semidiâmetro Solar de 2002 observados a leste do meridiano e sua média corrida a cada 150 pontos.**

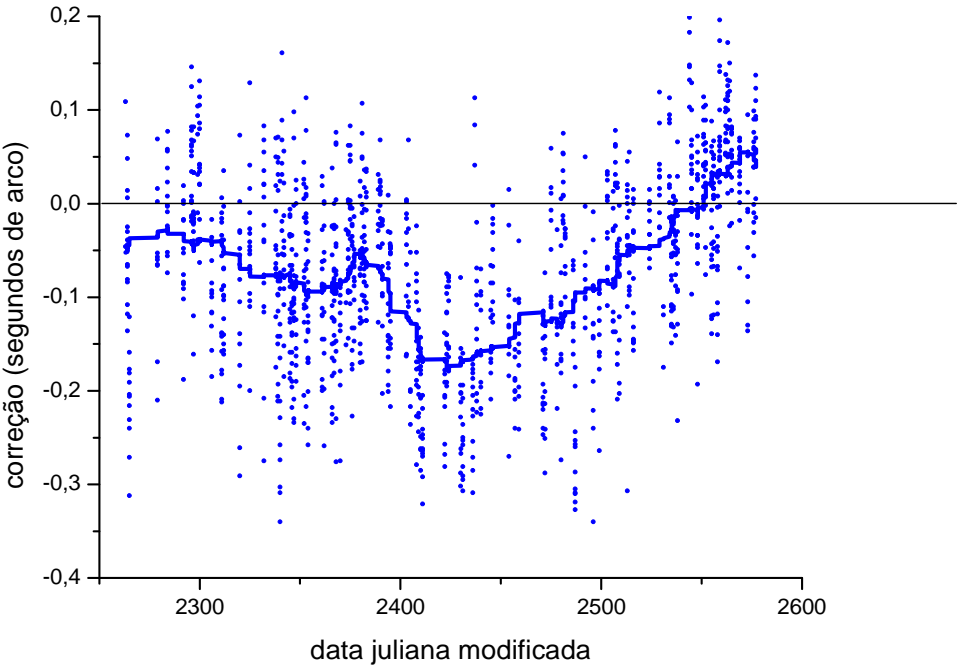

**Figura 28 – Correções da influência dos parâmetros nos valores de Semidiâmetro Solar de 2002 observados a oeste do meridiano e sua média corrida a cada 150 pontos.**

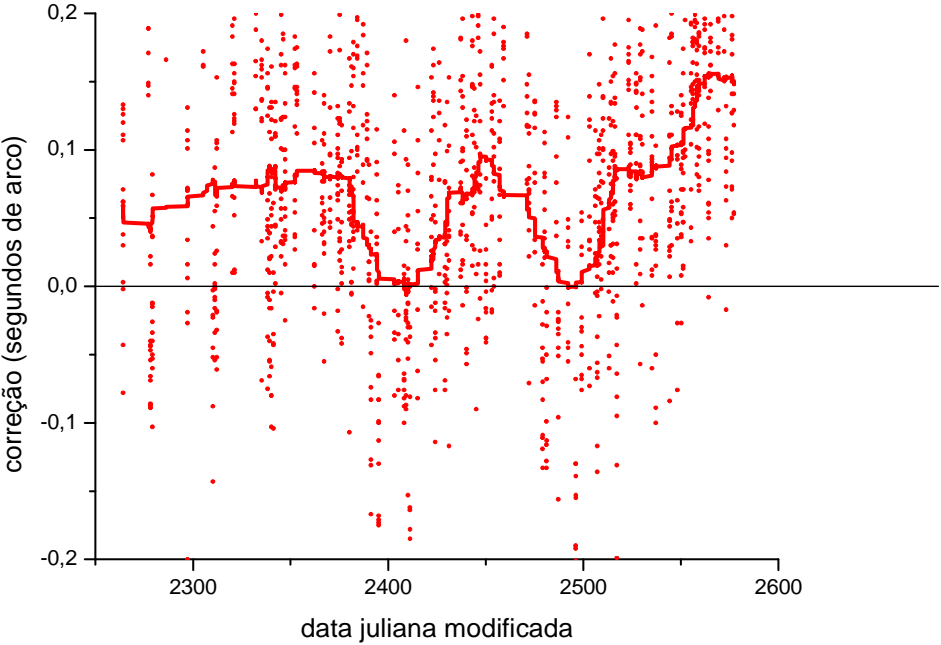

**Figura 29 – Histograma das correções da influência dos parâmetros aos valores observados de Semidiâmetro Solar a leste em 2002**

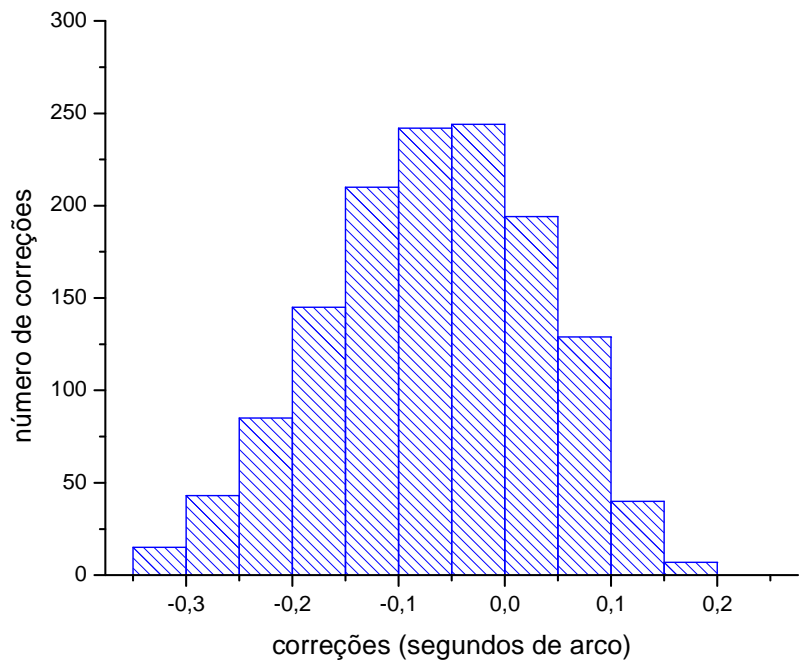

**Figura 30 – Histograma das correções da influência dos parâmetros aos valores observados de Semidiâmetro Solar a oeste em 2002**

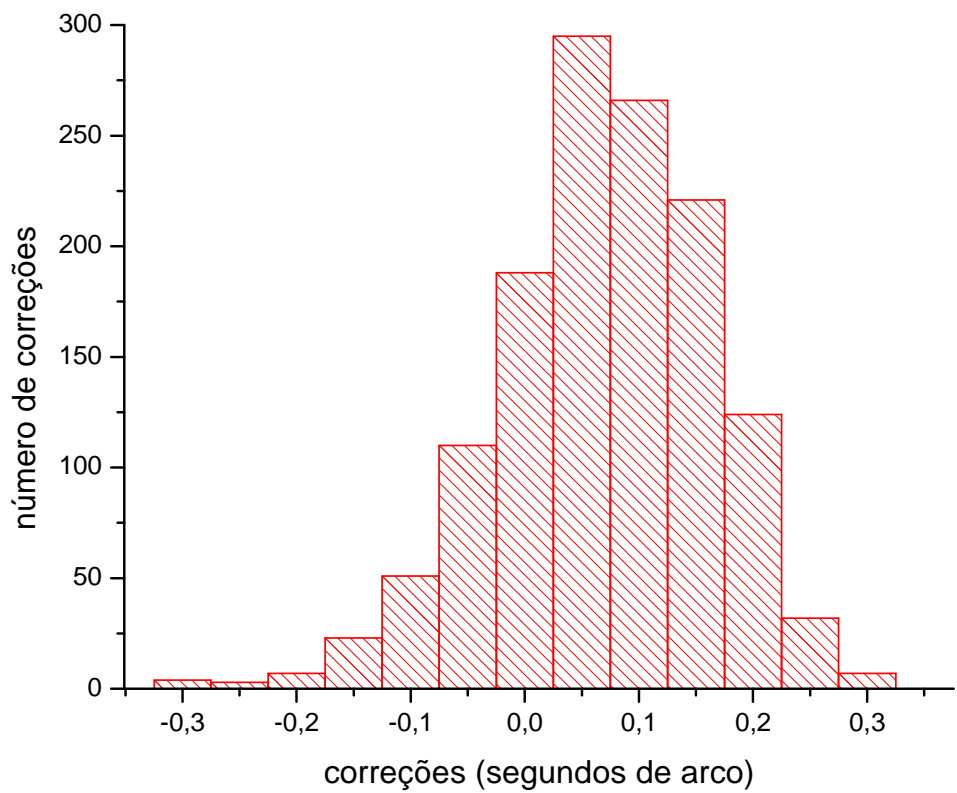

**Figura 31 – Média corrida de 150 pontos dos valores de Semidiâmetro Solar após as duas correções implementadas. A curva de leste é aquela que inicia e termina por baixo da outra.**

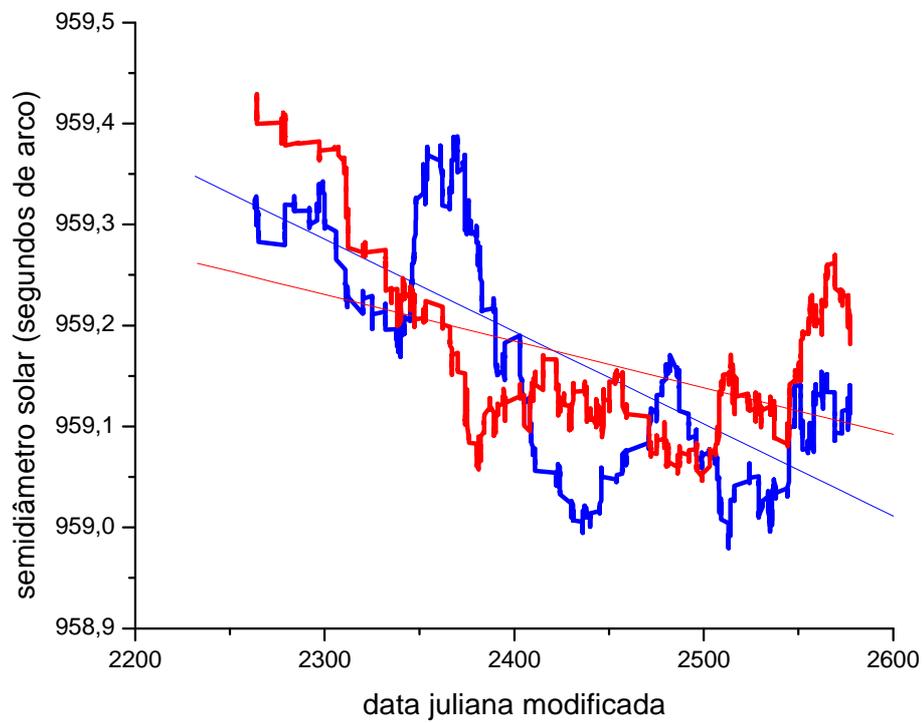

# CORREÇÃO DOS DADOS DE 2003

Os dados de 2003 se compõem de 2644 observações do Sol, sendo 1334 feitas a leste do meridiano local e 1310 a oeste deste. Estão incluídas nestes números, 181 observações a leste e 195 a oeste do meridiano que foram feitas nos meses de novembro e dezembro de 2002, e que foram adicionadas a esta série por força da análise anterior. A Figura 32 mostra a distribuição das observações ao longo dos meses de novembro de 2002 a dezembro de 2003, dividindo-as nas duas séries: os valores observados a leste, acima, e os observados a oeste do meridiano local, abaixo. As duas séries de valores do Semidiâmetro Solar podem ser vistas no gráfico da Figura 33 que mostra os valores em médias corridas de 150 pontos. Tal como em 2002, as duas séries apresentam-se destacadamente desviadas. Os valores a leste apresentam-se sempre acima dos valores a oeste. Foram traçadas duas retas que mostram a tendência linear das séries e que têm os seguintes valores: 0,2 milisegundos de arco por dia a leste e 0,1 milisegundos de arco por dia a oeste. Estes valores são bem diferentes, no entanto, no cômputo de um ano a sua diferença é de apenas quatro centésimos de segundo de arco o que é irrelevante.

As séries de observações do Semidiâmetro Solar tanto em 2002 como em 2003 apresentam valores a leste destacadamente acima dos valores a oeste. Tal fato não ocorreu com as séries anteriores que tinham as séries a leste e a oeste evoluindo dentro do mesmo espaço de valores. Isto passou a ocorrer depois que o Astrolábio Solar passou por reformas e certamente algum pequeno desajuste do alinhamento do prisma do instrumento vem causando esta separação. Entretanto como pôde ser visto nos resultados das correções impostas aos dados de 2002, tal anomalia foi significativamente reduzida quando impusemos as correções adequadas.

Analisamos as causas de erros da série de dados de 2003 utilizando a mesma abordagem utilizada para investigar os dados de 2002, isto é, dividindo a análise em duas etapas: uma primeira etapa de abordagem puramente estatística e uma segunda etapa analisando o comportamento das séries de valores em função de parâmetros de observação conhecidos e medidos durante as observações.

Na primeira parte da análise, utilizamos o mesmo procedimento que foi usado com sucesso para os dados de 2002. No presente caso, dividimos as duas séries de observações em 25 grupos temporalmente ao longo do ano. Os grupos se interpenetram, isto é, alguns pontos podem pertencer a mais de um grupo. Cada grupo foi escolhido de modo a perfazer no mínimo 30 dias de observações, ter início pelo menos doze dias após o início do grupo anterior e conter observações a leste e a oeste nos primeiros e últimos dias de sua duração de modo a se ter um bom cálculo da variação do Semidiâmetro Solar dentro do grupo. Dos 25 grupos, 22 abrangem 30 dias, um abrange 31 dias, um abrange 32 e um abrange 33 dias. Como na análise dos dados de 2002 a escolha do tamanho e da defasagem contempla aquela que obteve os períodos mais semelhantes em tamanho e defasagem em função da distribuição temporal das observações.

Da mesma forma que ocorreu com a série de 2002, decidimos retirar da série de 2003 o período final, neste caso, apenas o mês de dezembro. Embora as duas séries permaneçam separadas por pouco menos de quatro décimos de segundo de arco, a exclusão dos dados deste mês assegura às duas séries uma inclinação temporal ascendente bastante semelhante. Podemos anexar os dados do mês de dezembro aos dados de 2004 e estudá-los neste novo conjunto, sem que isto acarrete perdas para a análise. Assim procedendo, os grupos de dados foram reduzidos de 25 para 23, retirando-se da série estudada os dois grupos finais.

Tal como ocorre para os dados de 2002, as médias do Semidiâmetro Solar dentro de cada grupo de 2003 têm também uma distribuição próxima da normal com respeito aos desvios para a tendência linear temporal de cada série. As Figuras 34 e 35 mostram as distribuições das variações do Semidiâmetro dos grupos a leste e a oeste comparando-os com uma distribuição normal. Apenas um ponto se afasta da normal mais de 0,45 desvios padrão sendo o desvio padrão do afastamento igual a 0,26 a leste e 0,22 a oeste. Os valores de semidiâmetro são medidos em segundos de arco.

As médias dos 23 grupos de dados das séries reduzidas de observações do Semidiâmetro Solar de novembro de 2002 a novembro de 2003 podem ser vistas na Figura 36. Nela são mostradas as retas de tendência linear que se ajustam aos valores médios dos grupos. Para leste o coeficiente linear da reta é igual a 959,040 segundos de arco e o coeficiente angular é de +0,1274 milisegundos de arco por dia. Para oeste o coeficiente linear da reta é igual a 958,471 segundos de arco e o coeficiente angular é de +0,1994 milisegundos de arco por dia.

Os valores de correção para os pontos dentro de cada grupo foram calculados de acordo com a mesma proposta descrita para os dados de 2002. Tomaram por base os desvios da média de cada grupo às retas de tendência acima mencionadas. A Tabela V mostra a data de início e de final de cada grupo, os desvios de cada grupo para a reta de tendência e os valores de correção obtidos.

**Tabela V – Correções impostas aos grupos de dados de 2003 em função de seus desvios às retas de tendência.**

|  | Início do período | Final do período | Desvios | | Correções | |
|---|---|---|---|---|---|---|
|  | (data Juliana) | | (segundos de arco) | | | |
| 1 | 2596,5 | 2626,5 | -0,084 | -0,004 | -0,299 | -0,001 |
| 2 | 2613,5 | 2643,5 | -0,028 | -0,068 | 0,043 | 0,257 |
| 3 | 2627,5 | 2657,5 | -0,016 | -0,029 | 0,049 | 0,157 |
| 4 | 2641,5 | 2671,5 | -0,096 | 0,033 | -0,177 | -0,020 |
| 5 | 2655,5 | 2685,5 | 0,010 | -0,007 | -0,089 | -0,044 |
| 6 | 2669,5 | 2699,5 | -0,041 | 0,051 | 0,025 | 0,039 |
| 7 | 2683,5 | 2713,5 | -0,057 | 0,057 | 0,002 | 0,002 |
| 8 | 2697,5 | 2730,5 | 0,006 | -0,078 | 0,003 | 0,449 |
| 9 | 2718,5 | 2748,5 | -0,066 | 0,038 | -0,099 | -0,033 |
| 10 | 2730,5 | 2760,5 | -0,101 | 0,015 | -0,240 | -0,005 |
| 11 | 2746,5 | 2776,5 | 0,011 | 0,040 | 0,022 | 0,268 |
| 12 | 2759,5 | 2790,5 | 0,138 | -0,013 | -0,507 | -0,004 |
| 13 | 2774,5 | 2804,5 | 0,195 | -0,061 | -0,468 | -0,046 |
| 14 | 2786,5 | 2816,5 | 0,155 | -0,066 | -0,345 | -0,063 |
| 15 | 2800,5 | 2830,5 | 0,186 | 0,004 | -0,547 | 0,000 |
| 16 | 2816,5 | 2846,5 | 0,147 | 0,040 | -0,381 | -0,028 |
| 17 | 2841,5 | 2871,5 | -0,002 | 0,146 | 0,000 | 0,216 |
| 18 | 2872,5 | 2902,5 | 0,101 | 0,023 | -0,379 | -0,019 |
| 19 | 2886,5 | 2916,5 | 0,074 | -0,058 | -0,070 | -0,044 |
| 20 | 2901,5 | 2931,5 | -0,111 | -0,030 | -0,224 | -0,016 |
| 21 | 2914,5 | 2944,5 | -0,154 | 0,059 | -0,094 | -0,014 |
| 22 | 2928,5 | 2958,5 | -0,069 | -0,016 | -0,257 | -0,015 |
| 23 | 2940,5 | 2970,5 | -0,198 | -0,075 | -0,153 | -0,022 |

Estas correções foram implementadas aos valores observados de Semidiâmetro Solar. Como os valores podem estar dentro de mais de um grupo, a correção para cada valor é a média das correções dos grupos a que pertence. As correções a leste foram, na sua maioria, negativas ficando na faixa entre –0,49 e +0,05 segundos de arco com uma média de -0,19 segundos de arco. As correções a oeste, pelo contrário, foram quase todas positivas tendo ficado entre –0,05 e +0,26 segundos de arco com uma média de 0,03 segundos de arco. As Figuras 37 e 38 mostram o histograma das correções. Pode-se ver que as correções a leste foram bem maiores que as correções a oeste ficando, a grande maioria destas, entre –0,05 e 0,00 e, a grande maioria daquelas, entre –0,20 e 0,00 segundos de arco. Cabe, entretanto, notar que 400 pontos a leste, isto é, 28% do total, tiveram correções superiores a 0,3 segundos de arco. As Figuras 39 e 40 mostram estas correções distribuídas temporalmente onde se vê que elas se concentram em determinados períodos, onde as médias dos valores observados mais se afastaram das retas de tendência. A Figura 41 mostra o aspecto final da curva de Semidiâmetro Solar em função do tempo após esta correção. Nota-se que as duas curvas se aproximaram, mas, a principal mudança foi na tendência linear das curvas que passou a ser descendente.

Na segunda parte da análise, consideramos a influência dos parâmetros de observação conhecidos sobre os dados observados. No próximo item descrevemos este tratamento.

**Figura 32 – Número de observações do Semidiâmetro do Sol feitas de novembro de 2002 a dezembro de 2003. Valores acima do zero são observações a leste do meridiano. Valores abaixo do zero são observações a oeste.**

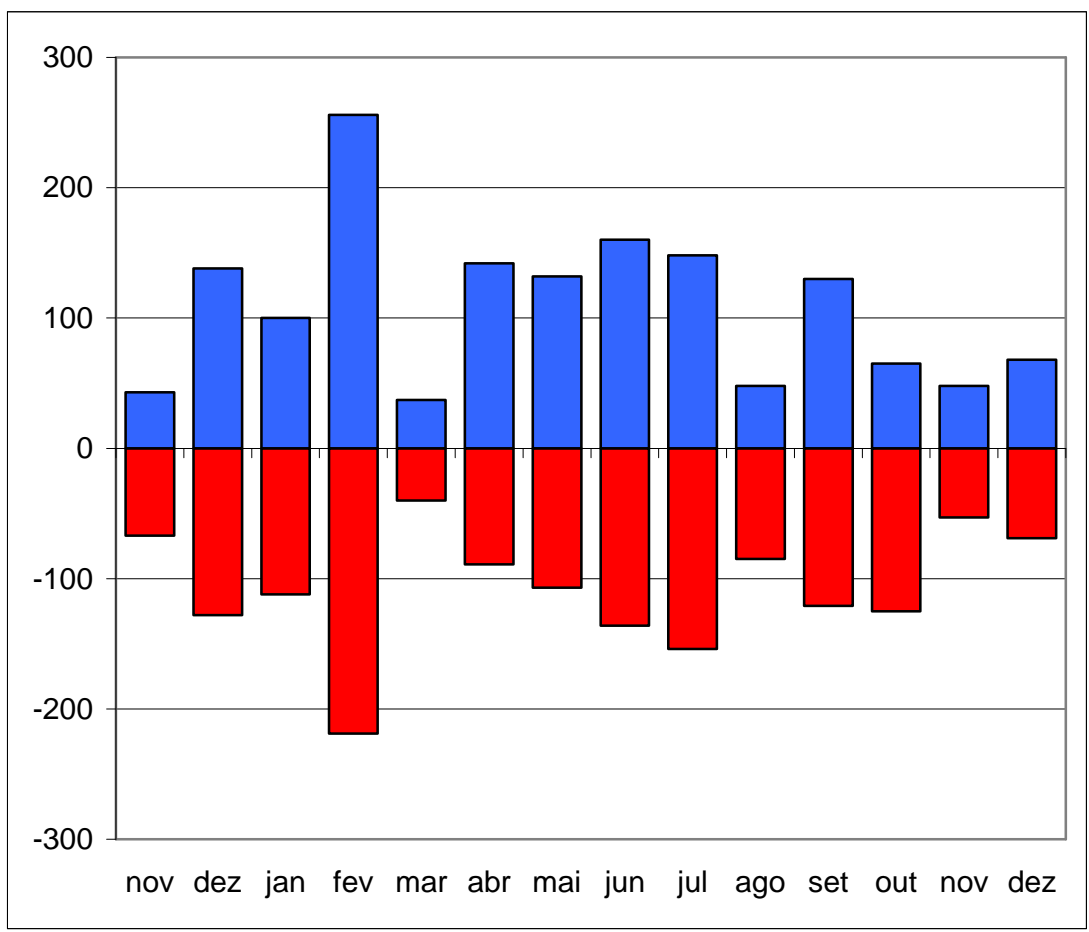

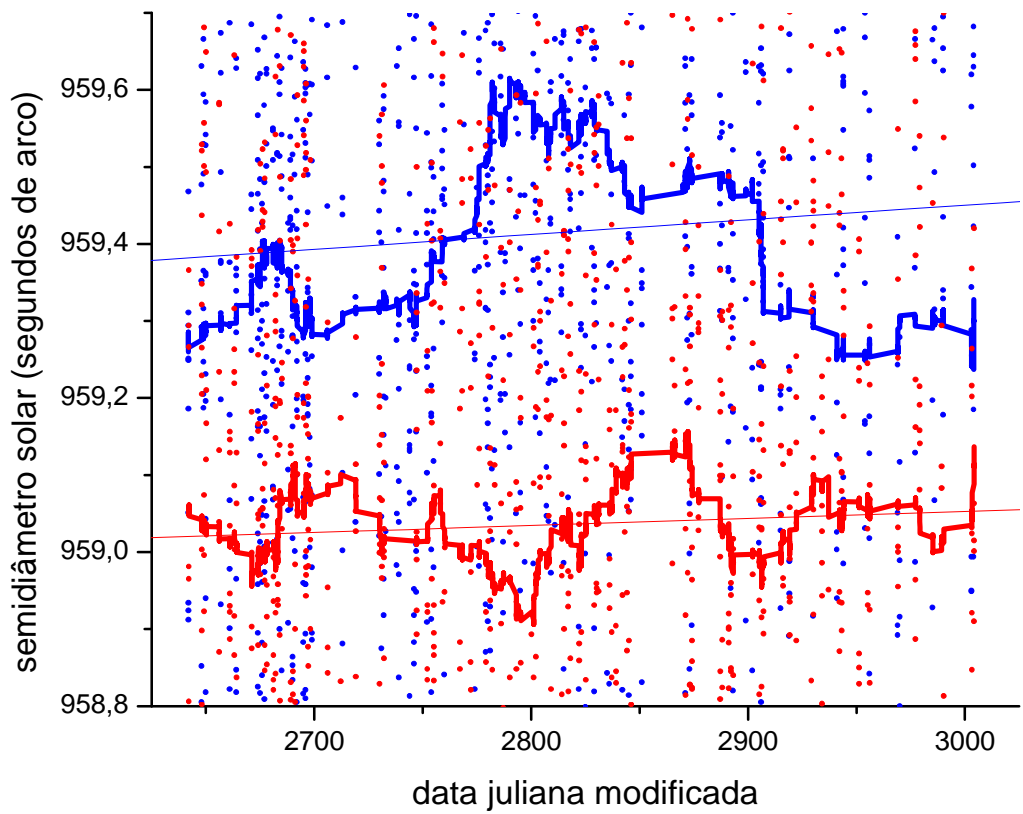

**Figura 33 – Os dados de Semidiâmetro Solar observados em 2003 (mais novembro e dezembro de 2002) e as retas que se ajustam às duas séries. Acima a série a leste e abaixo a série a oeste.**

**Figura 34 - Distribuições das médias de Semidiâmetro dos grupos a leste e comparação com a distribuição normal. À direita os desvios entre os pontos e a normal marcando-se um desvio padrão à esquerda e à direita da distribuição de desvios.**

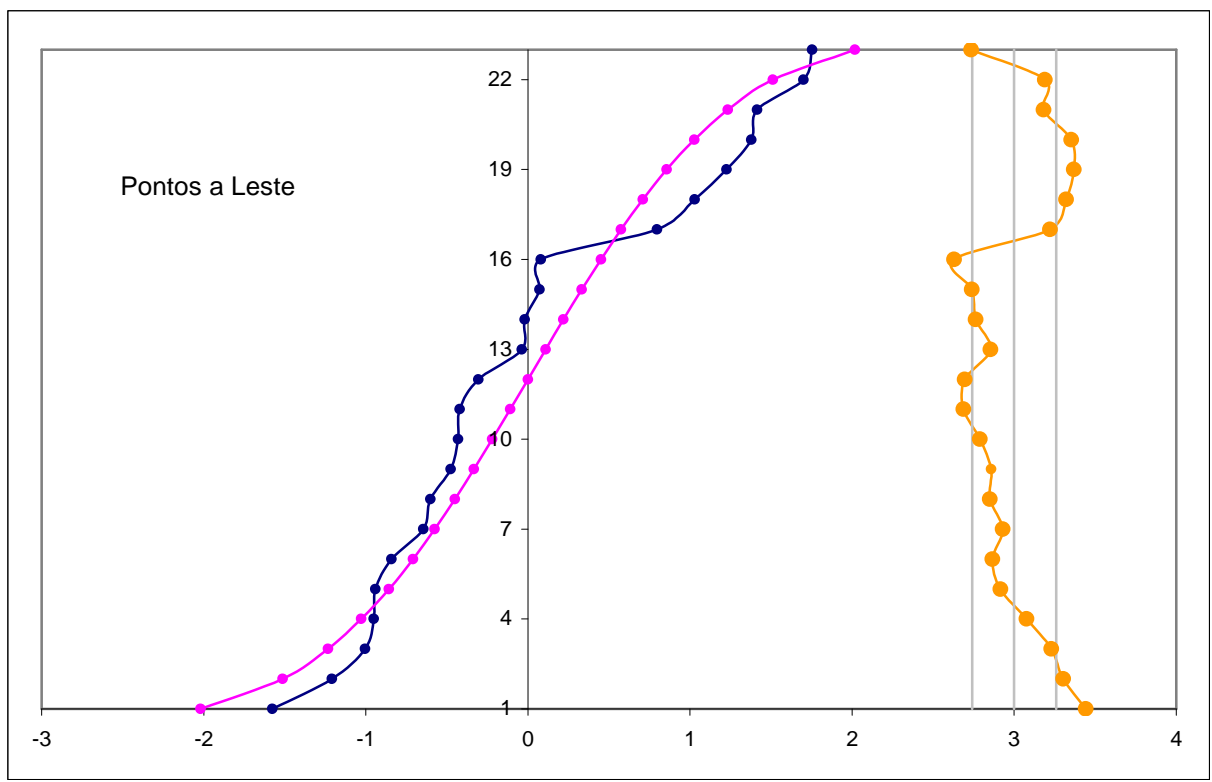

**Figura 35 - Distribuições das médias de Semidiâmetro dos grupos a oeste e comparação com a distribuição normal. À direita os desvios entre os pontos e a normal marcando-se um desvio padrão à esquerda e à direita da distribuição de desvios.**

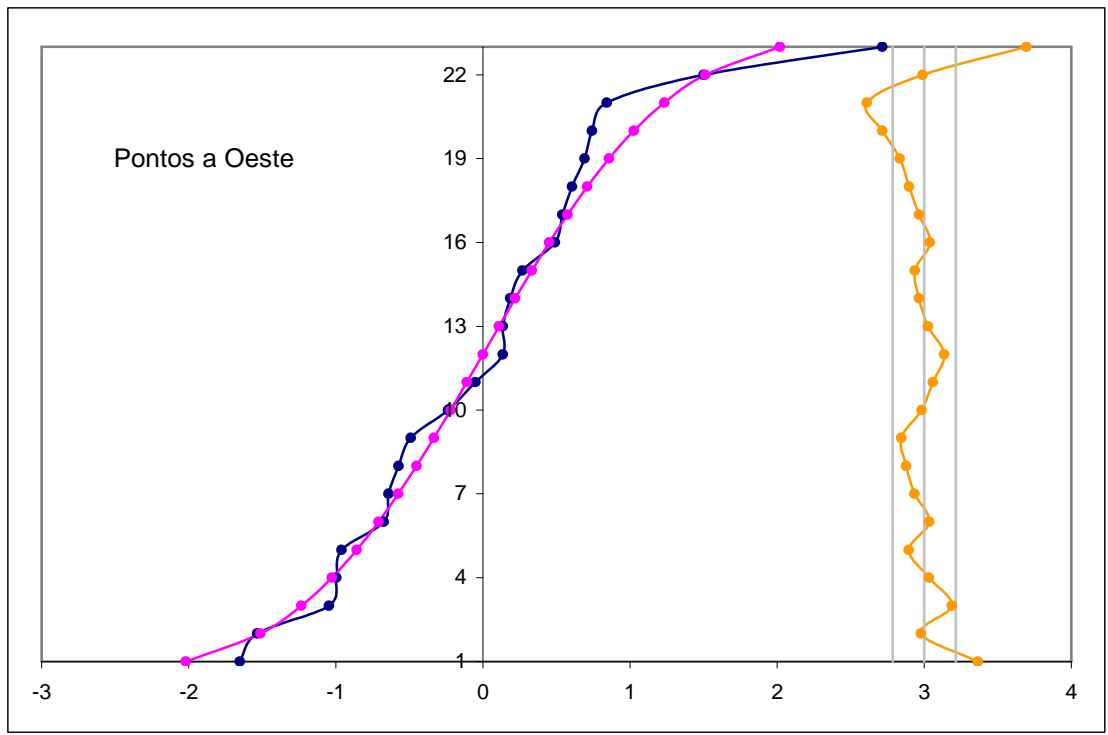

**Figura 36 – Média dos valores observados dentro de cada grupo. Mostra-se a abrangência temporal de cada grupo e uma linha de tendência para a série a leste (acima) e para a série a oeste (em baixo). As linhas de tendência têm sua equação exibida.**

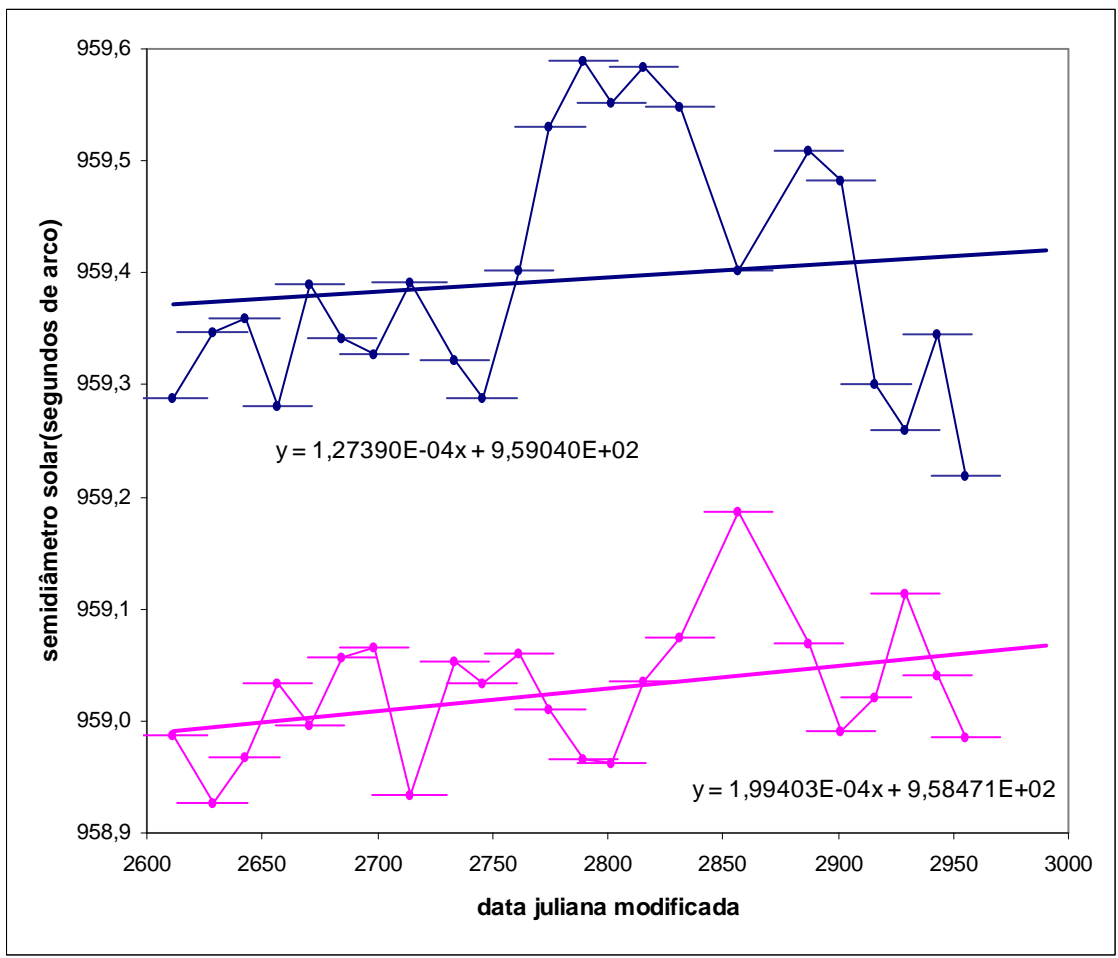

**Figura 37 - Histograma da primeira correção feita aos valores de 2003 observados a leste.**

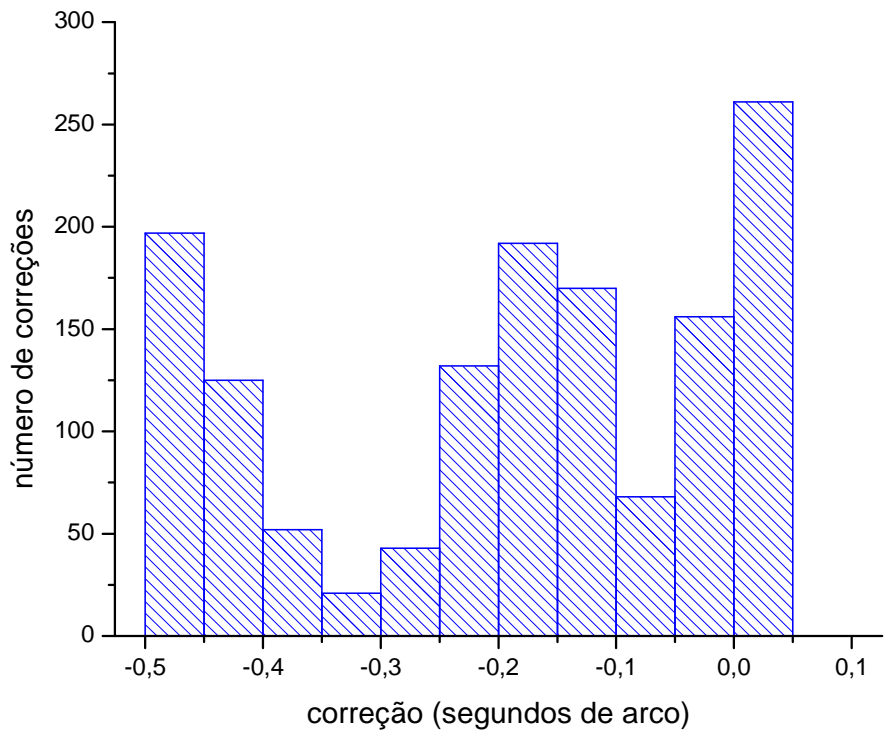

**Figura 38 - Histograma da primeira correção feita aos valores de 2003 observados a oeste.**

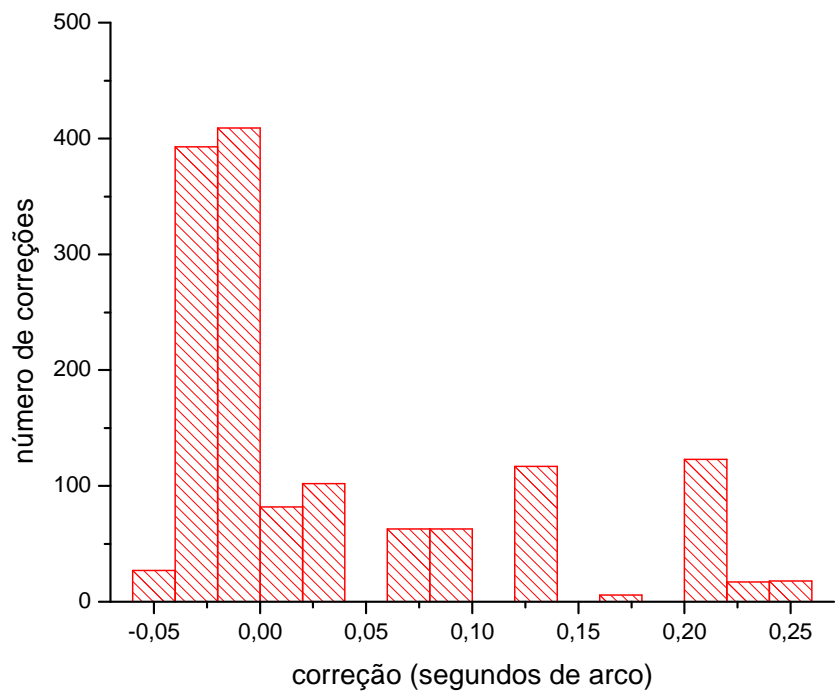

**Figura 39 – Primeira correção imposta aos valores de 2003 a leste em função do tempo.**

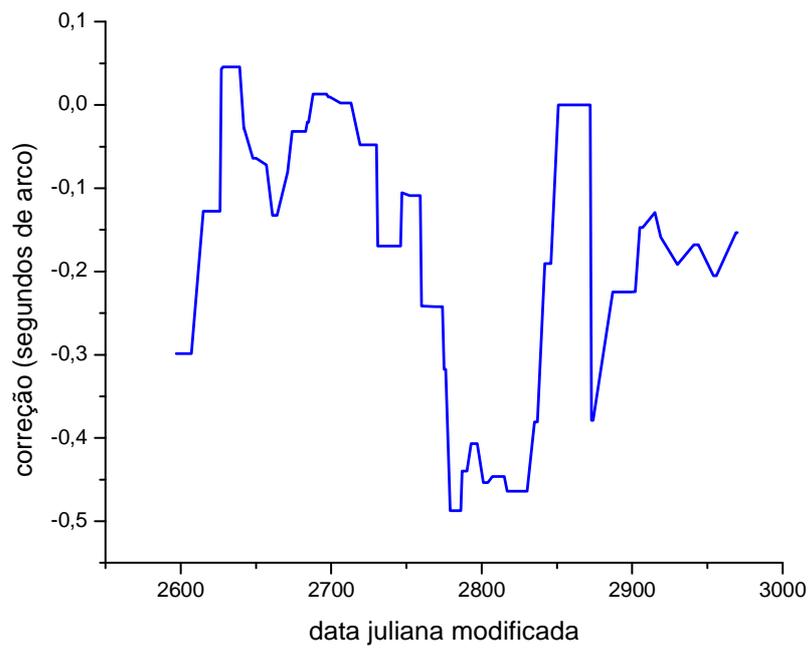

**Figura 40 – Primeira correções imposta aos valores de 2003 a oeste em função do tempo.**

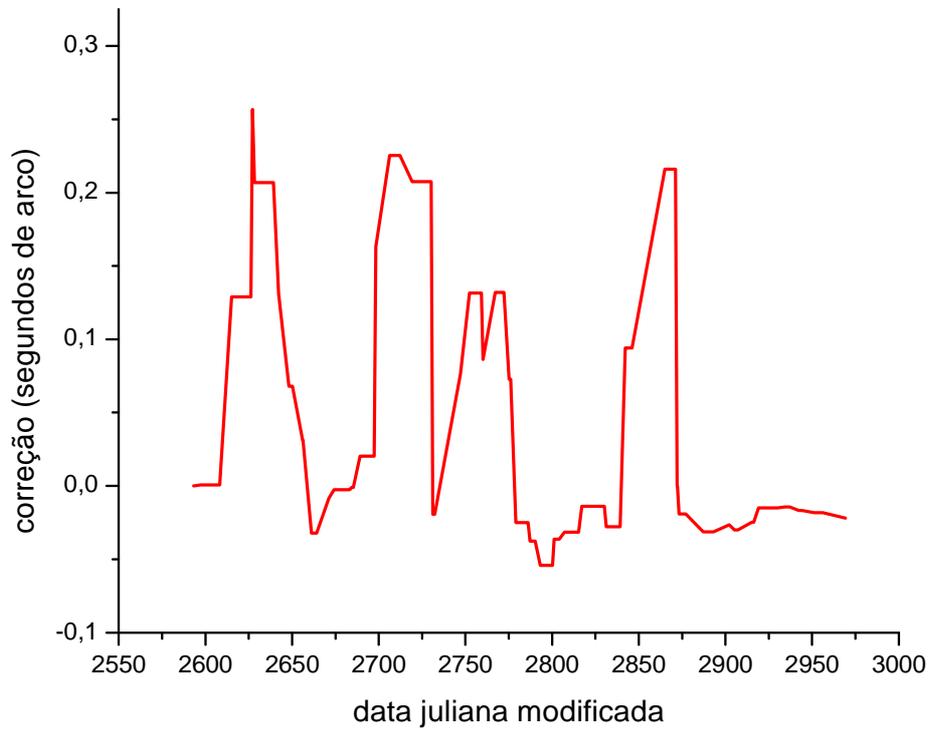

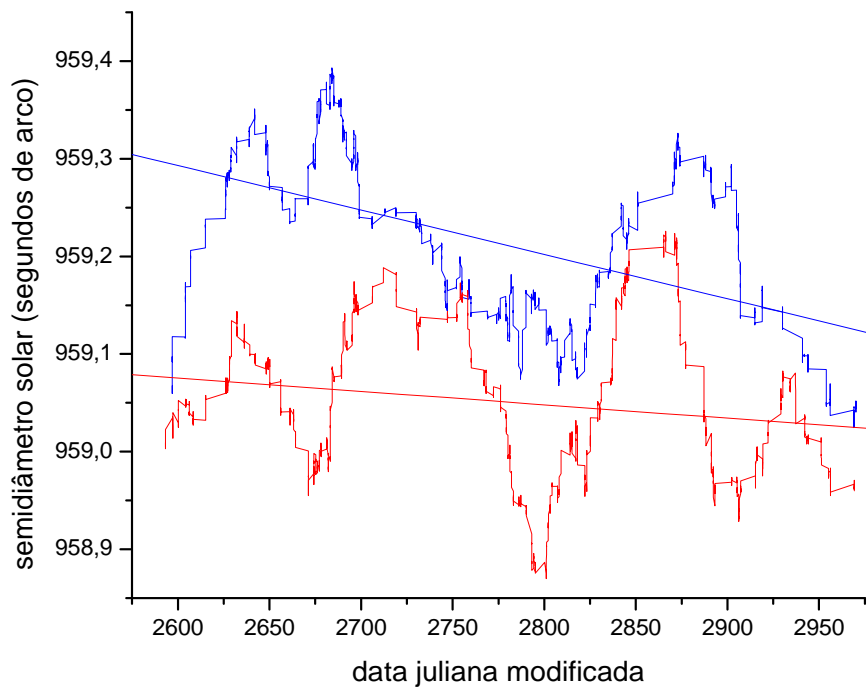

**Figura 41 – Semidiâmetro Solar em função do tempo após a primeira correção dos dados de 2003. A curva superior é dos valores observados a leste, a inferior dos valores observados a oeste.**

# CORREÇÃO DOS PARÂMETROS DAS OBSERVAÇÕES PARA 2003.

Os mesmos parâmetros utilizados para avaliar os dados de 2002, foram utilizados para avaliar também os de 2003, isto é: a data Juliana, a distância zenital do Sol, o azimute de observação do Sol, o parâmetro de Fried, o desvio padrão do ajuste da parábola ao bordo direto do Sol, o desvio padrão do ajuste da parábola ao bordo refletido do Sol, a temperatura do ar no instante médio da observação, a variação da temperatura durante a observação e a pressão atmosférica no instante médio da observação. Cada um destes parâmetros pode separadamente ou em conjunto com outros, influir e alterar os valores medidos. As influências destes parâmetros foram verificadas separadamente e em conjunto por meio de ajustes lineares dos pontos observados aos parâmetros propostos e verificando-se o valor do coeficiente angular da relação linear bem como o desvio padrão do ajuste. Os pontos foram assim verificados, no seu conjunto total, nos subconjuntos a leste e a oeste e num conjunto total onde se tomaram os pontos a leste e os pontos a oeste com sinais invertidos, procurando-se assim estabelecer a existência de assimetria nas influências de observações em lados opostos. O ajuste de retas aos dados foi executado com a utilização da técnica de mínimos quadrados. Os dados de Semidiâmetro Solar utilizados nesta analise foram os valores observados e corrigidos pela análise anterior.

A análise selecionou para correção, os mesmos quatro parâmetros selecionados para corrigir os dados de 2002 e que são: o fator de Fried, o desvio padrão do ajuste da parábola ao bordo direto do Sol, a temperatura média do ar durante a observação e a diferença de temperatura durante a observação. Como estes parâmetros influenciam de maneira diferente os valores observados a leste e os valores observados a oeste, foram calculados valores diferentes para cada um dos lados. A Tabela VI mostra os resultados, nela são vistos os coeficientes angulares das retas que se ajustam aos dados do Semidiâmetro Solar tomados em função dos parâmetros analisados. Os valores estão em segundos de arco e correspondem ao desvio que cada um dos parâmetros causa ao Semidiâmetro Solar. Os valores se referem aos parâmetros normalizados, isto é, diminuídos de sua média e divididos por seu desvio padrão. Os desvios acarretados às observações são os produtos dos valores da tabela multiplicados pelo parâmetro normalizado.

**Tabela VI – Influência dos parâmetros medidos aos valores observados do Semidiâmetro Solar. Valores em segundos de arco. Parâmetros normalizados.**

| PARÂMETRO | VALORES A LESTE | VALORES A OESTE |
|---|---|---|
| **Fator de Fried** | −0,011 | 0,009 |
| **Desvio do Bordo** | 0,140 | 0,049 |
| **Temperatura** | 0,076 | −0,063 |
| **Diferença de Temperatura** | 0,031 | −0,012 |

Aos valores de Semidiâmetro corrigidos da primeira analise devem ser somados os desvios calculados acima com sinal invertido e multiplicados pelos respectivos parâmetros normalizados. Assim procedendo, estamos retirando das observações a influência introduzida a elas pelo parâmetro em questão, as quatro correções, assim calculadas, devem ser somadas para a obtenção da correção total do Semidiâmetro Solar.

As correções obtidas nesta segunda etapa para os valores a leste variam entre –0,265 e 0,357 e têm a média de 0,004, para os valores a oeste variam entre -0,162 e 0,239 e têm média de 0,041. As Figuras 42 e 43 mostram as correções feitas aos pontos ao longo do ano e sua média corrida de 150 pontos. As Figuras 44 e 45 mostram os histogramas destas correções onde se pode ver que apenas muito poucos pontos tiveram correções maiores que 0,2 segundos de arco para mais ou para menos. A Figura 46 mostra o resultado final das correções sobre os dados de 2003. As curvas a leste e a oeste permanecem ainda separadas em muitos de seus pontos, mas, agora, sua diferença é bem menor. A tendência linear mostra uma evolução descendente de -0,910 milisegundos de arco por dia a leste e de -0,600 milisegundos de arco por dia a oeste.

**Figura 42 – Correções da influência dos parâmetros aos valores de Semidiâmetro Solar de 2003 observados a leste do meridiano e sua média corrida a cada 150 pontos.**

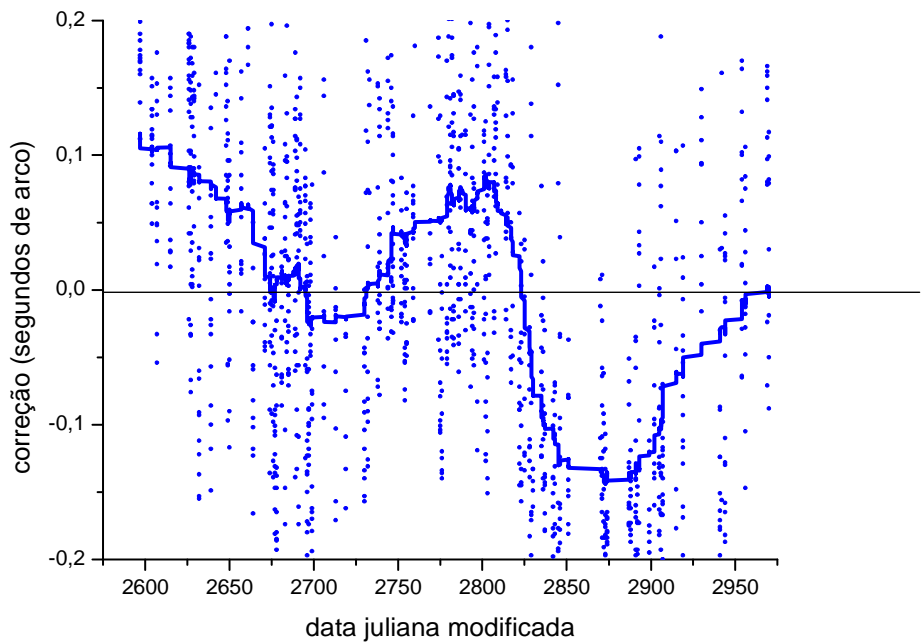

**Figura 43 – Correções da influência dos parâmetros aos valores de Semidiâmetro Solar de 2003 observados a oeste do meridiano e sua média corrida a cada 150 pontos.**

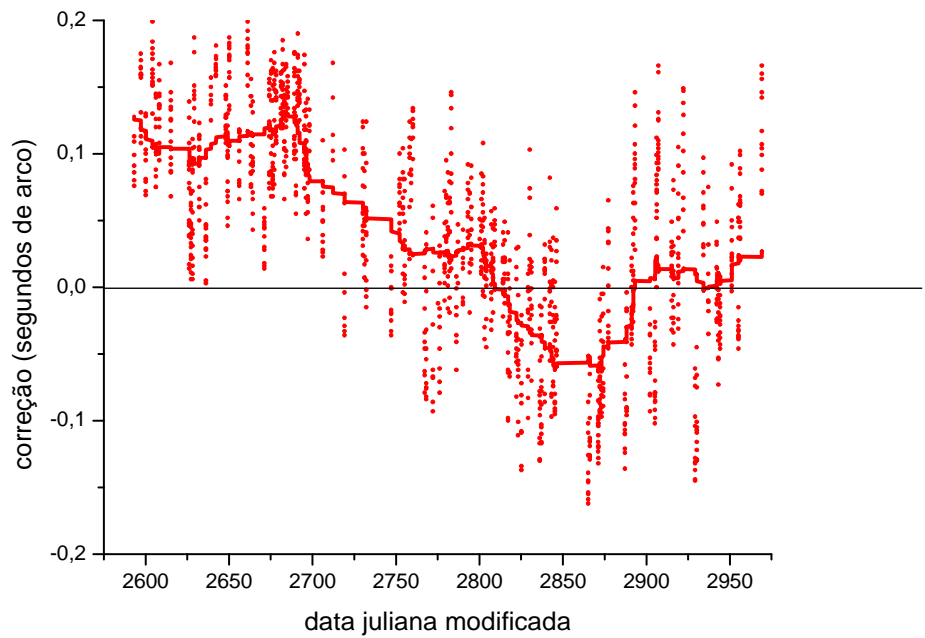

**Figura 44 – Histograma das correções da influência dos parâmetros aos valores observados de Semidiâmetro Solar a leste em 2003.**

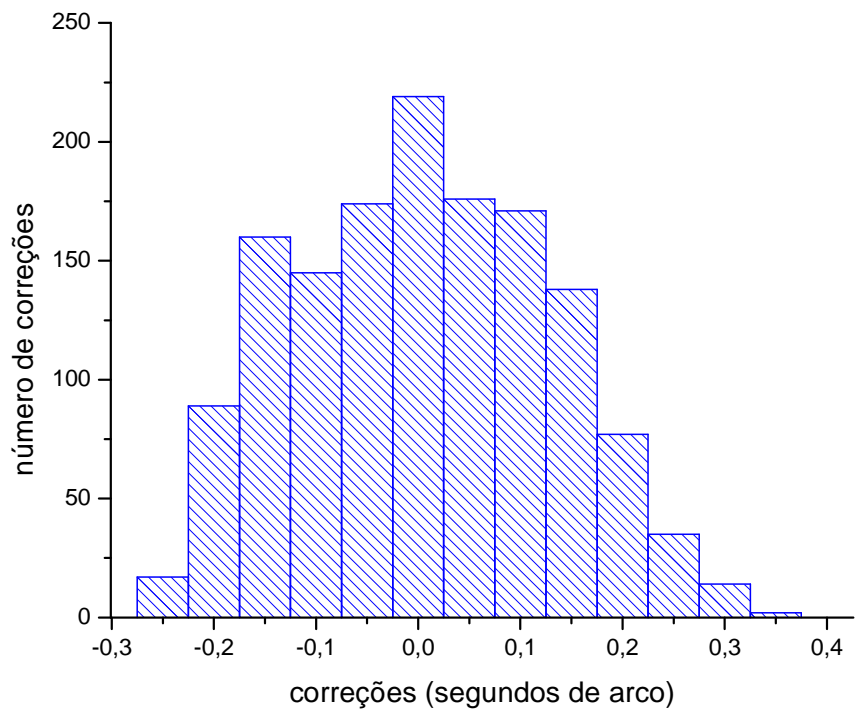

**Figura 45 – Histograma das correções da influência dos parâmetros aos valores observados de Semidiâmetro Solar a oeste em 2003.**

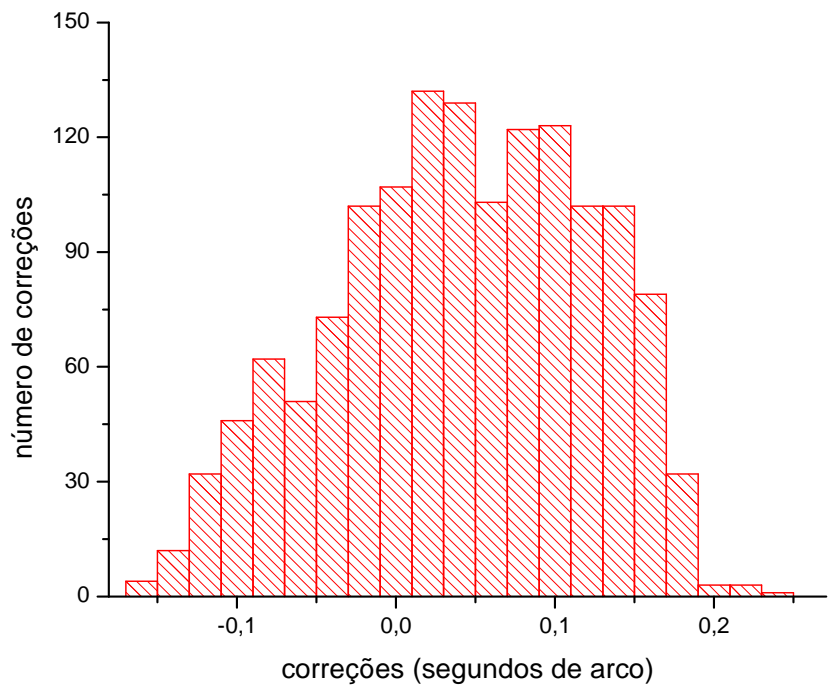

**Figura 46 – Média corrida de 150 pontos dos valores de Semidiâmetro Solar após as correções implementadas. A curva de leste é a de cima e a de oeste a de baixo.**

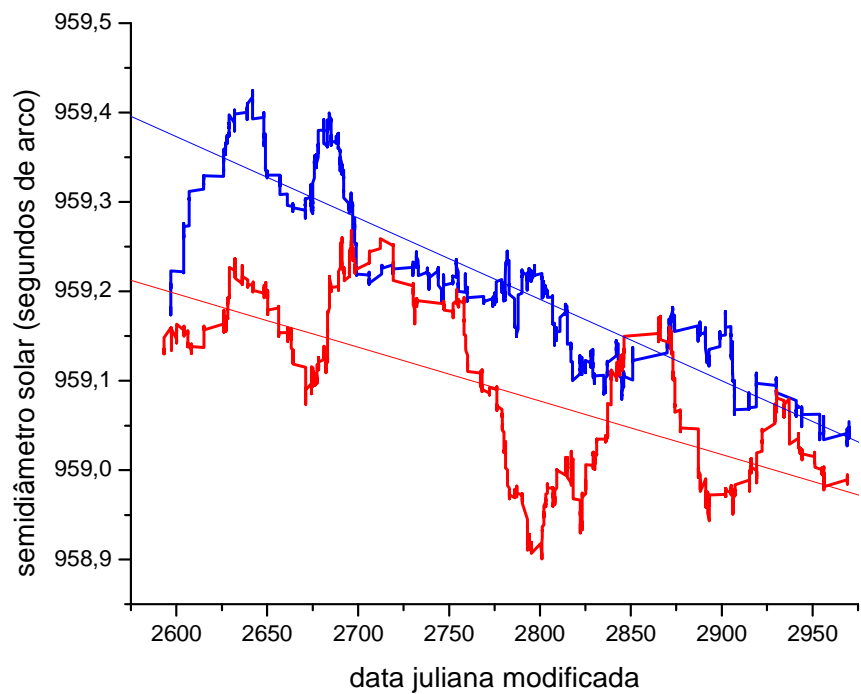

**CONJUNTO DE DADOS CORRIGIDOS DE 1998 A 2003.**

Temos um total de 16.523 observações do Semidiâmetro Solar que abrangem um período que se estende desde março de 1998 até novembro de 2003. Estas observações foram tratadas e corrigidas o que nos dá uma coleção de dados da qual foram retirados erros de observação. Embora sejam valores ainda bastante dispersos por uma série de outros erros ainda presentes, de sua média podemos retirar muitas informações. As séries a leste e a oeste que até então foram tratadas em separado foram juntadas em uma única coleção de dados distribuídos temporalmente. O número de observações a cada mês de 1998 até 2003 está mostrado na Figura 47. Pode-se ver que não há dados em janeiro e fevereiro de 1998, em outubro e novembro de 2001 e em dezembro de 2003. Os pontos apresentam-se bem distribuídos ao longo de todo o período, com uma média de 230 pontos por mês. Apenas quatro meses têm número de pontos inferior a 100 (além dos cinco antes citados com zero pontos) e apenas cinco os têm superior a 400.

A Tabela VII mostra que os valores corrigidos do Semidiâmetro Solar são bastante uniformes. O número de pontos por ano é semelhante. Convém lembrar que em 2001 houve uma parada por dois meses. As médias são próximas e revelam tão somente a variação temporal da série total e, principalmente, os desvios padrões são muito semelhantes.

**Tabela VII– Número de pontos, média e desvio padrão para cada ano dos valores corrigidos de Semidiâmetro Solar. Valores em segundos de arco.**

| ANO | número de pontos | Média | desvio padrão |
|---|---|---|---|
| **1998** | 2633 | 959,097 | 0,553 |
| **1999** | 3603 | 959,062 | 0,585 |
| **2000** | 2876 | 959,184 | 0,553 |
| **2001** | 1936 | 959,199 | 0,578 |
| **2002** | 2968 | 959,317 | 0,553 |
| **2003** | 2507 | 959,142 | 0,534 |

A média corrida de 500 pontos da série completa destes valores é apresentada na Figura 48. A curva é nitidamente crescente, apesar de picos e vales locais, até atingir um máximo em 2002 a partir do qual a curva passa a ser decrescente. Pode-se ver o período sem dados de outubro a dezembro de 2001 que corresponde aos três meses em que o Astrolábio Solar foi desativado para manutenção e re-espelhamento do filtro. Considerando que a média é corrida de 500 pontos e que o desvio padrão dos pontos é de 0,567, pode-se estimar a incerteza em torno desta curva em 0,025 segundos de arco. Com os dados corrigidos agora disponíveis podemos estudar as correlações entre as variações do Semidiâmetro e alguns parâmetros da atividade solar, e analisar a figura do Sol, uma vez que para cada valor observado do Semidiâmetro Solar temos também a latitude solar que foi observada.

**Figura 47 – Distribuição mensal da série completa de valores observados e corrigidos do Semidiâmetro Solar de 1998 a 2003.**

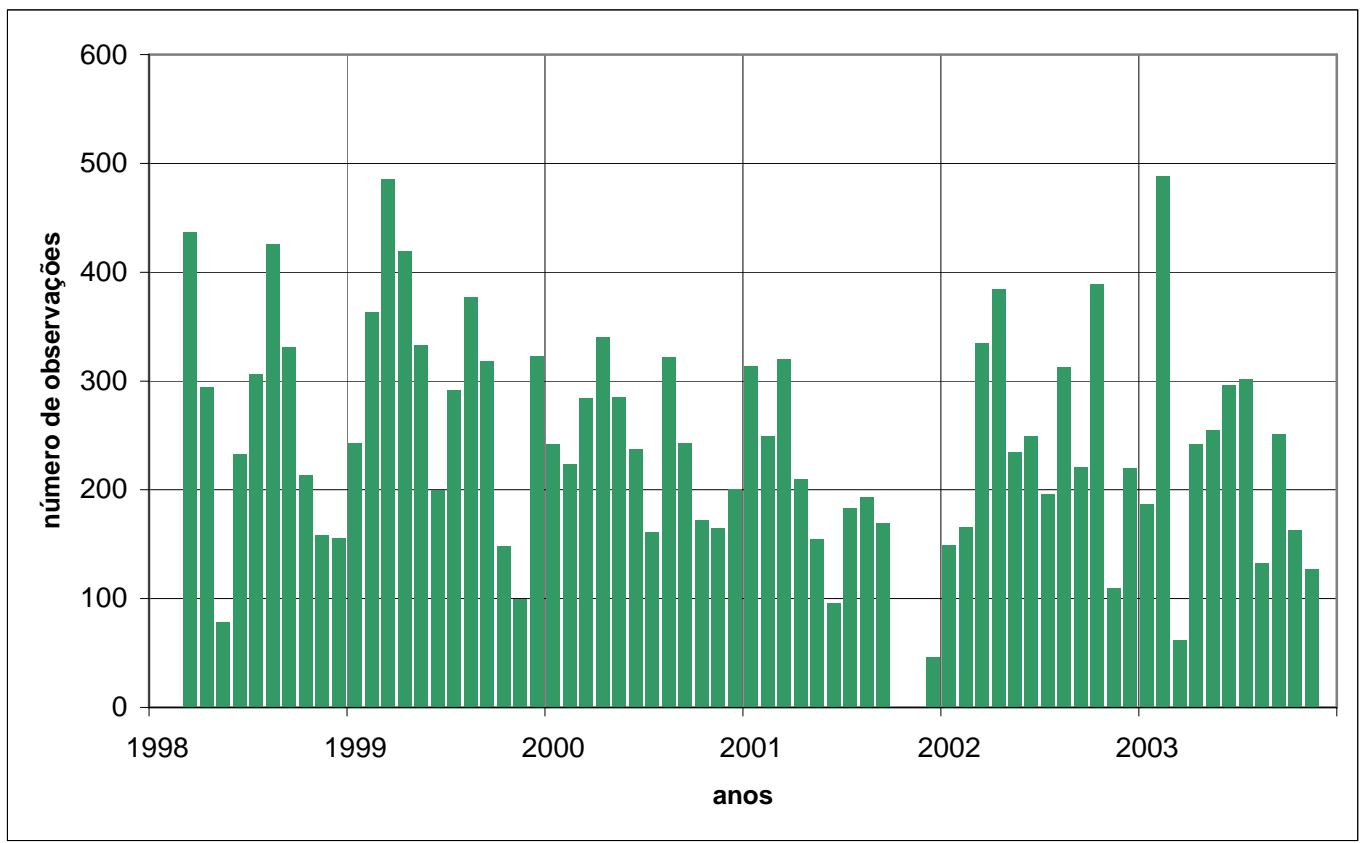

**Figura 48 – Média corrida da série completa de valores observados e corrigidos do Semidiâmetro Solar de 1998 a 2003.**

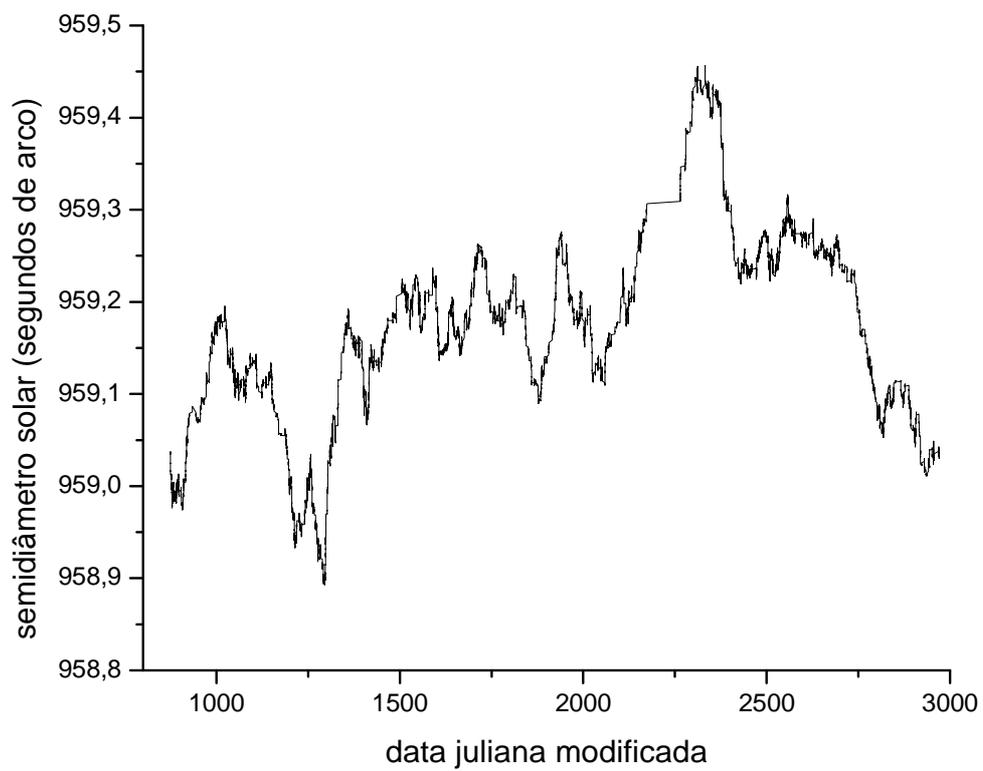

# COMPARAÇÕES COM ÍNDICES DA ATIVIDADE SOLAR.

**Abordagem Inicial**

Comparamos a série do Semidiâmetro Solar com cinco séries de parâmetros da atividade solar do mesmo período. São eles: Índice de Flares, Irradiância, Campo Magnético Integrado, Contagem de Manchas e Fluxo Rádio em 10,7cm. Os valores foram obtidos do National Geophysical Data Center – NGDC. Estes parâmetros foram escolhidos porque indicam a ocorrência de fenômenos possivelmente interligados tais como: variação da superfície emissora e do fluxo; interferência do campo magnético na formação de manchas e de flares; e correntes originadas pelo campo magnético das manchas que dão origem ao Fluxo Rádio em 10,7cm. A comparação entre as séries pode ser graficamente apresentada através de séries alisadas. As médias corridas de Semidiâmetro Solar são tomadas a cada 500 pontos e as demais séries com médias corridas de 100 pontos. Fizemos assim porque estas séries são muito ruidosas sendo impossível visualizá-las sem promover a média corrida. A média corrida de 30 pontos para os índices de atividade solar e de 300 para o semidiâmetro solar, que significam uma média em torno de um mês, também se mostrou ruidosa. O número de pontos aglutinados para fazer média foi escolhido de modo a se ter tipicamente uma média mensal. Cada uma das curvas dos parâmetros solares foi normalizada, isto é, diminuída de sua média e dividida pelo desvio padrão, e depois adequadas à comparação com a curva de Semidiâmetro Solar, multiplicando-se os valores normalizados pelo desvio padrão da curva de Semidiâmetro Solar e acrescentando-lhes a média desta. As Figuras 49 a 53 apresentam estas comparações. O aspecto geral das figuras mostra que há grande semelhança entre as curvas desenhadas, há uma tendência crescente inicial e uma tendência decrescente final nos pares em cada gráfico, há a presença de dois picos que variam de intensidade conforme a figura considerada. A curva de Semidiâmetro Solar tem o primeiro pico pouco definido. Há em todas as curvas dois picos menores e crescentes antes do primeiro pico mais forte e entre eles há sempre vales bem evidentes. Há entre os dois picos mais fortes um vale bem pronunciado. Depois do segundo pico mais forte há ainda outro pico e entre eles um vale bem definido. Estes detalhes ocorrem todos mais ou menos na mesma época, com alguma defasagem que varia um parâmetro a outro. Quando se olha para os detalhes, as semelhanças começam a desaparecer, por exemplo, as alturas dos picos e as profundidades dos vales são bem diferentes, há eventos secundários que não aparecem em todos os gráficos.

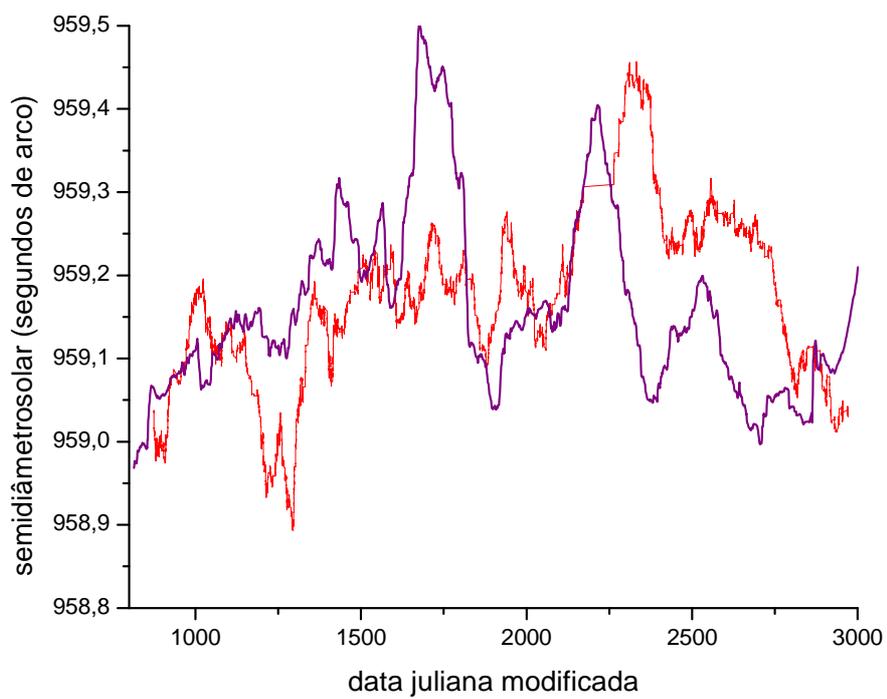

**Figura 49 – Médias corridas comparadas das curvas de Semidiâmetro Solar e de Índice de Flares de 1998 a 2003. O Semidiâmetro é representado pela curva em vermelho.**

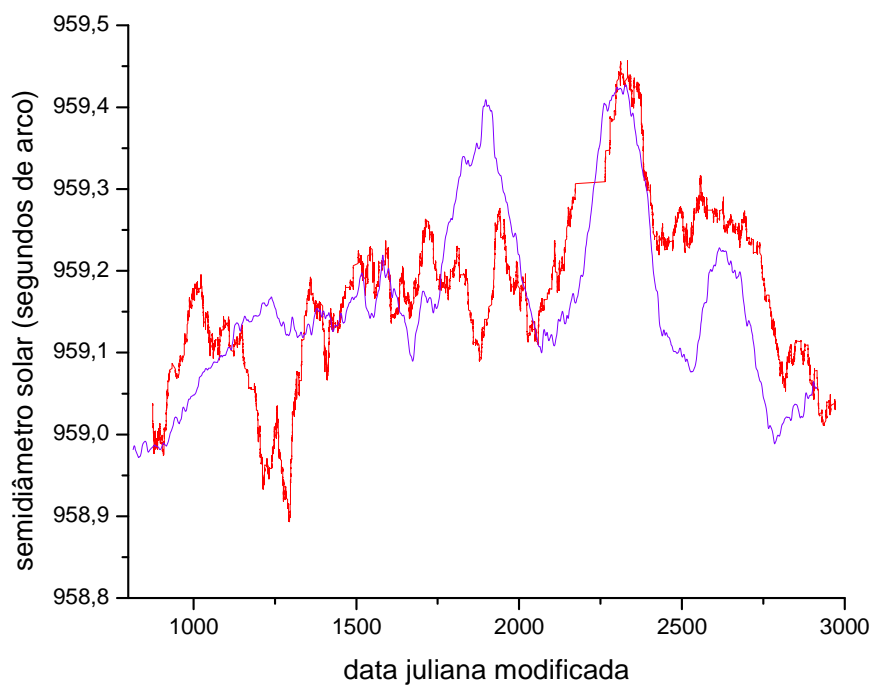

**Figura 50 – As médias corridas comparadas das curvas de Semidiâmetro Solar e de Irradiância de 1998 a 2003. O Semidiâmetro é representado pela curva em vermelho.**

**Figura 51 – Médias corridas comparadas das curvas de Semidiâmetro
Solar e de Campo Magnético Integrado de 1998 a 2003.
O Semidiâmetro é representado pela curva em vermelho.**

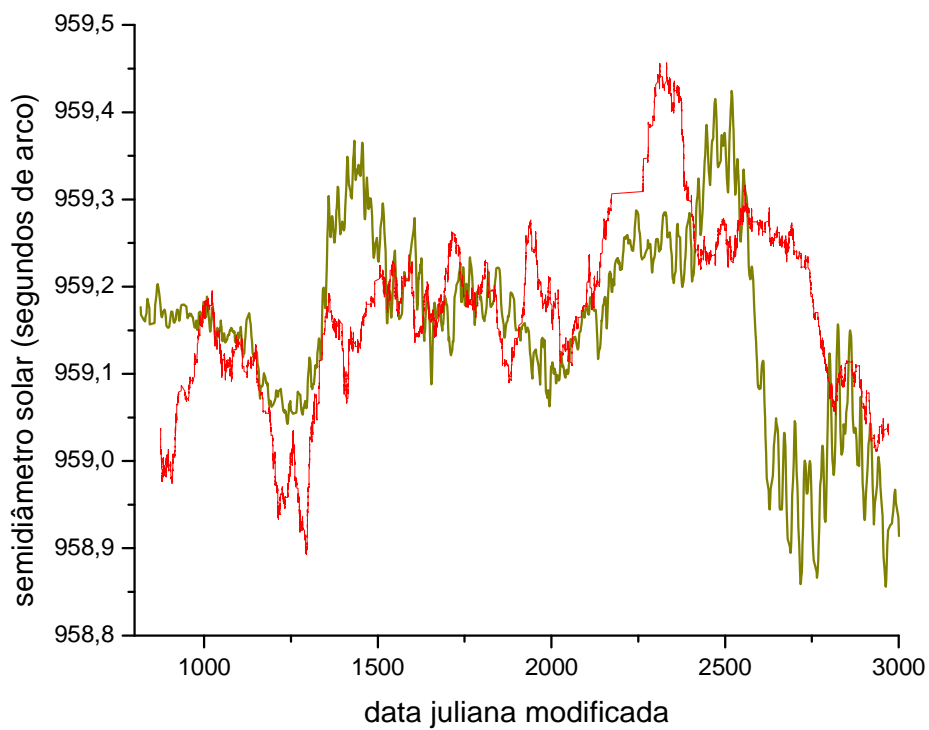

**Figura 52 – Médias corridas comparadas das curvas de Semidiâmetro Solar e de Contagem de Manchas de 1998 a 2003. O Semidiâmetro é representado pela curva em vermelho.**

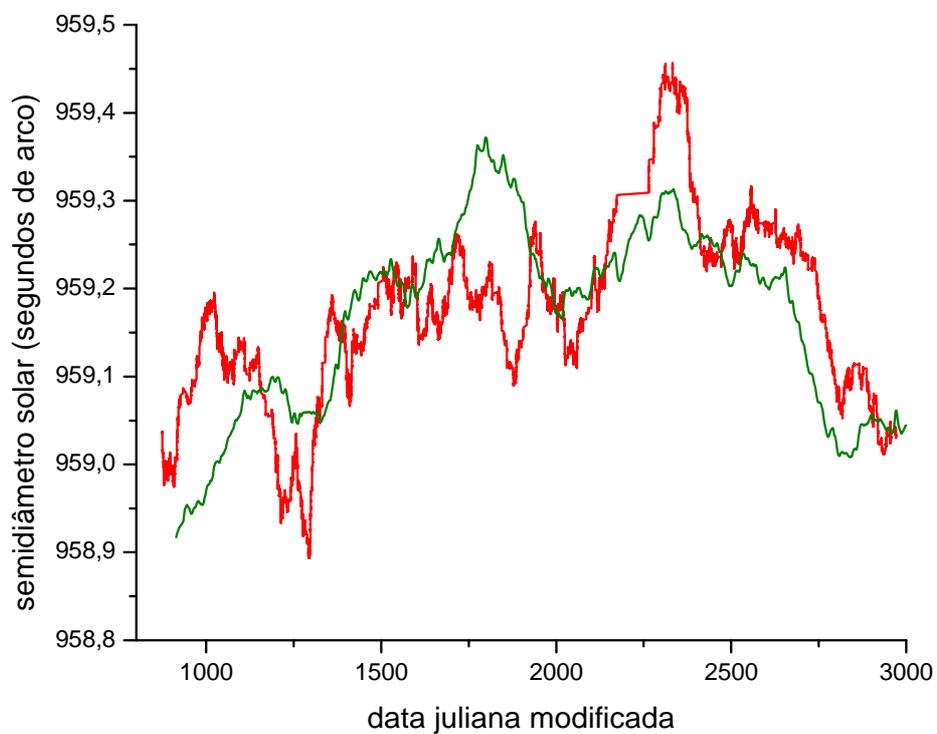

**Figura 53 – Médias corridas comparadas das curvas de Semidiâmetro Solar e de Fluxo Rádio em 10,7cm de 1998 a 2003. O Semidiâmetro é representado pela curva em vermelho.**

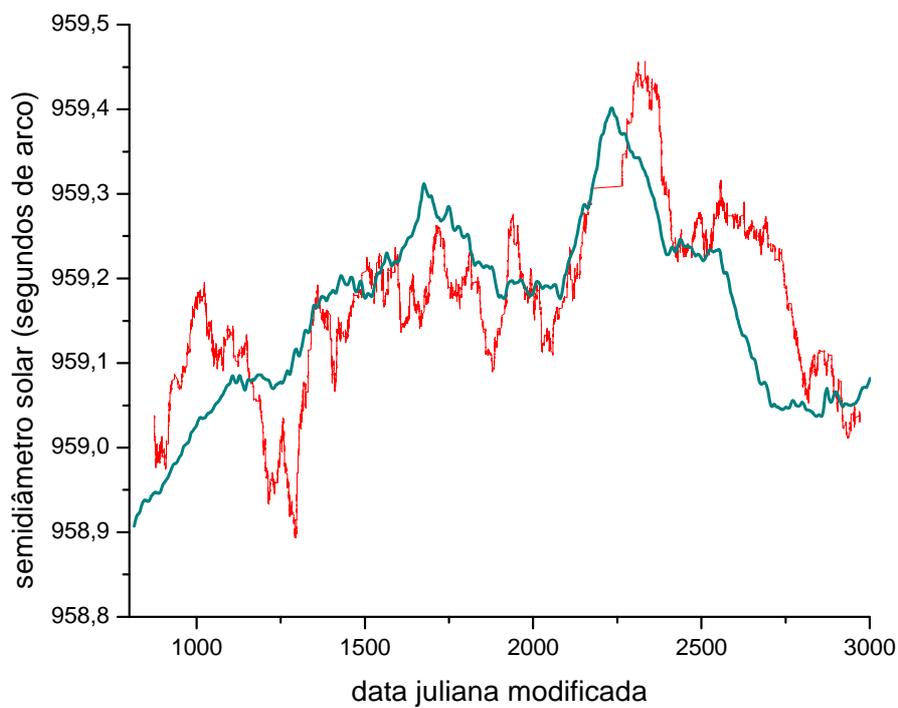

**Correlações.**

As semelhanças encontradas entre as curvas comparadas dos diversos parâmetros de atividade do Sol ilustram haver correlações físicas entre elas. Para calcular estas correlações dividimos o intervalo de dados de março de 1998 a novembro de 2003 em períodos e tomamos a média dos valores em cada período. O número de períodos é variado de forma contínua entre 6 e 72, isto é, períodos variando desde o tamanho de um ano até o tamanho de um mês. Assim fazendo obtivemos para cada dupla de parâmetros uma coleção de 67 correlações. No cálculo das correlações utilizamos o método de Pearson, que computa o coeficiente de correlação e o nível de significância para dois vetores com o mesmo número de elementos. Para comparação calculamos também da mesma forma as correlações entre o Semidiâmetro Solar e séries de números randômicos escolhidos de duas maneiras: primeiramente com números randômicos distribuídos de forma uniforme entre dois limites e depois com números randômicos distribuídos de forma gaussiana. Como as séries de atividades solares escolhidas têm um valor médio a cada dia, também foram atribuídos valores diários às séries de valores randômicos.

Observando estas correlações constatamos que todas elas têm valores maiores quando se comparam períodos mais longos e, na medida em que consideramos períodos menores as correlações diminuem. Isto revela que as correlações crescem quando se olha para o aspecto geral das curvas e diminuem quanto mais se olha para os detalhes. A Tabela VIII mostra para os parâmetros comparados a maior correlação, a média, a mínima, a mediana e a média do maior quartil. Indica ainda o número de valores (do total de 67) em que a correlação foi superior a 0,8, o número de valores em que ela foi superior a 0,6 e em que foi superior a 0,4. As duas últimas linhas mostram estes dados para as correlações entre o Semidiâmetro Solar e as duas distribuições randômicas.

Esta tabela mostra que há entre alguns parâmetros uma forte correlação o que indica uma interação física entre eles. Os parâmetros melhor correlacionados, como já era esperado, são a Contagem de Manchas com o Fluxo Rádio. São muito bem correlacionados: o Índice de Flares com a Contagem de Manchas, o Índice de Flares com o Fluxo Rádio, o Semidiâmetro com o Fluxo Rádio, o Semidiâmetro com a Irradiância e a Irradiância com o Fluxo Rádio. São razoavelmente bem correlacionados: o Semidiâmetro com a Contagem de Manchas, o Campo Magnético com o Fluxo Rádio, o Campo Magnético com a Contagem de Manchas e a Irradiância com a Contagem de Manchas. A Figura 54 mostra, para todas as duplas de parâmetros, a faixa de abrangência de suas correlações. As Figuras 55 a 69 mostram os gráficos das correlações entre todas as duplas de parâmetros em função do número de períodos. Para cada correlação foi também calculada sua significância cujo complemento para 1,0 é mostrado como erro de medida nestes gráficos.

**Tabela VIII – A maior, a média, a menor, a mediana e a média do maior quartil das correlação entre os parâmetros indicados. E os números de correlações superiores a 0,8 a 0,6 e a 0,4 de um total de 67.**

| Parâmetros | | max | med | min | mid | mq | >.8 | >.6 | >.4 |
|---|---|---|---|---|---|---|---|---|---|
| Semidiâmetro | Flares | 0,43 | 0,16 | 0,00 | 0,14 | 0,29 | 0 | 0 | 1 |
| Semidiâmetro | Irradiância | 0,73 | 0,57 | 0,44 | 0,56 | 0,67 | 0 | 22 | 67 |
| Semidiâmetro | C.Magnético | 0,54 | 0,32 | 0,15 | 0,33 | 0,45 | 0 | 0 | 13 |
| Semidiâmetro | Manchas | 0,68 | 0,46 | 0,28 | 0,45 | 0,57 | 0 | 5 | 44 |
| Semidiâmetro | Rádio | 0,83 | 0,62 | 0,47 | 0,60 | 0,75 | 2 | 32 | 67 |
| Flares | Irradiância | 0,74 | 0,03 | -0,18 | -0,05 | 0,31 | 0 | 3 | 3 |
| Flares | C.Magnético | 0,68 | 0,26 | -0,07 | 0,25 | 0,44 | 0 | 1 | 9 |
| Flares | Manchas | 0,89 | 0,70 | 0,62 | 0,68 | 0,79 | 6 | 67 | 67 |
| Flares | Rádio | 0,81 | 0,66 | 0,60 | 0,65 | 0,71 | 1 | 66 | 67 |
| Irradiância | C.Magnético | 0,63 | 0,11 | -0,12 | 0,07 | 0,33 | 0 | 1 | 5 |
| Irradiância | Manchas | 0,95 | 0,39 | 0,18 | 0,35 | 0,64 | 4 | 7 | 22 |
| Irradiância | Rádio | 0,99 | 0,55 | 0,34 | 0,52 | 0,75 | 5 | 22 | 55 |
| C.Magnético | Manchas | 0,68 | 0,41 | 0,21 | 0,41 | 0,57 | 0 | 5 | 34 |
| C.Magnético | Rádio | 0,63 | 0,43 | 0,26 | 0,42 | 0,55 | 0 | 2 | 39 |
| Manchas | Rádio | 0,97 | 0,92 | 0,90 | 0,91 | 0,94 | 67 | 67 | 67 |
| Semidiâmetro | uniforme | 0,07 | 0,00 | -0,06 | 0,00 | 0,03 | 0 | 0 | 0 |
| Semidiâmetro | gaussiana | 0,06 | 0,00 | -0,06 | 0,00 | 0,03 | 0 | 0 | 0 |

**Figura 54 – Faixa de abrangência das correlações entre as duplas indicadas de parâmetros da atividade solar.**

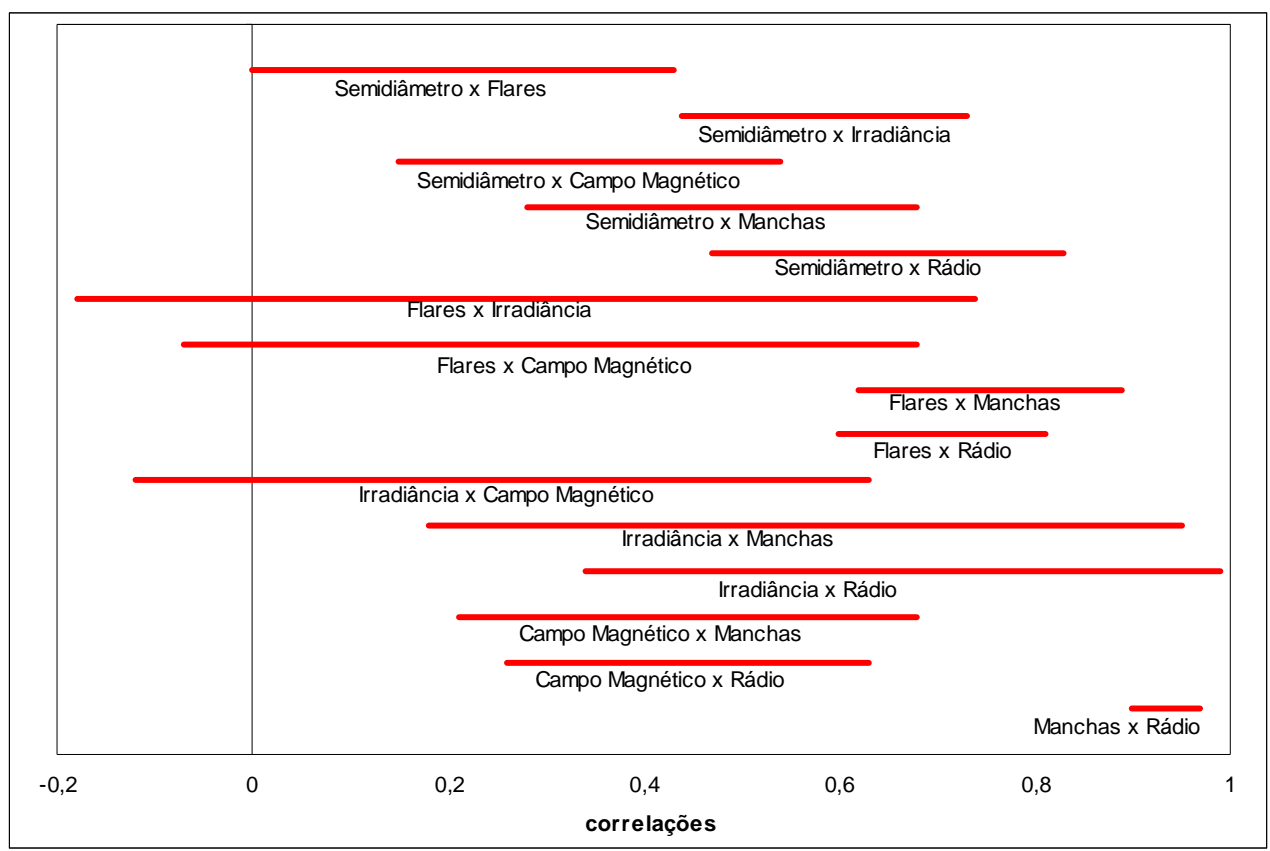

**Figura 55 – Correlações entre Semidiâmetro Solar e Índice de Flares para os períodos considerados.**

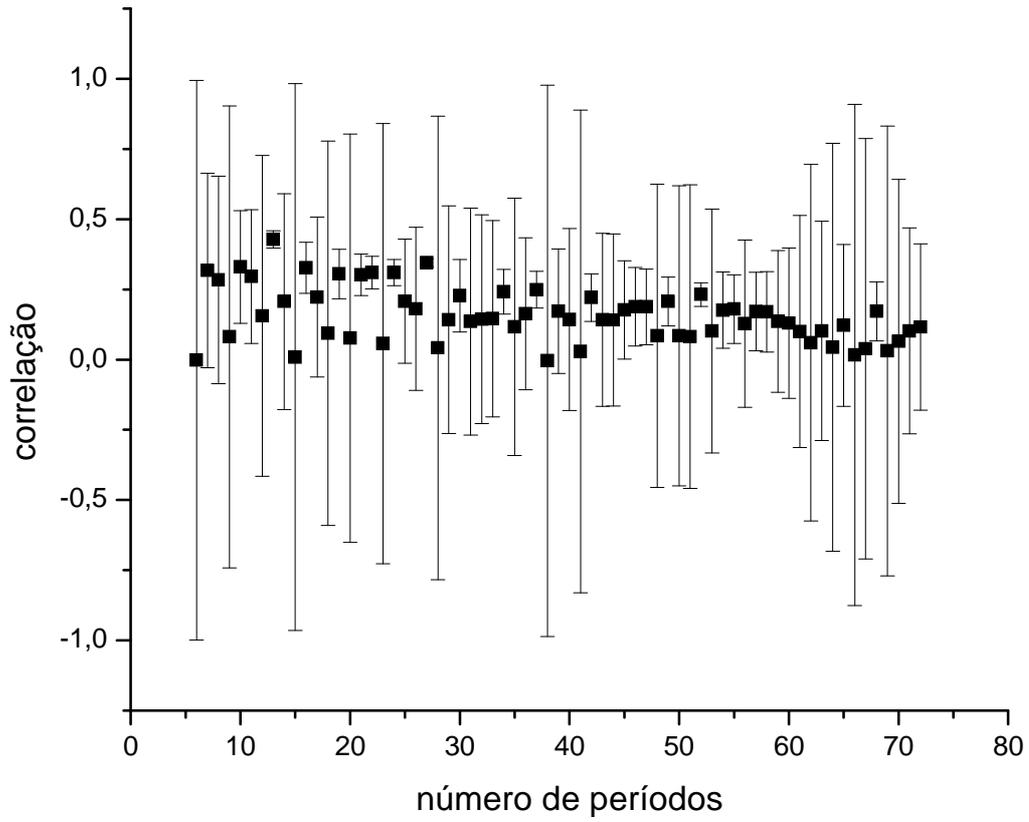

**Figura 56 – Correlações entre Semidiâmetro Solar e Írradiância para os períodos considerados.**

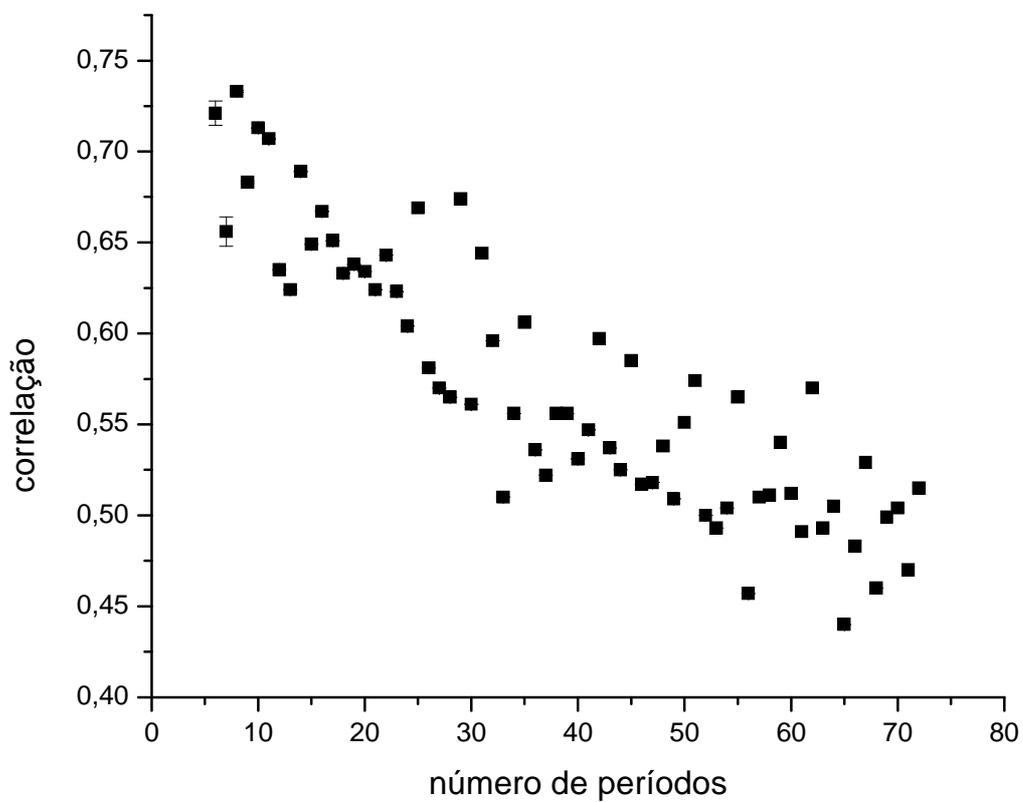

**Figura 57 – Correlações entre Semidiâmetro Solar e Campo Magnético para os períodos considerados.**

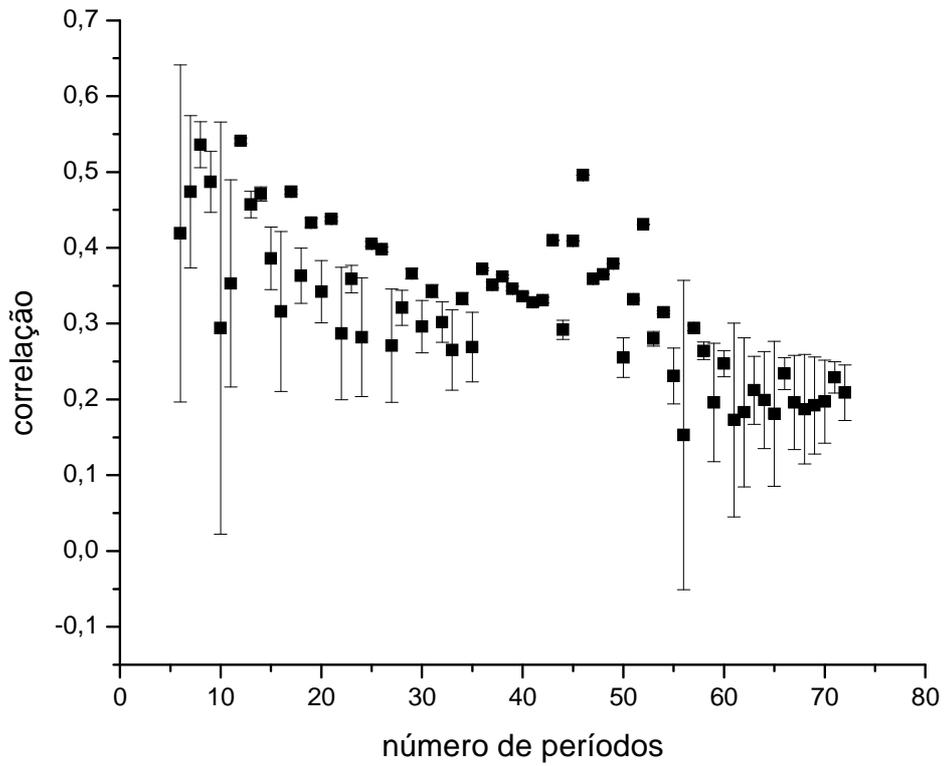

**Figura 58 – Correlações entre Semidiâmetro Solar e
Contagem de Manchas para os períodos considerados.**

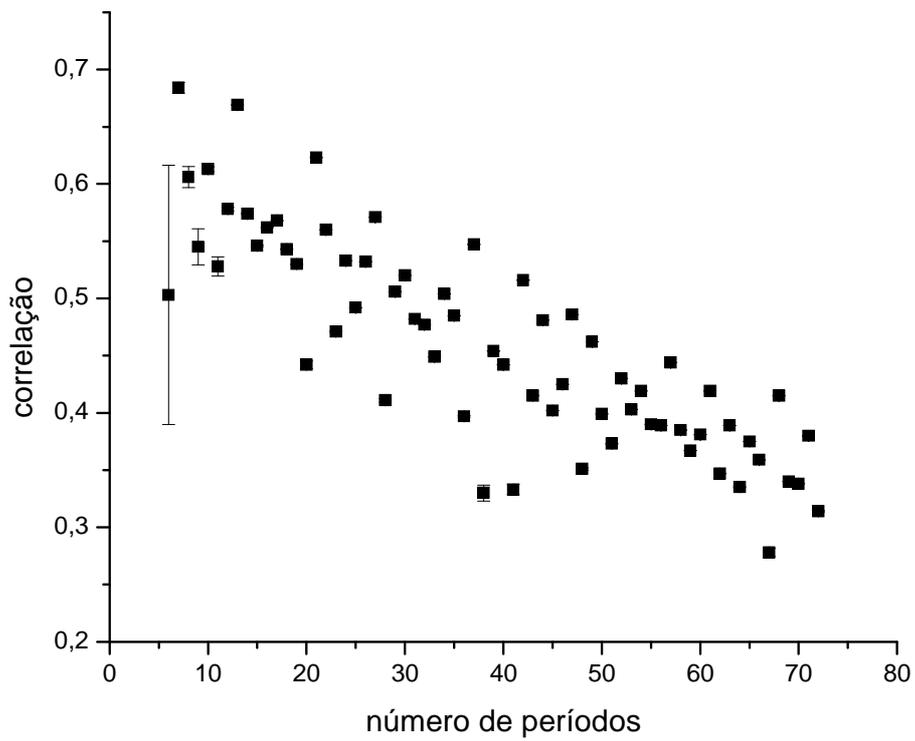

**Figura 59 – Correlações entre Semidiâmetro Solar e
Fluxo de Rádio para os períodos considerados.**

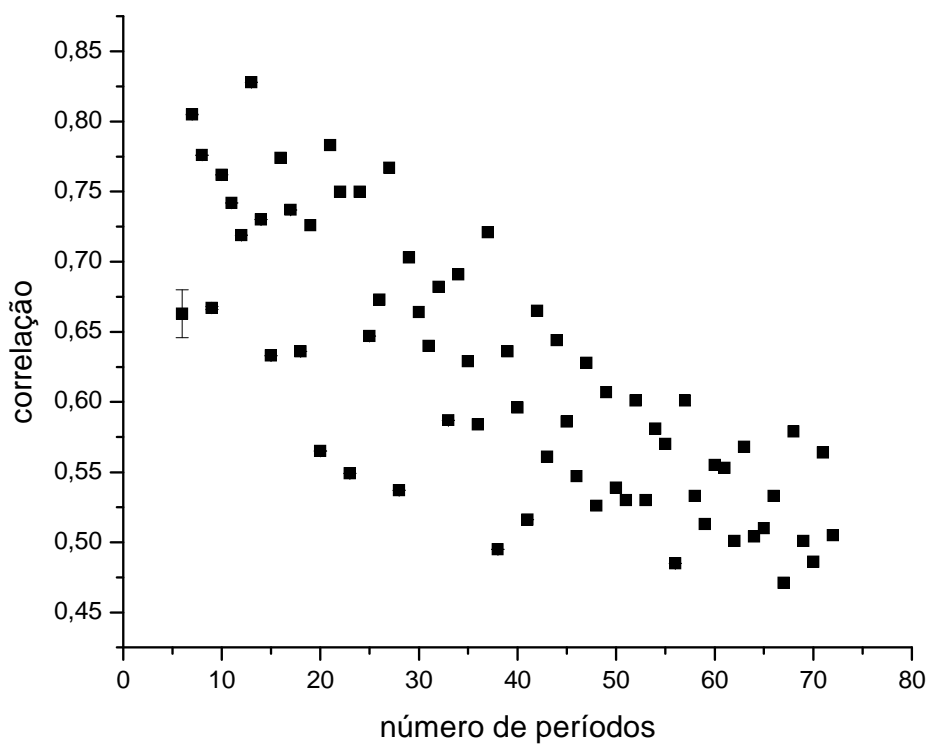

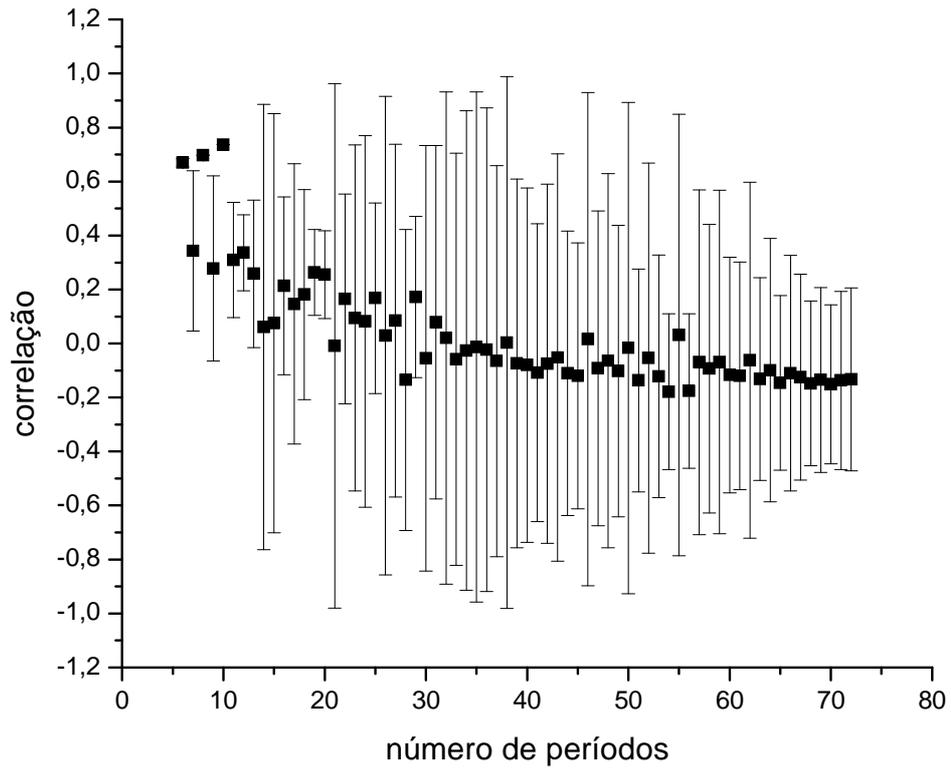

**Figura 60 – Correlações entre Índice de Flares e Irradiância para os períodos considerados.**

**Figura 61 – Correlações entre Índice de Flares e Campo Magnético para os períodos considerados.**

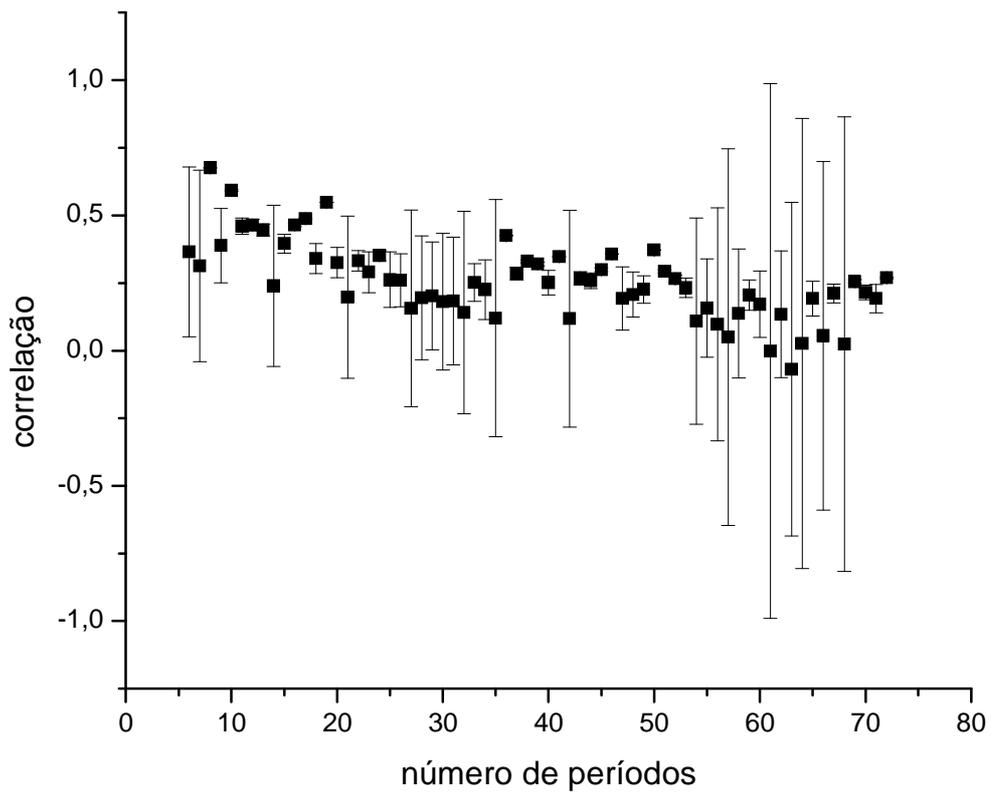

**Figura 62 – Correlações entre Índice de Flares e
Contagem de Manchas para os períodos considerados.**

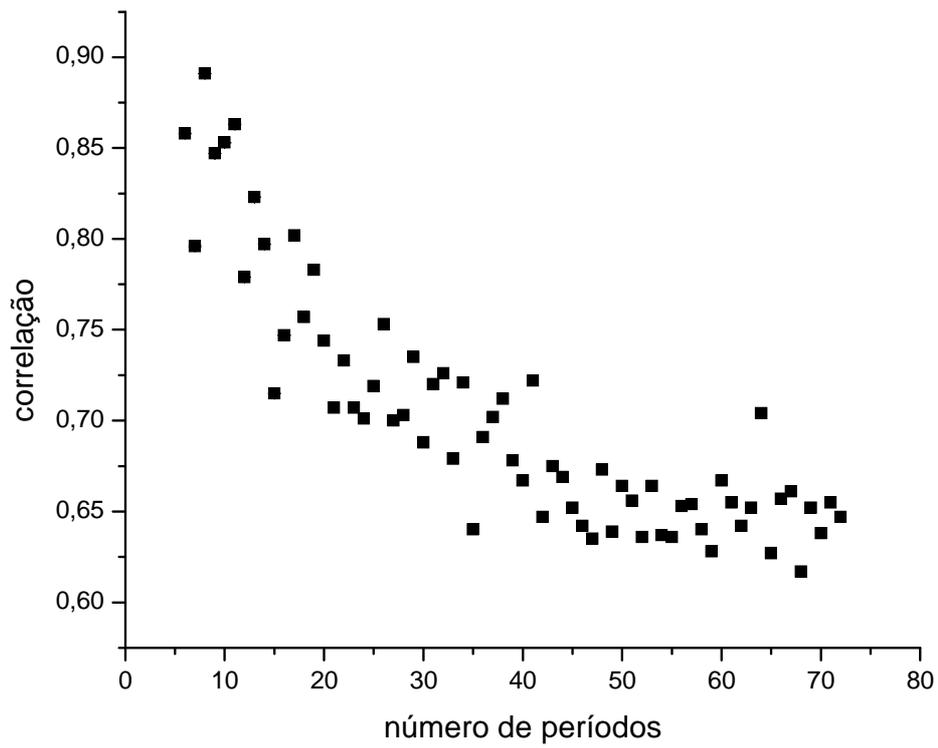

**Figura 63 – Correlações entre Índice de Flares e Fluxo Rádio para os períodos considerados.**

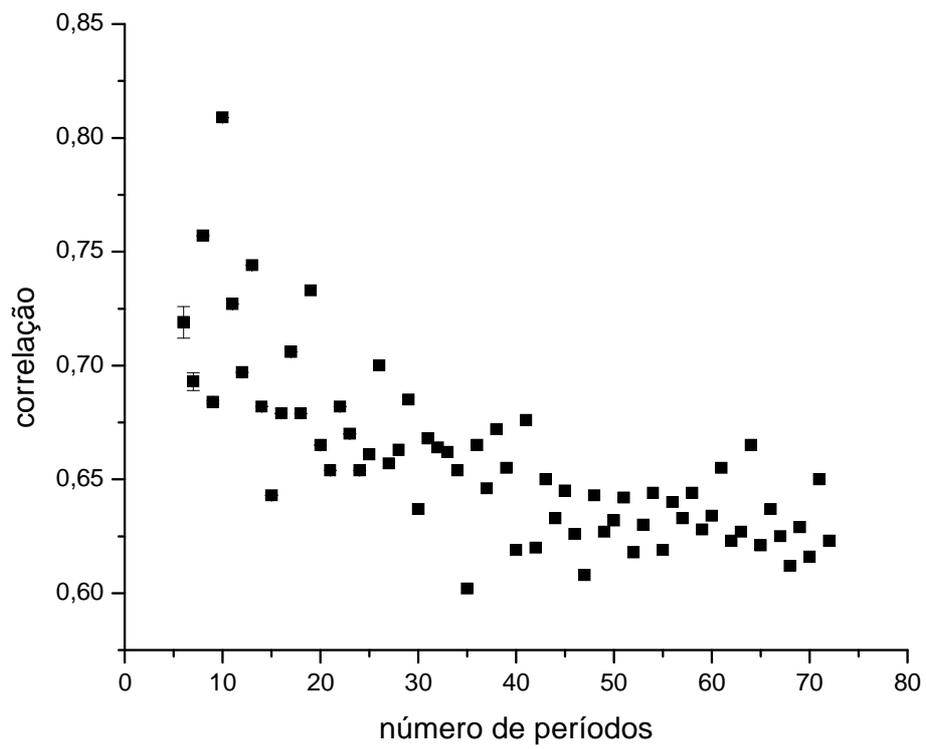

**Figura 64 – Correlações entre Irradiância e Campo Magnético para os períodos considerados.**

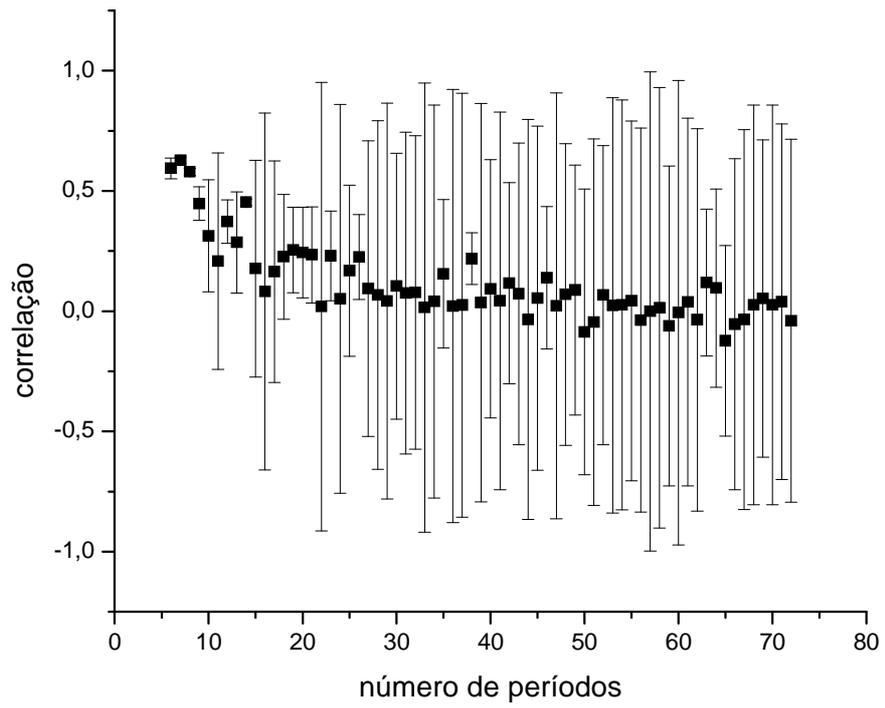

**Figura 65 – Correlações entre Irradiância e Contagem de Manchas para períodos considerados.**

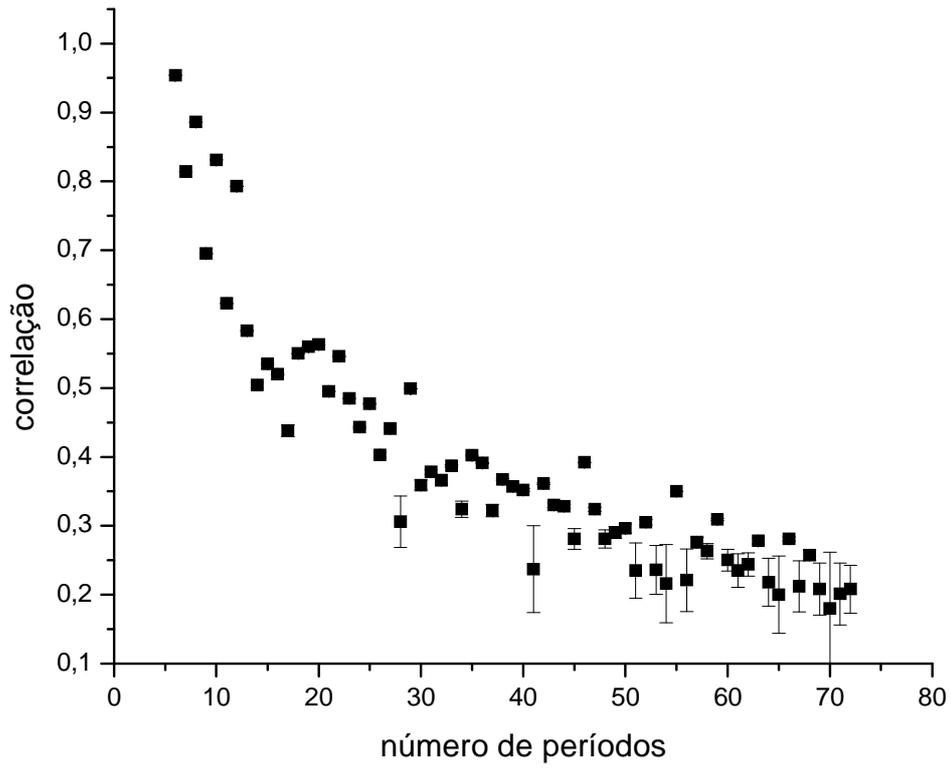

**Figura 66 – Correlações entre Irradiância e Fluxo Rádio para os períodos considerados.**

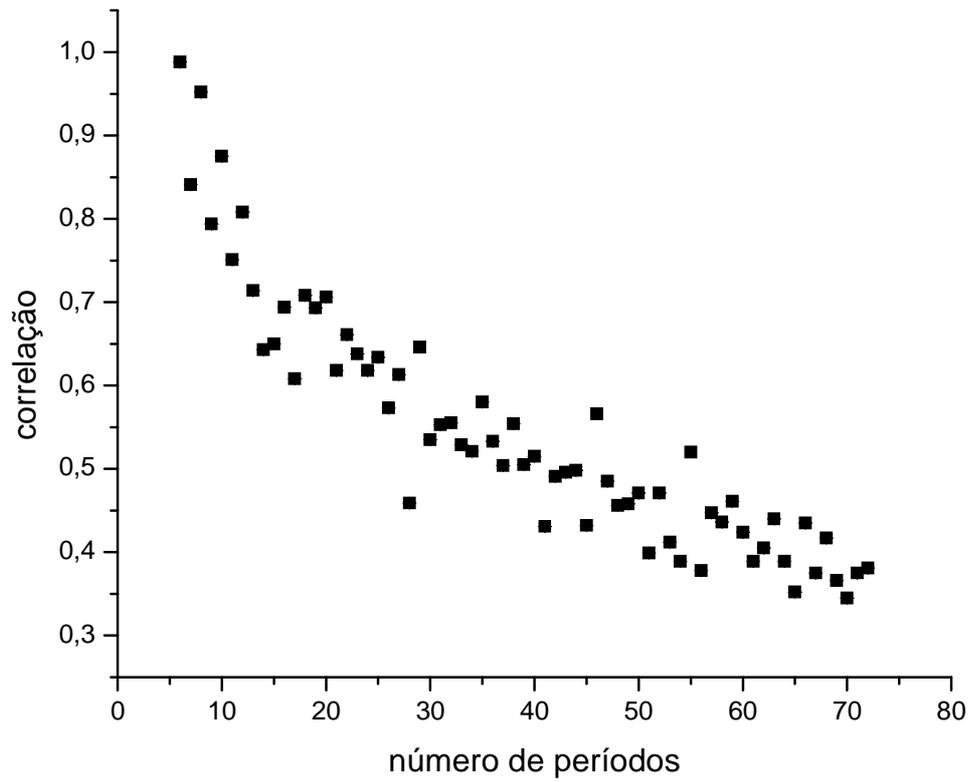

**Figura 67 – Correlações entre Campo Magnético e Contagem de Manchas para os períodos considerados.**

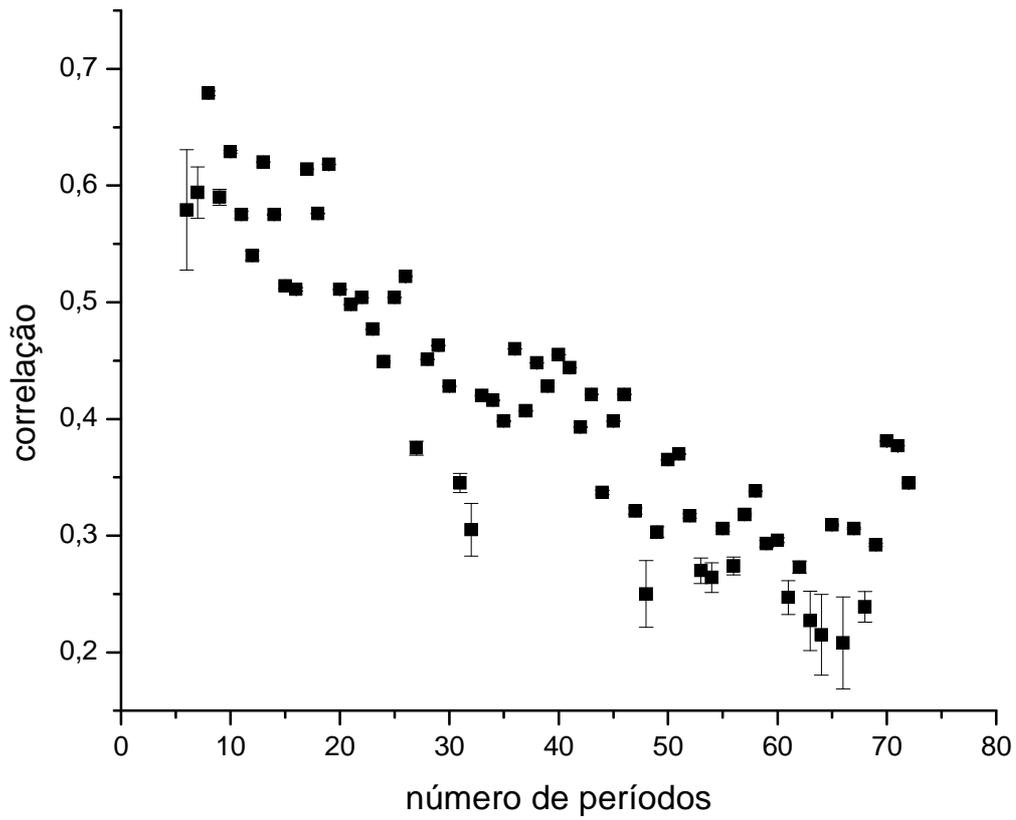

**Figura 68 – Correlações entre Campo Magnético e
Fluxo Rádio para os períodos considerados.**

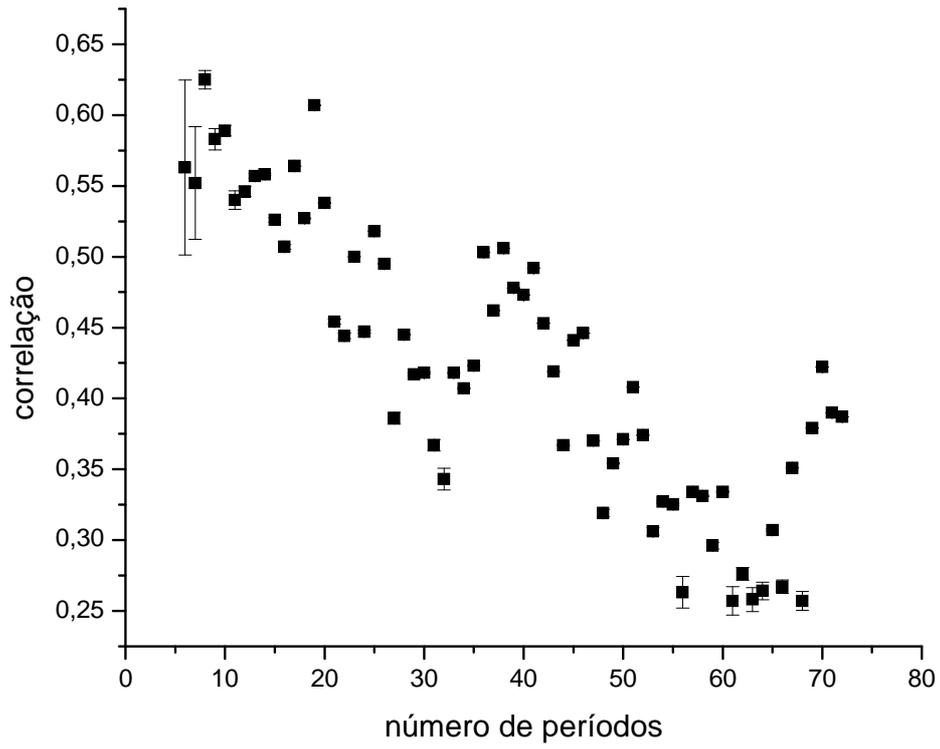

**Figura 69 – Correlações entre Contagem de Manchas e Fluxo Rádio para os períodos considerados.**

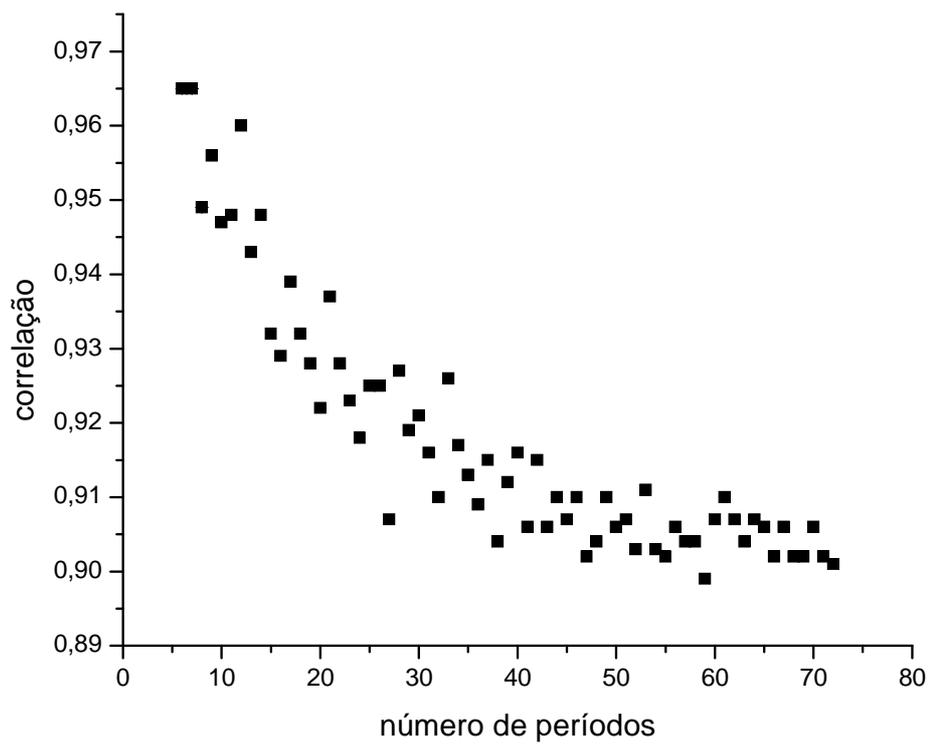

**Correlações com Defasagem.**

Várias correlações estatisticamente significantes são obtidas para o Semidiâmetro e os estimadores da atividade solar. No entanto, a única próxima da unidade é aquela entre a Contagem de Manchas e Fluxo Rádio (já esperada). As correlações são mais complexas e sua descrição demanda uma busca de fase. Graficamente isto é indicado nas figuras que comparam o Semidiâmetro Solar e outros parâmetros. Há diferenças nas datas de picos e de vales com valores que parecem ser constantes. Isto sugere que pode haver ainda uma maior coincidência entre as curvas se adiantarmos uma das curvas temporalmente em relação à outra. Os dois parâmetros podem ter uma conexão física, um deles, porém, com atraso em relação ao outro. Levando em conta este fato calculamos também correlações entre cada dupla de parâmetros considerando defasagens nas datas. Fizemos as defasagens variarem de –360 dias até +360 dias, calculando as correlações a cada dez dias de defasagem. Continuamos considerando ainda a divisão em períodos variando entre 6 e 72.

As Figuras 70 a 84 mostram para cada dupla de parâmetros todas as correlações para períodos entre 6 e 72 em função da defasagem entre os parâmetros. Em cada figura foi também traçada uma curva média. Os valores se concentram em uma faixa, destacando-se alguns valores acima e, por vezes, poucos valores abaixo. As correlações variam bastante na medida em que se defasam os parâmetros. Para quatro destas duplas as maiores correlações se dão quando a defasagem é zero: Semidiâmetro com Irradiância (notar comentário adiante), Índice de Flares com Contagem de Manchas, Índice de Flares com Fluxo Rádio e Contagem de Manchas com Fluxo Rádio. Para duas duplas as maiores correlações se deram próximo de zero: Campo Magnético com Contagem de Manchas e Campo Magnético com Fluxo Rádio. Para cinco duplas as correlações máximas ocorreram para uma defasagem em torno de cem dias: Semidiâmetro com Índice de Flares, Semidiâmetro com Campo Magnético, Semidiâmetro com Contagem de Manchas, Semidiâmetro com Fluxo Rádio e Irradiância com Contagem de Manchas. Para Irradiância com Fluxo Rádio as máximas ocorreram numa defasagem de 50 dias. Para Índice de Flares com Irradiância elas se deram para uma defasagem de –140 dias. As duplas: Índice de Flares com Campo Magnético e Irradiância com Campo Magnético têm mais de um máximo bem definido. A curva representativa da correlação entre Semidiâmetro com Irradiância sugere a existência de um segundo máximo para além dos 360 dias de defasagem. As defasagens aqui citadas se referem ao segundo parâmetro em relação ao primeiro, isto é, um número positivo significa atrasar o segundo parâmetro do número de dias lido no gráfico enquanto que um número negativo significa adiantar o segundo parâmetro.

Algumas duplas apresentam alguns máximos secundários. A Tabela IX mostra, para a defasagem onde a média das correlações foi máxima, a maior correlação, a média, a mínima, a mediana e a média do maior quartil. Indica ainda o número de valores, do total de 67, em que a correlação foi superior a 0,8, o número de valores em que ela foi superior a 0,6 e em que foi superior a 0,4.

**Tabela IX – Para o melhor caso, a defasagem em dias, a maior, a média, a menor, a mediana e a média do maior quartil das correlações entre os parâmetros indicados. E os números de correlações superiores a 0,8 a 0,6 e a 0,4 de um total de 67.**

| Parâmetros | | defas | max | med | min | mid | mq | >.8 | >.6 | >.4 |
|---|---|---|---|---|---|---|---|---|---|---|
| Semidiâmetro | Flares | 90 | 0,63 | 0,44 | 0,29 | 0,44 | 0,44 | 0 | 2 | 47 |
| Semidiâmetro | Irradiância | 0 | 0,73 | 0,57 | 0,44 | 0,56 | 0,65 | 0 | 23 | 67 |
| Semidiâmetro | C.Magnético | 140 | 0,70 | 0,39 | 0,22 | 0,36 | 0,39 | 0 | 4 | 28 |
| Semidiâmetro | Manchas | 100 | 0,76 | 0,60 | 0,45 | 0,59 | 0,67 | 0 | 29 | 67 |
| Semidiâmetro | Rádio | 90 | 0,86 | 0,73 | 0,61 | 0,72 | 0,80 | 10 | 67 | 67 |
| Flares | Irradiância | -140 | 0,94 | 0,62 | 0,42 | 0,57 | 0,65 | 10 | 28 | 67 |
| Flares | C.Magnético | -280 | 0,68 | 0,33 | 0,09 | 0,31 | 0,23 | 0 | 2 | 17 |
| Flares | Manchas | 0 | 0,89 | 0,70 | 0,62 | 0,68 | 0,79 | 7 | 67 | 67 |
| Flares | Rádio | 0 | 0,81 | 0,66 | 0,60 | 0,65 | 0,70 | 1 | 67 | 67 |
| Irradiância | C.Magnético | -200 | 0,61 | 0,41 | 0,23 | 0,42 | 0,40 | 0 | 1 | 37 |
| Irradiância | Manchas | 110 | 0,99 | 0,67 | 0,49 | 0,65 | 0,76 | 8 | 45 | 67 |
| Irradiância | Rádio | 50 | 0,99 | 0,72 | 0,59 | 0,70 | 0,81 | 8 | 66 | 67 |
| C.Magnético | Manchas | 10 | 0,69 | 0,43 | 0,26 | 0,41 | 0,45 | 0 | 6 | 35 |
| C.Magnético | Rádio | -20 | 0,63 | 0,44 | 0,25 | 0,44 | 0,45 | 0 | 3 | 44 |
| Manchas | Rádio | 0 | 0,97 | 0,92 | 0,90 | 0,91 | 0,92 | 67 | 67 | 67 |

As correlações entre os parâmetros são, nos casos onde o máximo se dá para alguma defasagem, maiores que aquelas da Tabela VIII. Considerando valores médios, o valor que têm o maior aumento é o da dupla: Índice de Flares com Irradiância que teve um aumento de 0,590. Outras três duplas tiveram a média de suas correlações aumentada em mais de 0,200: Irradiância com Campo Magnético, Semidiâmetro com Índice de Flares e Irradiância com Contagem de Manchas. Mais cinco duplas de parâmetros tiveram aumentos relevantes nas médias de suas correlações. A Tabela X mostra as diferenças nas correlações quando se consideram defasagens, para os mesmos indicadores estatísticos considerados nas duas tabelas anteriores.

**Tabela X – Diferenças entre as correlações quando se considera defasagem: a melhor defasagem em dias, a maior, a média, a menor, a mediana e a média do maior quartil das correlação entre os parâmetros indicados. E dos números de correlações superiores a 0,8 a 0,6 e a 0,4 de um total de 67.**

| Parâmetros | | defas | max | med | min | mid | Mq | >.8 | >.6 | >.4 |
|---|---|---|---|---|---|---|---|---|---|---|
| Semidiâmetro | Flares | 90 | 0,20 | 0,28 | 0,29 | 0,30 | 0,15 | 0 | 2 | 46 |
| Semidiâmetro | C.Magnético | 140 | 0,16 | 0,07 | 0,07 | 0,03 | -0,06 | 0 | 4 | 15 |
| Semidiâmetro | Manchas | 100 | 0,08 | 0,14 | 0,17 | 0,14 | 0,10 | 0 | 24 | 23 |
| Semidiâmetro | Rádio | 90 | 0,03 | 0,11 | 0,14 | 0,12 | 0,05 | 8 | 35 | 0 |
| Flares | Irradiância | -140 | 0,20 | 0,59 | 0,60 | 0,62 | 0,33 | 10 | 25 | 64 |
| Flares | C.Magnético | -280 | 0,00 | 0,07 | 0,16 | 0,06 | -0,21 | 0 | 1 | 8 |
| Irradiância | C.Magnético | -200 | -0,02 | 0,30 | 0,35 | 0,35 | 0,07 | 0 | 0 | 32 |
| Irradiância | Manchas | 110 | 0,04 | 0,27 | 0,31 | 0,30 | 0,11 | 4 | 38 | 45 |
| Irradiância | Rádio | 50 | 0,00 | 0,17 | 0,25 | 0,18 | 0,06 | 3 | 44 | 12 |
| C.Magnético | Manchas | 10 | 0,01 | 0,02 | 0,05 | 0,00 | -0,12 | 0 | 1 | 1 |
| C.Magnético | Rádio | -20 | -0,01 | 0,01 | -0,01 | 0,02 | -0,10 | 0 | 1 | 5 |

**Figura 70 – Correlações entre Semidiâmetro Solar e Índice de Flares para 6 a 72 períodos em função da defasagem em dias do segundo parâmetro.**

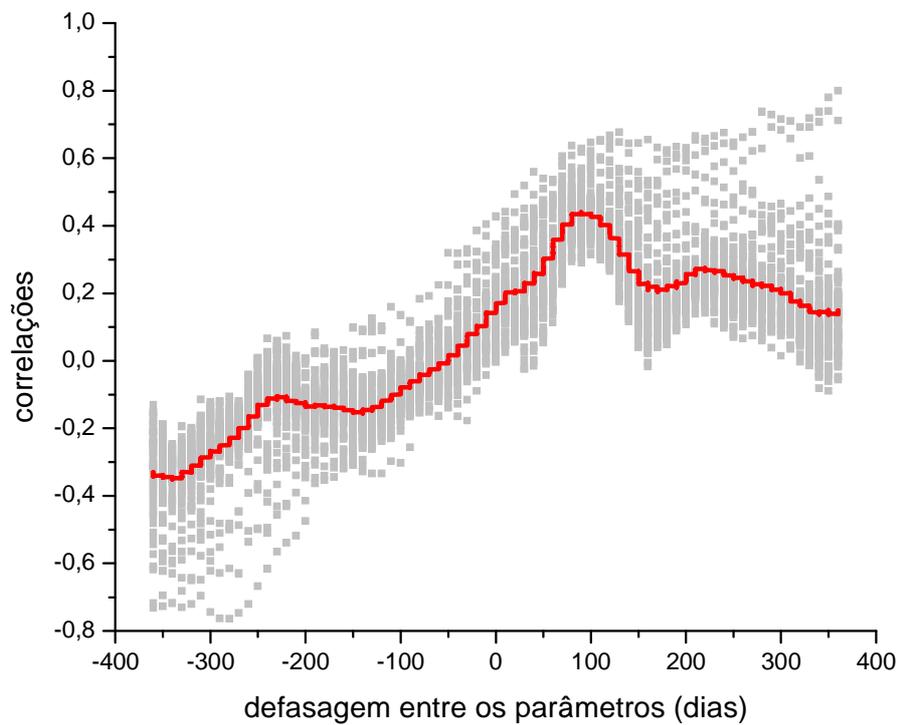

**Figura 71 – Correlações entre Semidiâmetro Solar e Irradiância para
6 a 72 períodos em função da defasagem em dias do segundo parâmetro.**

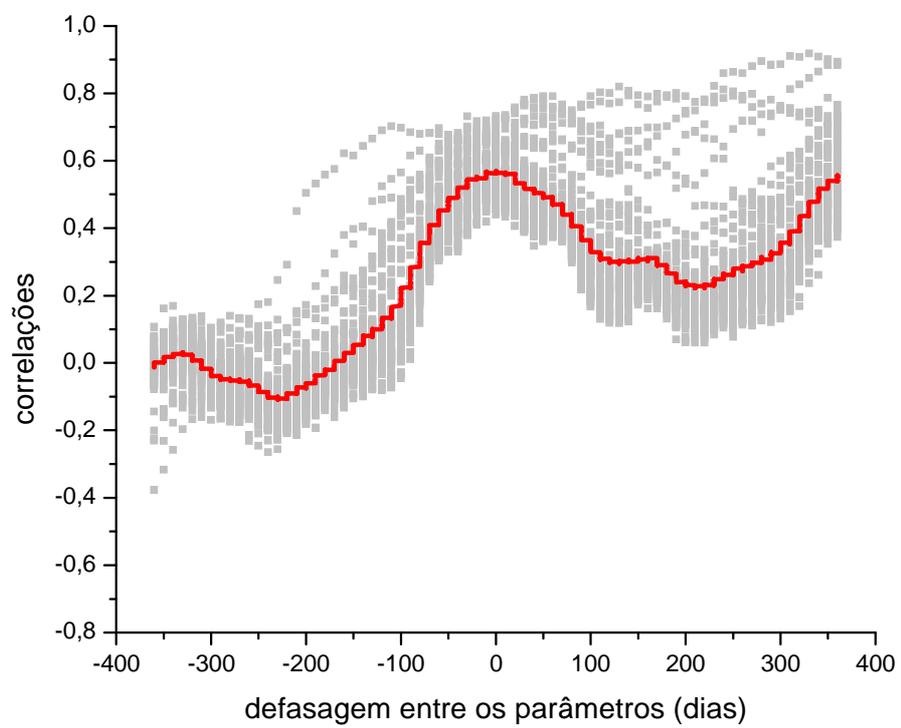

**Figura 72 – Correlações entre Semidiâmetro Solar e Campo Magnético para 6 a 72 períodos em função da defasagem em dias do segundo parâmetro.**

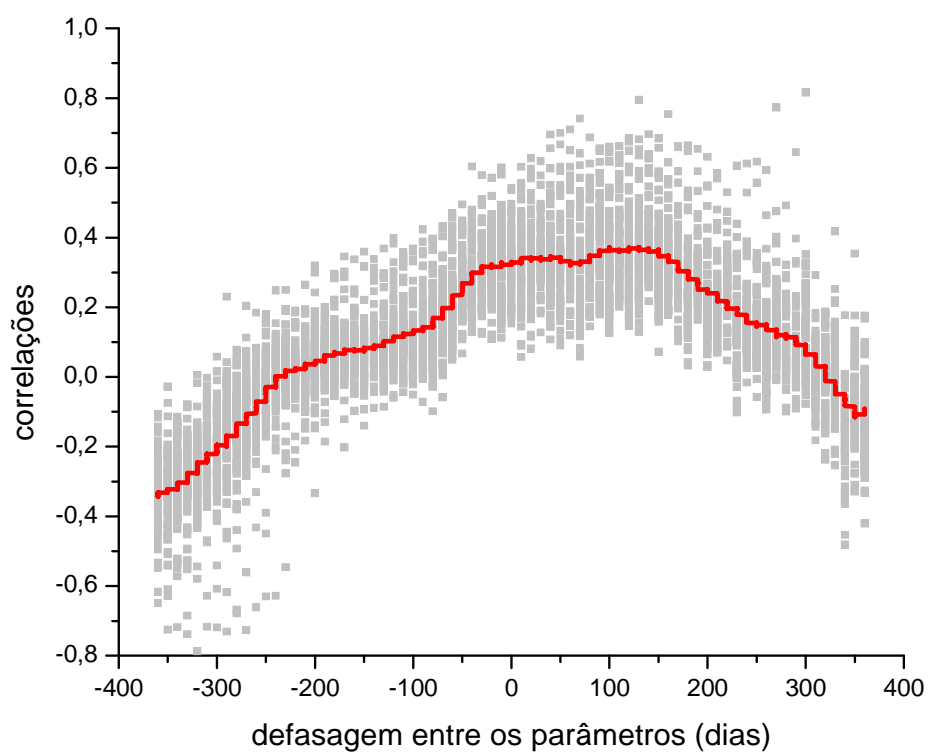

**Figura 73 – Correlações entre Semidiâmetro Solar e Contagem de Manchas para 6 a 72 períodos em função da defasagem em dias do segundo parâmetro.**

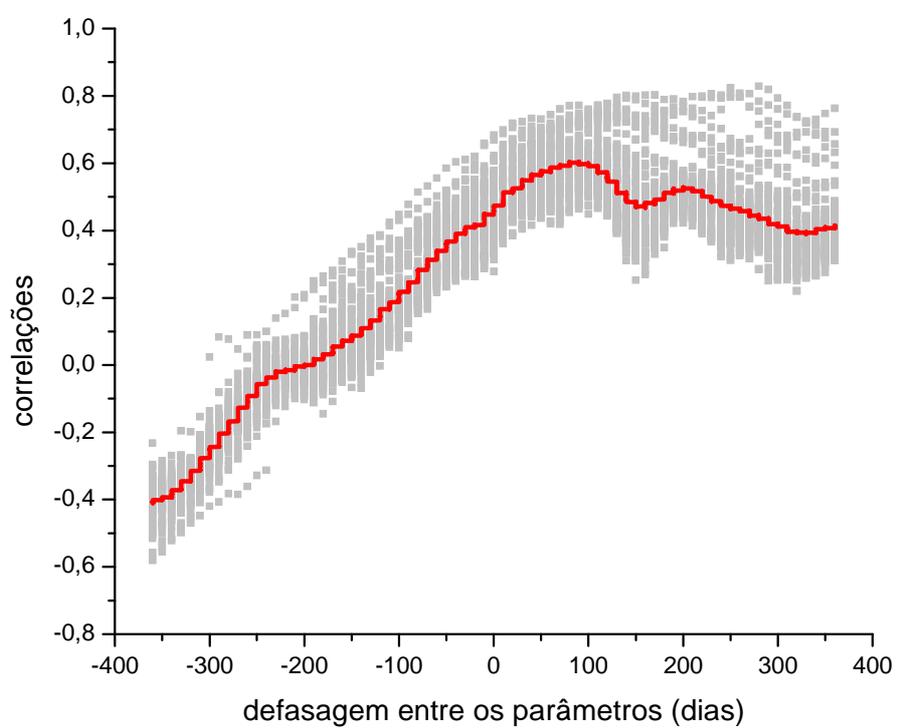

**Figura 74 – Correlações entre Semidiâmetro Solar e Fluxo Rádio para 6 a 72 períodos em função da defasagem em dias do segundo parâmetro.**

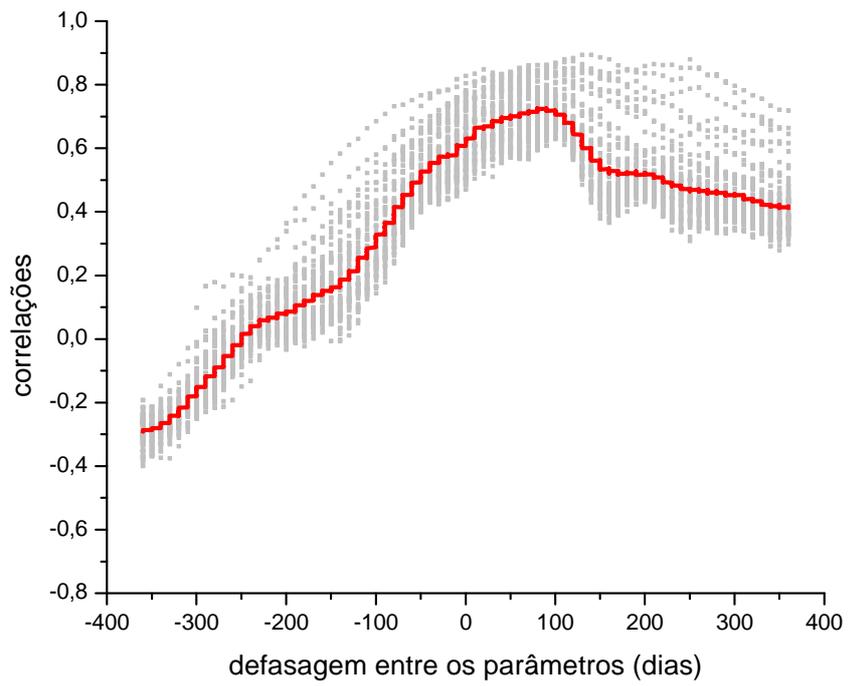

**Figura 75 – Correlações entre Índice de Flares e Irradiância para 6 a 72 períodos em função da defasagem em dias do segundo parâmetro.**

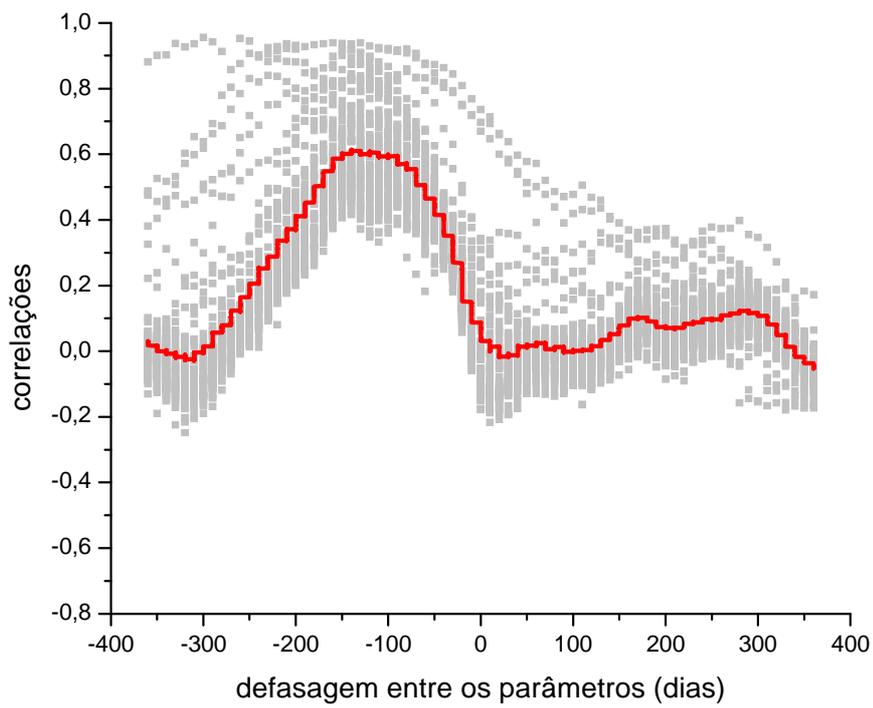

**Figura 76 – Correlações entre Índice de Flares e Campo Magnético para 6 a 72 períodos em função da defasagem em dias do segundo parâmetro.**

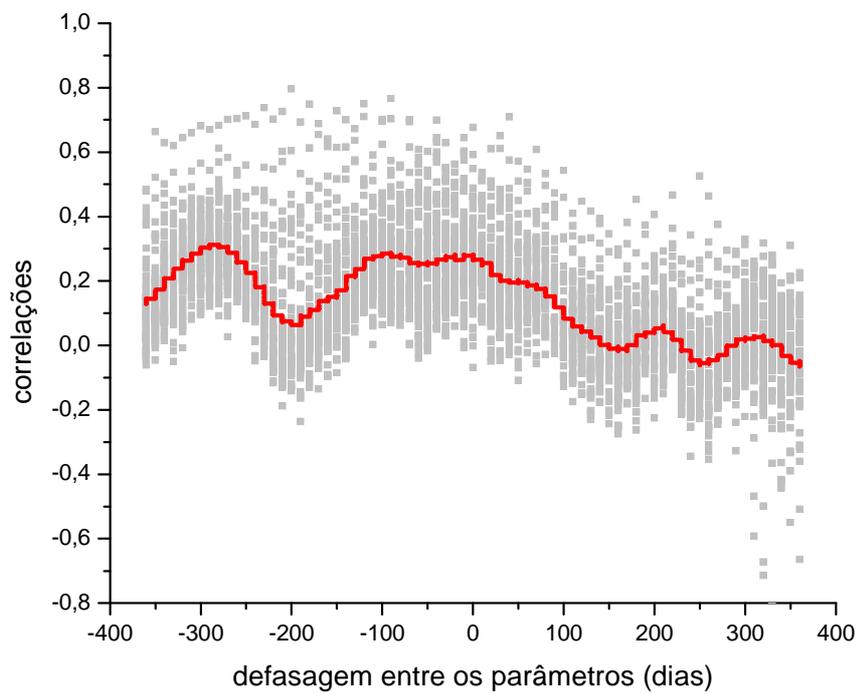

**Figura 77 – Correlações entre Índice de Flares e Contagem de Manchas para 6 a 72 períodos em função da defasagem em dias do segundo parâmetro.**

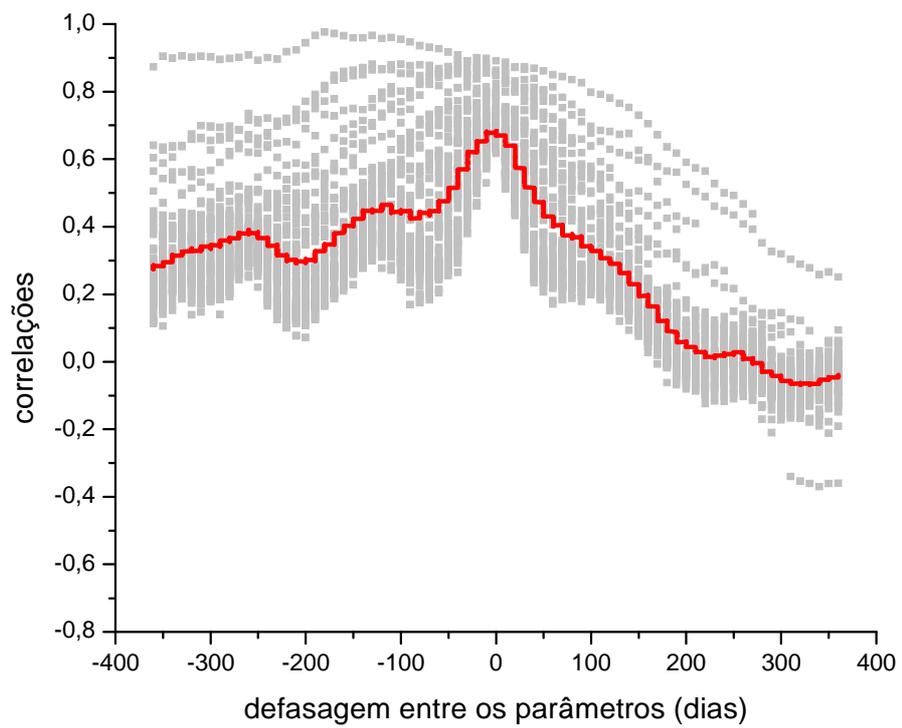

**Figura 78 – Correlações entre Índice de Flares e Fluxo Rádio para 6 a 72 períodos em função da defasagem em dias do segundo parâmetro.**

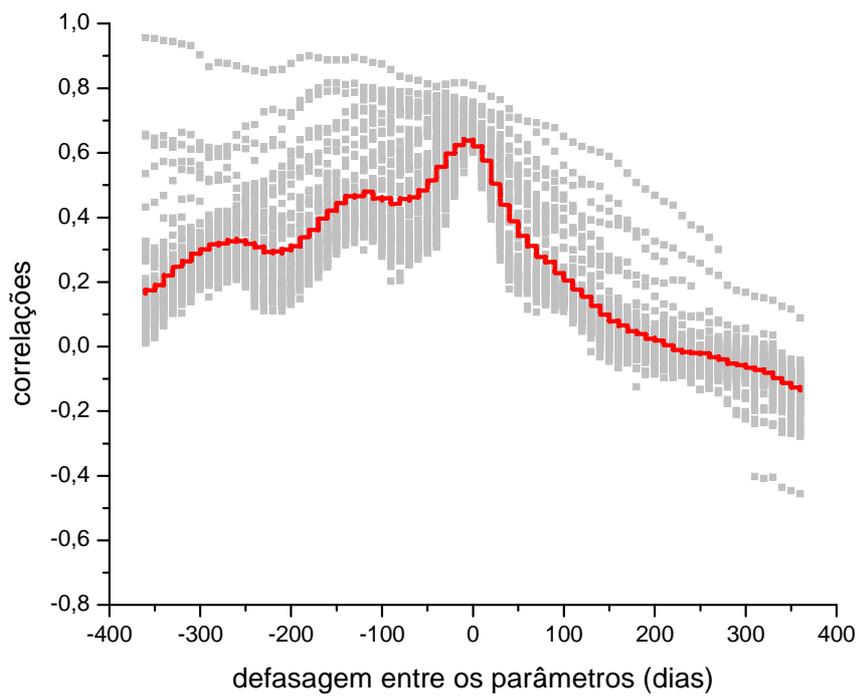

**Figura 79 – Correlações entre Irradiância e Campo Magnético para 6 a 72 períodos em função da defasagem em dias do segundo parâmetro.**

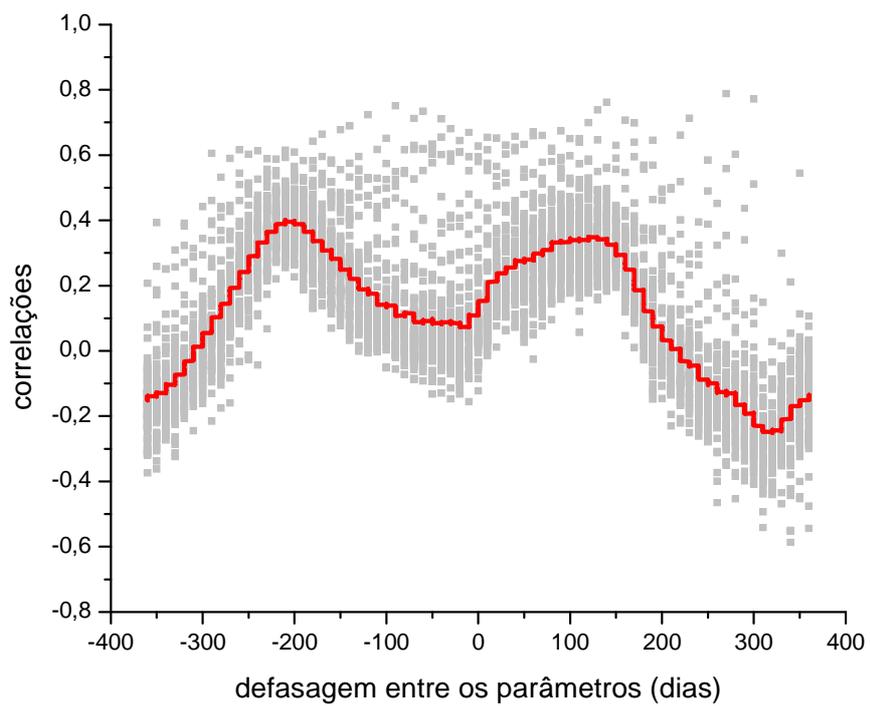

**Figura 80 – Correlações entre Irradiância e Contagem de Manchas para 6 a 72 períodos em função da defasagem em dias do segundo parâmetro.**

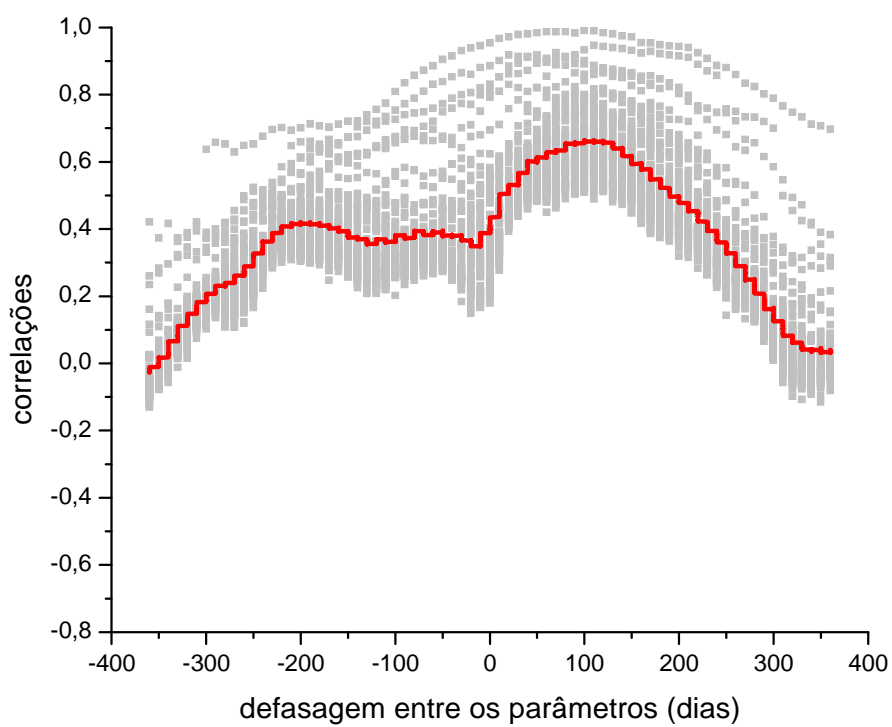

**Figura 81 – Correlações entre Irradiância e Fluxo Rádio para 6 a 72 períodos em função da defasagem em dias do segundo parâmetro.**

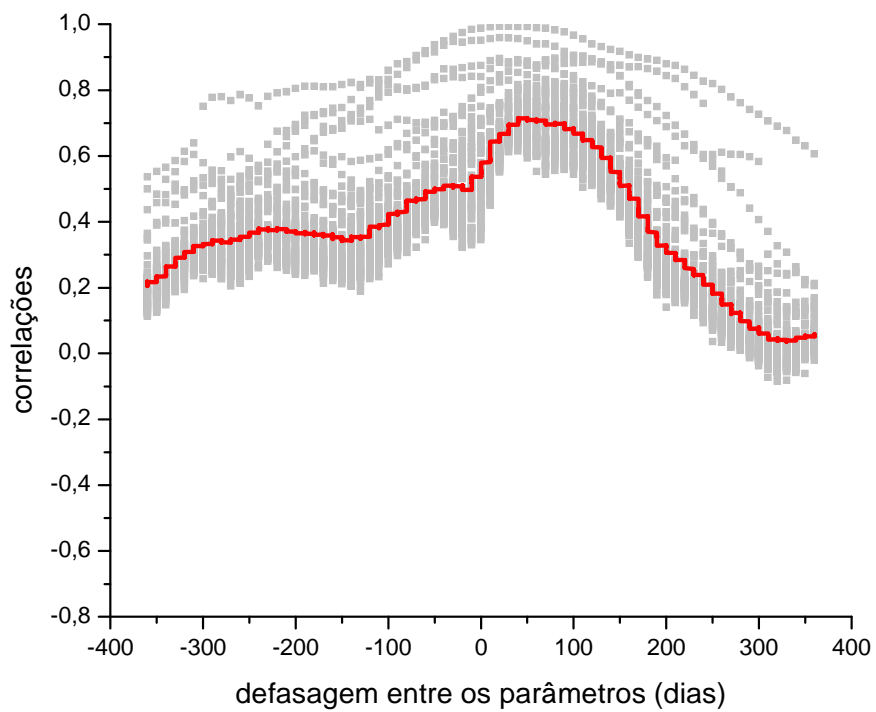

**Figura 82 – Correlações entre Campo Magnético e Contagem de Manchas para 6 a 72 períodos em função da defasagem em dias do segundo parâmetro.**

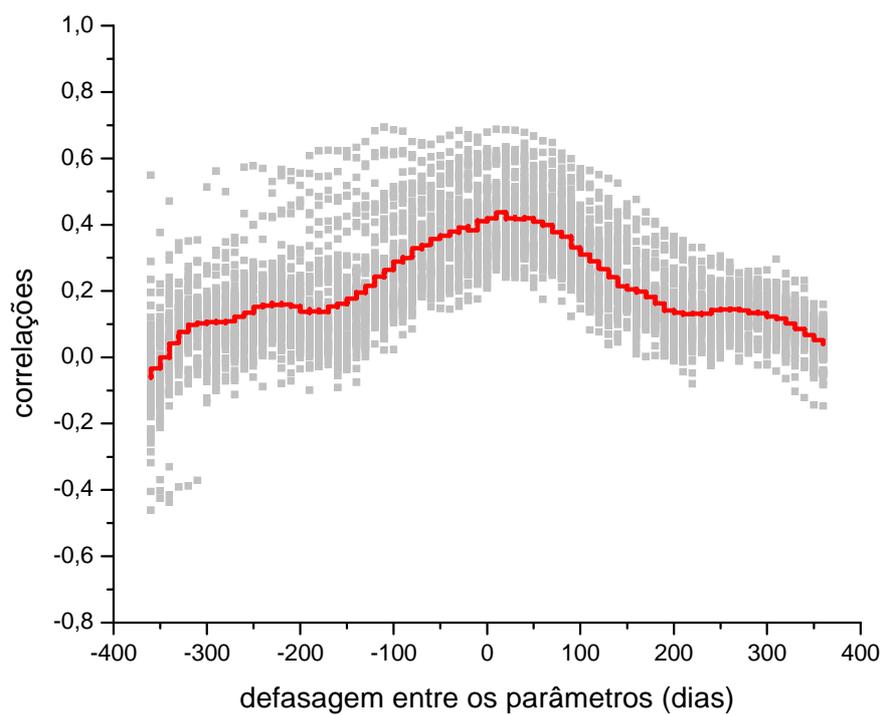

**Figura 83 – Correlações entre Campo Magnético e Fluxo Rádio para 6 a 72 períodos em função da defasagem em dias do segundo parâmetro.**

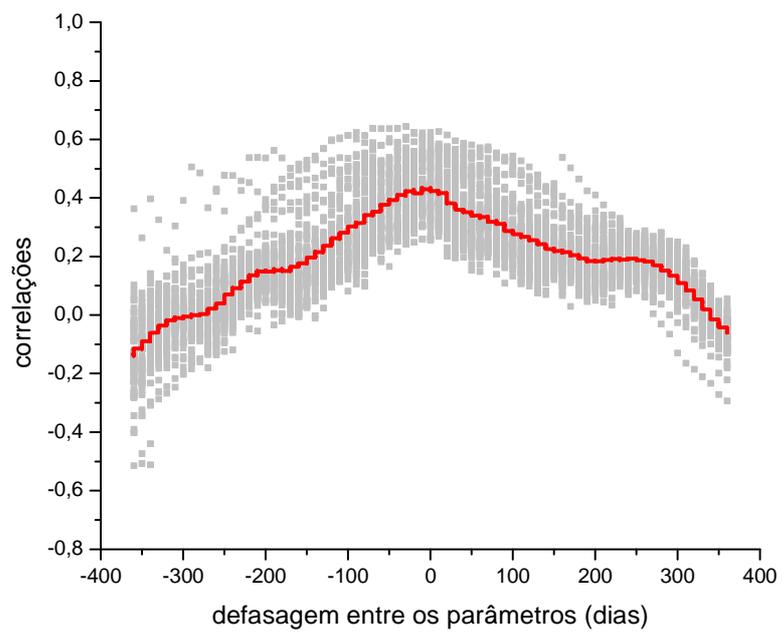

**Figura 84 – Correlações entre Contagem de Manchas e Fluxo Rádio para 6 a 72 períodos em função da defasagem em dias do segundo parâmetro.**

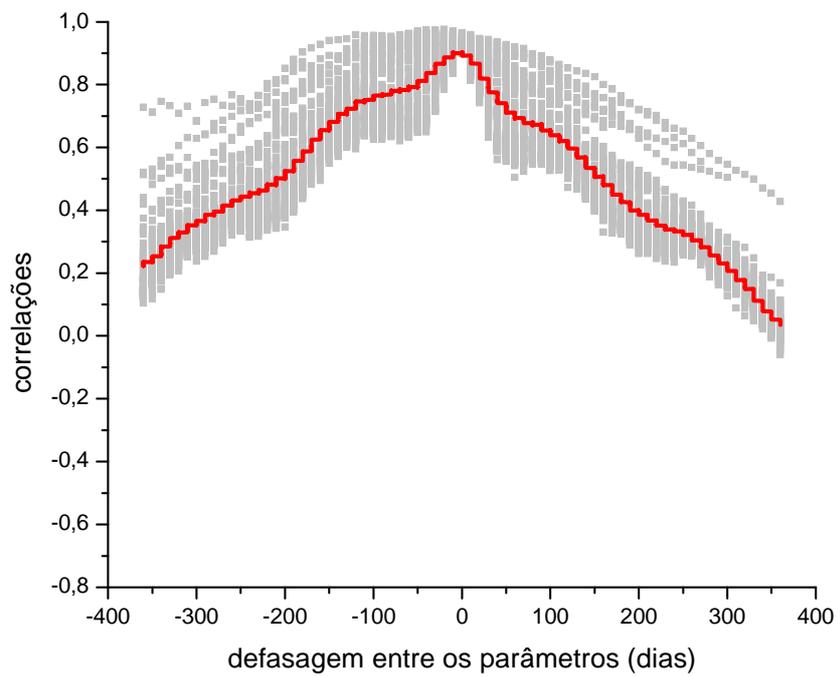

**Correlações examinando os picos de atividade.**

Há dois picos bem definidos de atividades do Sol que aparecem em todos os indicadores. Eles se situam entre as datas Julianas modificados de 1709,0 e 1919,0 e entre 2213,0 e 2378,0. Se retirarmos as medidas do Semidiâmetro e da Irradiância relativas àqueles intervalos, verificamos que o máximo das correlações se desloca significativamente da condição de simultaneidade, mantendo o patamar entre 0,5 e 0,6. Verificamos que para a série completa, a moda dos valores de defasagem de maior correlação ocorre em torno de zero como mostra a Figura 85. Entretanto para a série restrita (sem os picos), esta moda ocorre para valores em torno de 100 dias como mostra a Figura 86. Isto nos diz que há uma resposta diferente do Semidiâmetro Solar para os diferentes regimes de atividade solar, quais sejam elas as atividades moduladas pelo ciclo ou atividades de pico. Nas atividades de ciclo a variação do Semidiâmetro precede à variação da Irradiância, e o faz com cerca de cem dias de diferença. Nas atividades de pico a variação do Semidiâmetro é imediata. Isso pode explicar as diferenças encontradas por diferentes pesquisadores nas correlações entre os dois índices, enquanto alguns encontram correlações positivas, outros encontram anticorrelações. Quando se têm observações bem distribuídas e freqüentes é possível observar todos os detalhes, incluindo aí as variações que ocorrem durante os picos de atividades. Entretanto quando se observa por períodos não abrangentes, durante limitadas datas do ano, observam-se apenas os efeitos gerais das variações. A Figura 85 também nos mostra isto, para poucas divisões do período estudado o gráfico mostra uma tendência crescente dos valores da moda da maior correlação.

**Figura 85 – Moda da defasagem das maiores correlações entre Semidiâmetro Solar e Irradiância quando se considera a série completa de dados das atividades.**

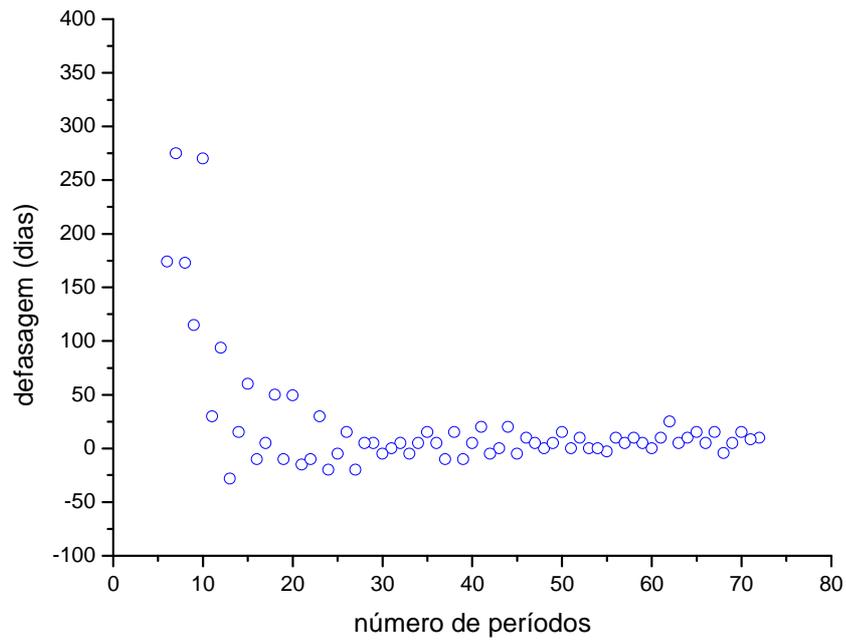

**Figura 86 – Moda da defasagem das maiores correlações entre Semidiâmetro Solar e Irradiância quando se desconsidera os picos de atividade da série de dados.**

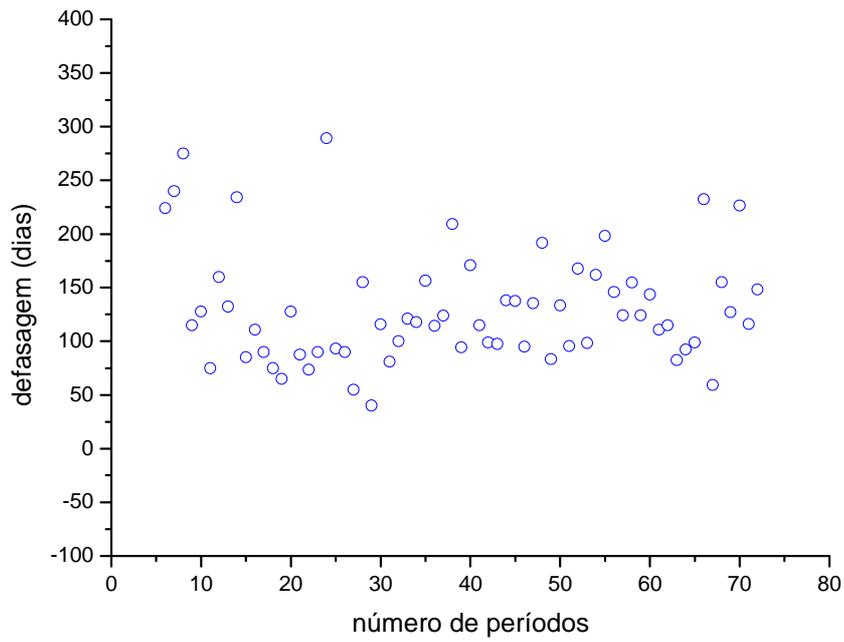

# FIGURA DO SOL.

Nosso segundo objetivo é estudar a chamada "Figura do Sol". Diferentemente de como os antigos pensavam o Sol não é um círculo perfeito. Tampouco tem a forma de uma elipse, sua forma tem uma complexidade ainda não perfeitamente compreendida (Lefebvre et Rozelot, 2003). As observações feitas com o Astrolábio Solar nos impõem uma grande limitação: observamos passagens do Sol por um almicantarado e assim determinamos o Semidiâmetro Solar em uma única direção, a direção vertical de observação. Desta maneira, a cada momento de cada dia só é possível medir o Semidiâmetro em uma determinada latitude solar. Na medida em que as horas do dia correm e principalmente, os dias do ano, é possível observar outras heliolatitudes. A Figura 87 mostra como as latitudes do Sol foram observadas a leste e a oeste em função dos dias do ano em 2002. O eixo de rotação do Sol se inclina de maneira diferente, em relação à vertical local, nas diferentes épocas do ano, variando lentamente de dia para dia e havendo uma pequena mudança ao longo das horas do dia. As latitudes solares mais baixas são observadas no começo e no final do ano e as latitudes mais altas observadas no meio do ano. Certas latitudes são observadas durante um maior número de dias que outras.

Os dados de observação dos seis anos disponíveis fazem um conjunto de 16.523 valores que cobrem inúmeras vezes todas as heliolatitudes. A Figura 88 mostra o número de observações do Semidiâmetro Solar feitas no período de estudo por faixas de cinco graus de latitude solar. Pode-se ver que há um maior número de observações até 10 graus e entre 60 e 75 graus. Todas as faixas têm mais de 480 valores observados exceto as faixas acima de 80 graus. Assim, a figura do Sol pode ser bem estudada até 80 graus de latitude. A evolução da figura deve ser estudada com cuidado uma vez que há épocas certas para se observar cada faixa de latitudes como exemplifica a Figura 89. Nela aparecem os valores observados para a faixa de heliolatitudes de zero a cinco graus. Pode-se ver que há períodos onde se concentram as observações e outros onde não há observações.

Inicialmente colocamos todos os valores de Semidiâmetro em função da latitude solar e obtivemos a Figura 90. Fizemos a média corridas de 500 pontos do Semidiâmetro Solar e, temos então, o lugar médio do Semidiâmetro de cada latitude do Sol durante o período de março de 1998 a novembro de 2003. Uma linha reta marca a tendência geral que se mostra descendente como se espera de uma esfera fluida em rotação. Em torno da reta os valores flutuam ora acima ora abaixo e há cinco pontos com uma depressão profunda em torno de 0,1 segundos de arco em relação a seus vizinhos, eles estão em torno de zero graus, 5 graus, 12 graus, 29 graus e 72 graus. Acima da reta se destacam três pontos, a 9 graus, a 38 graus e a 50 graus. Devemos lembrar que durante o período de estudo a figura do Sol sofreu muitas mudanças em todas as suas latitudes, por isso, este valores mostram apenas uma média e convém examinar como estas variações ocorrem, ou seja, como o Sol varia em cada uma de suas latitudes ao longo do tempo. Para isso, dividimos as latitudes do Sol em doze faixas de 7,5 graus e calculamos a média do Semidiâmetro Solar em cada uma destas faixas para cada um dos anos do estudo (1998 a 2003). Não pudemos fazer para a última faixa que vai de 82,5 a 90,0 graus porque ela contém menos de 30 pontos em alguns dos anos o que oferece pouca precisão aos cálculos. A Figura 91 mostra em seis curvas anuais, o Semidiâmetro médio de cada faixa. Podemos ver que de um modo geral todas as latitudes se mantém iguais de 1998 a 1999, crescem de 1999 a 2000 se mantêm iguais entre 2000 e 2001, tornam a crescer entre 2001 e 2002 e depois caem de 2002 para 2003. Cada faixa de latitude tem, entretanto, sua evolução própria. Para vermos como se comporta cada faixa construímos as Figuras 92 a 94 que mostram a evolução temporal do Semidiâmetro médio de cada faixa ao longo dos seis anos. Das onze faixas, sete têm diminuição suave do Semidiâmetro, ou seja, menos de 0,15 segundos de arco de 1998 para 1999 e quatro têm crescimento suave. Todas têm crescimento de 1999 para 2000, oito de maneira suave. De 2000 para 2001, uma tem crescimento substancial, seis têm crescimento suave, duas tem diminuição suave e duas têm diminuição substancial. De 2001 para 2002, quatro têm crescimento substancial, seis têm crescimento suave e uma tem diminuição suave do Semidiâmetro. De 2002 para 2003 todas diminuem, três suavemente. A Tabela VIII mostra estas variações de Semidiâmetro.

A Figura 95 compara a média dos módulos das variações anuais cujos valores estão na última coluna da Tabela VIII, desta figura podemos dizer que há no Sol duas faixas de latitudes: de 22,5 a 30,0 graus e de 45,0 a 52,5 graus que são sujeitas a variações suaves de Semidiâmetro. Há duas faixas de latitudes: de 0,0 a 7,5 graus e de 52,5 a 67,5 graus que são sujeitas a variações medianas de Semidiâmetro e três faixas: 7,5 a 22,5 graus, 30,0 a 45,0 graus e 67,5 a 82,5 graus sujeitas a variações fortes do Semidiâmetro.

Cabe ressaltar que as variações de Semidiâmetro solar aqui observadas não têm qualquer relação com os estudos de heliossimologia, primeiramente porque não temos condição de medir variações do semidiâmetro ao longos das longitudes do Sol, além disto as variações observadas na heliossismologia contemplam uma faixa temporal bem distante da nossa (variações da ordem de minutos).

**Tabela VIII – Variação do Semidiâmetro médio solar entre dois anos consecutivos para faixas de heliolatitude de 7,5 graus de largura.**

| Latitudes (graus) | 1998 a 1999 | 1999 a 2000 | 2000 a 2001 | 2001 a 2002 | 2002 a 2003 | média dos módulos |
|---|---|---|---|---|---|---|
| 0,0 a 7,5 | -0,061 | 0,057 | 0,110 | 0,087 | -0,135 | 0,090 |
| 7,5 a 15,0 | -0,028 | 0,161 | 0,081 | 0,185 | -0,223 | 0,136 |
| 15,0 a 22,5 | 0,024 | 0,210 | -0,169 | 0,195 | -0,307 | 0,181 |
| 22,5 a 30,0 | 0,022 | 0,178 | 0,032 | 0,015 | -0,063 | 0,062 |
| 30,0 a 37,5 | -0,067 | 0,172 | 0,183 | -0,040 | -0,161 | 0,125 |
| 37,5 a 45,0 | -0,136 | 0,232 | 0,055 | 0,037 | -0,258 | 0,144 |
| 45,0 a 52,5 | 0,019 | 0,062 | -0,038 | 0,046 | -0,061 | 0,045 |
| 52,5 a 60,0 | 0,107 | 0,006 | 0,032 | 0,094 | -0,187 | 0,085 |
| 60,0 a 67,5 | -0,077 | 0,122 | 0,058 | 0,013 | -0,157 | 0,085 |
| 67,5 a 75,0 | -0,094 | 0,127 | -0,068 | 0,204 | -0,151 | 0,129 |
| 75,0 a 82,5 | -0,103 | 0,066 | -0,230 | 0,534 | -0,180 | 0,223 |

**Figura 87 – As latitudes do Sol observadas durante 2002 a leste e a oeste.**

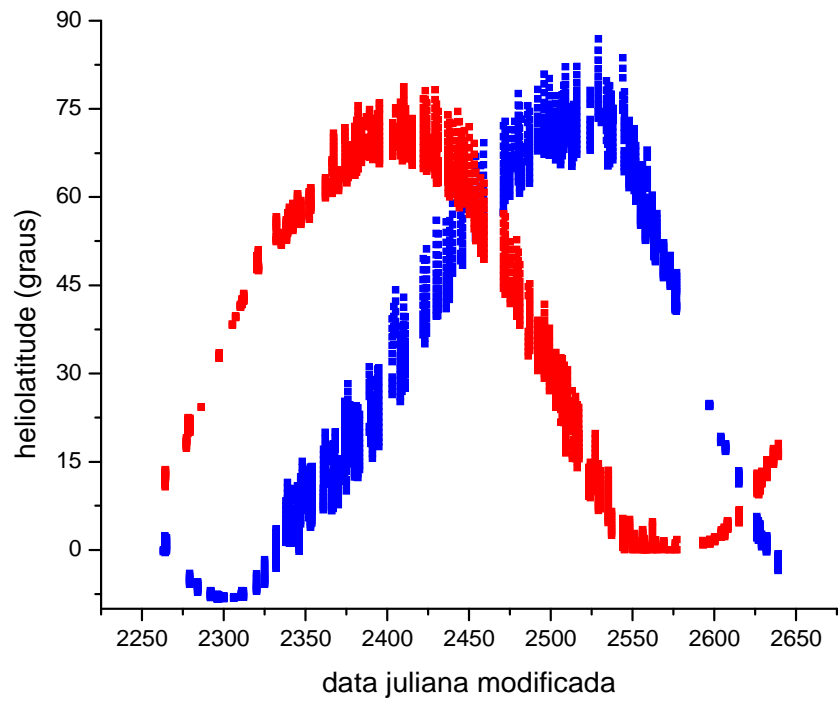

**Figura 88 - Número de observações do Semidiâmetro Solar por faixas de latitude solar.**

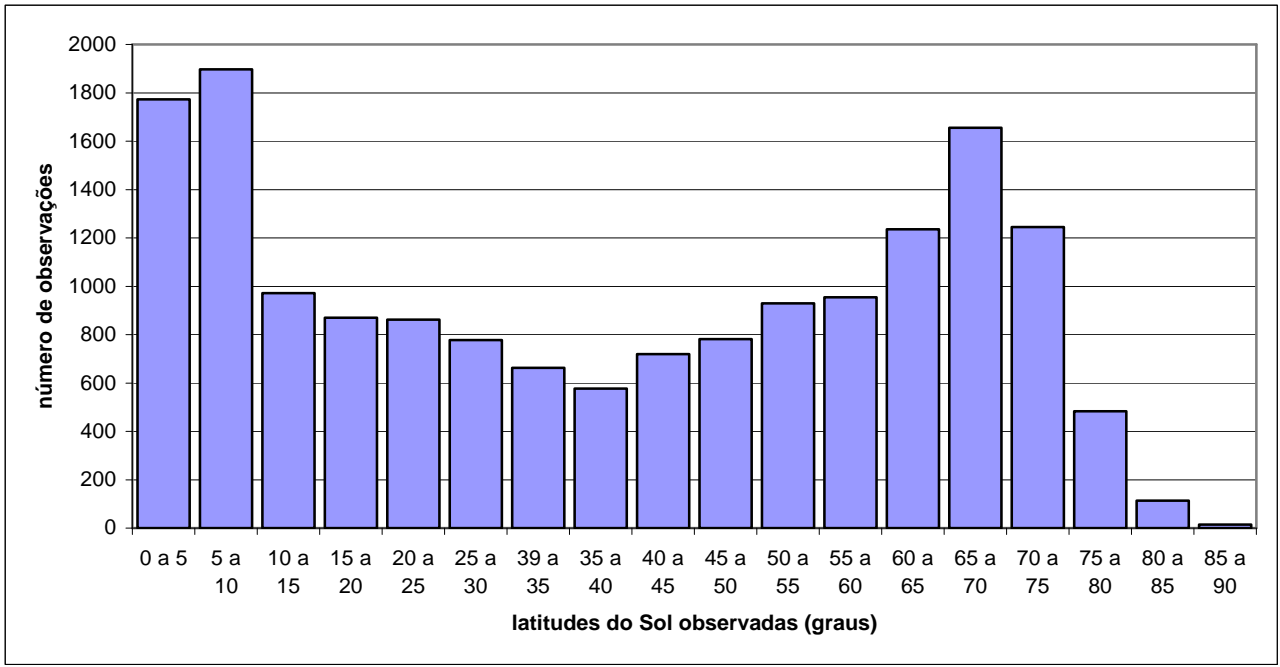

**Figura 89 - As datas em que o Semidiâmetro Solar foi observado na faixa de zero a cinco graus de latitude solar.**

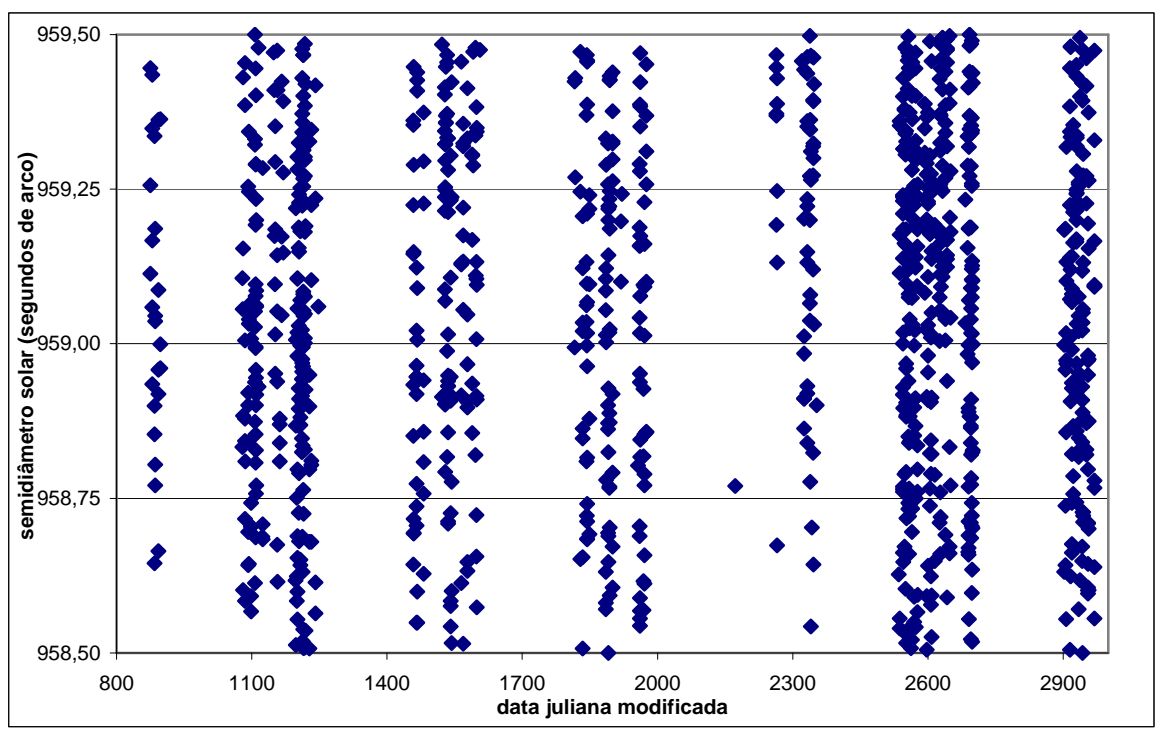

**Figura 90 – A Figura do Sol – média corridas de 150 pontos dos valores de Semidiâmetro Solar em função das latitudes do Sol onde foram observados.**

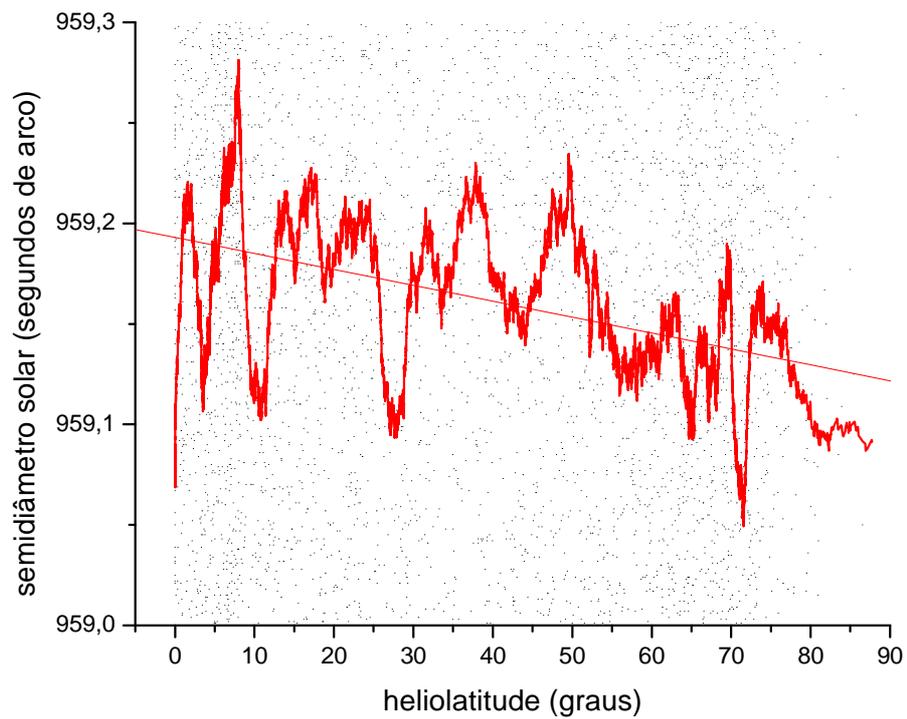

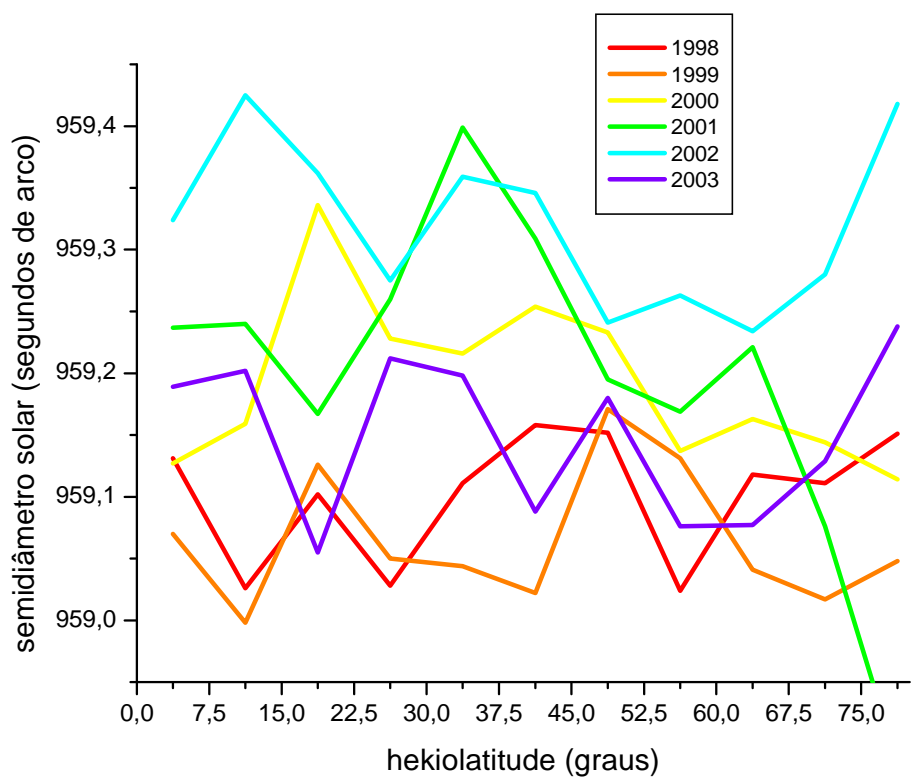

**Figura 91 – Média do Semidiâmetro por faixas de 7,5 graus de latitude solar para os anos de 1998 a 2003.**

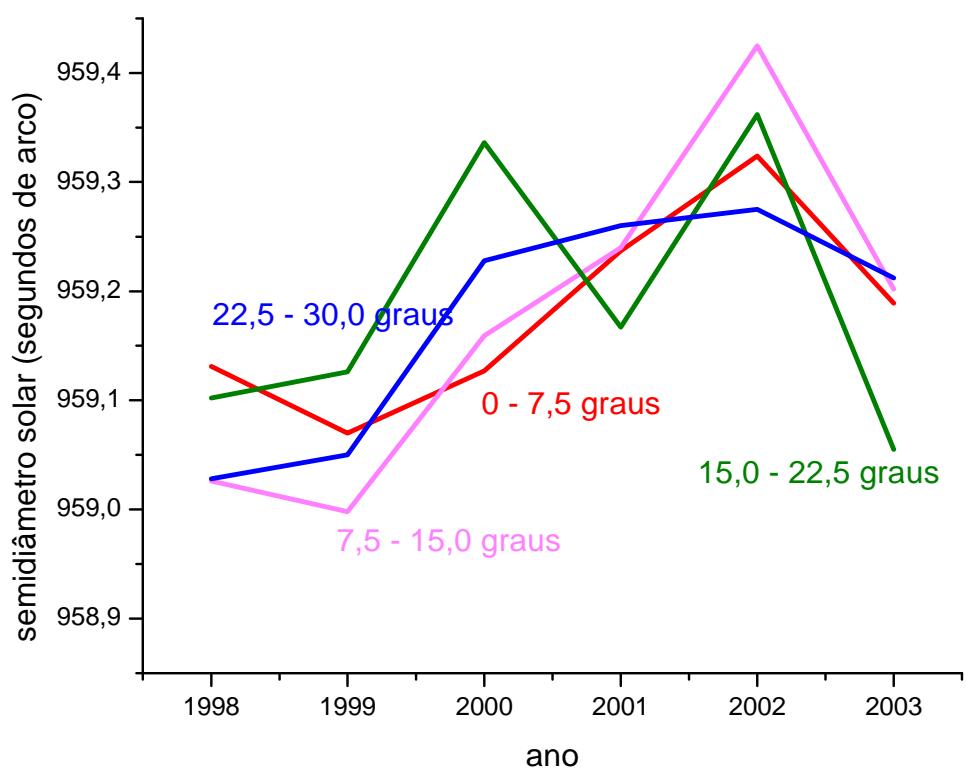

**Figura 92 – Evolução temporal da média do Semidiâmetro de quatro faixas de latitudes baixas do Sol. A incerteza é sempre menor que 0,05 segundos de arco.**

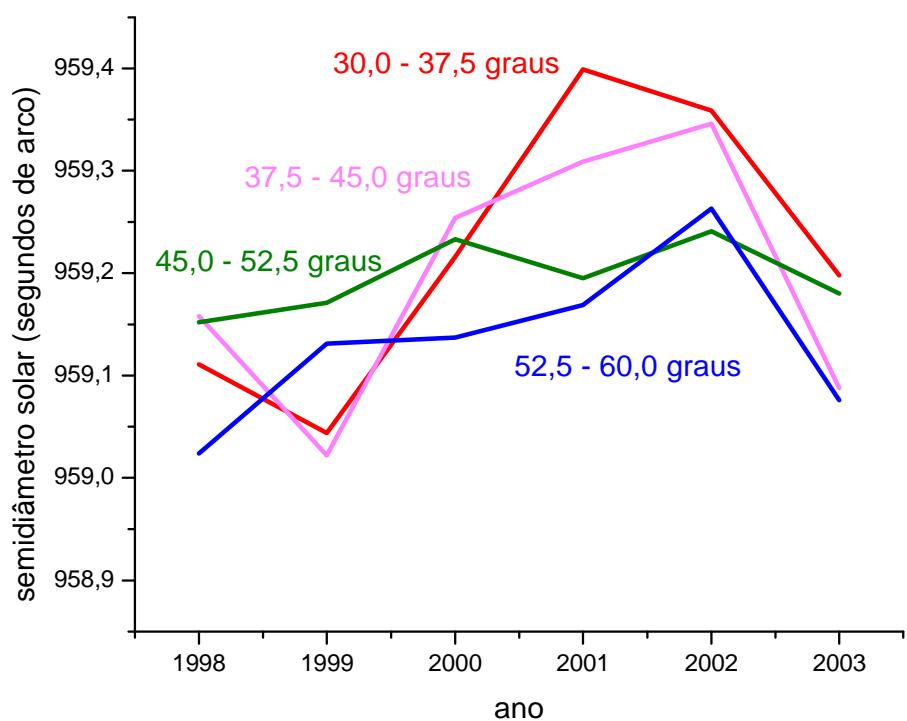

**Figura 93 – Evolução temporal da média do Semidiâmetro de quatro faixas de latitudes medianas do Sol. A incerteza é sempre menor que 0,06 segundos de arco.**

**Figura 94 – Evolução temporal da média do Semidiâmetro de três faixas de latitudes altas do Sol. A incerteza é sempre menor que 0,09 segundos de arco para a faixa mais próxima do pólo e sempre menor que 0,04 para as outras duas.**

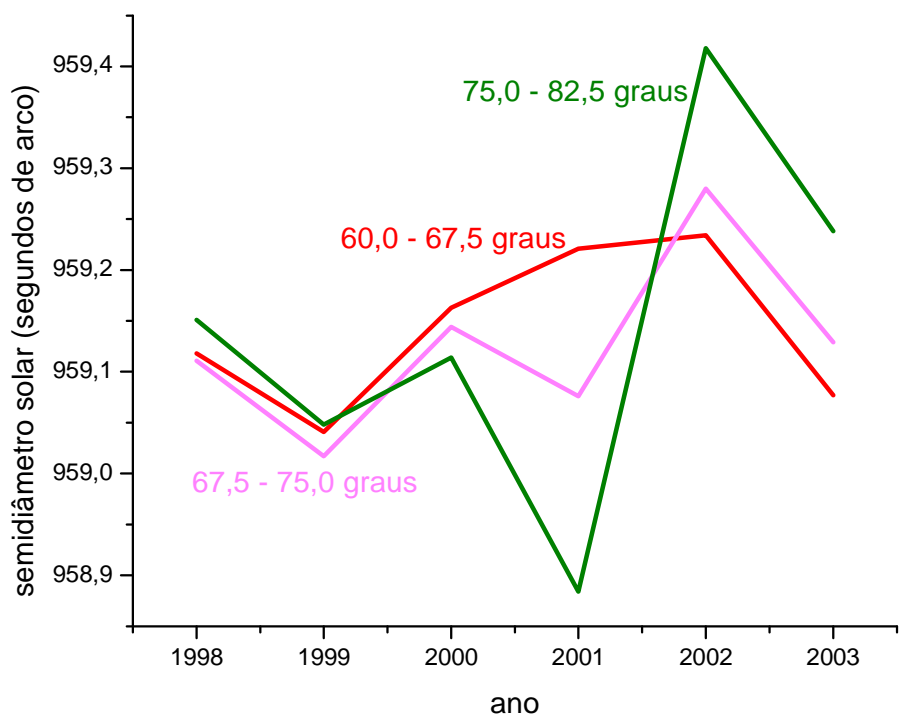

**Figura 95 – Média do módulo da variação anual de Semidiâmetro Solar para cada faixa de heliolatitude. A intensidade da cor diferencia as de variação suave, mediana e forte.**

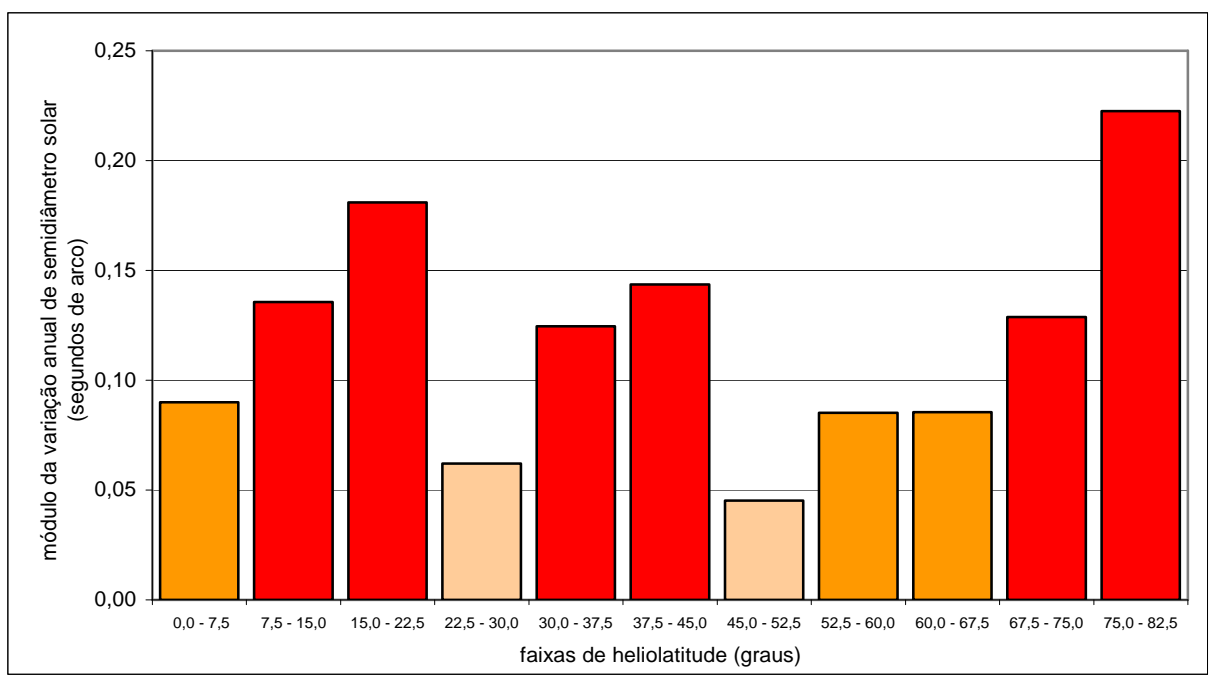

# CONCLUSÕES.

**Correção de dados**

Os dados das observações do Semidiâmetro do Sol feitas no ON durante os anos de 1998 a 2003 foram tratados de modo a se retirar deles influências externas introduzidas por alguns fatores ligados ao instrumento de observação ou ao modo de como dele se utiliza. Ao se retirar estas influências, obtivemos uma série de valores menos contaminados para a análise, bem como, para serem posteriormente utilizados em outras investigações.

As correções efetuadas foram sempre inferiores ao desvio padrão dos valores observados. Isto garante a integridade dos dados observados. As correções foram sempre pouco superiores para os valores observados a leste que para os observados a oeste. Isto ocorre porque há maior estabilidade das temperaturas à tarde de que pelas manhãs.

Quando conhecemos os parâmetros que contribuem para os erros de observação podemos retirar sua influência por meio de uma correlação linear entre estes parâmetros e os valores observados. Quando não os conhecemos podemos lançar mão de algum modelo estatístico para retirar os erros da série observada.

Apesar das correções terem sido feitas em etapas diversas, utilizando métodos diferentes e feita por pesquisadores diferentes, os valores finais corrigidos exibem uniformidade. Os desvios padrões das séries corrigidas são bastante semelhantes.

**Correlações entes Semidiâmetro e outros parâmetros.**

As variações do Semidiâmetro do Sol exibem uma correlação alta com as variações da Irradiância, com as variações da Contagem de Manchas e com as variações do Fluxo Rádio em 10,7cm. Exibem alguma correlação com o Campo Magnético Integrado, e pouca ou nenhuma correlação com o Índice de Flares.

Há uma correlação muito forte entre a Contagem de Manchas e o Fluxo Rádio, já esperada. Há correlações altas entre o Índice de Flares e a Contagem de Manchas, entre o Índice de Flares e o Fluxo Rádio e entre a Irradiância e o Fluxo Rádio.

Todas as correlações são mais fortes quando se compara períodos maiores, e mais fracas quando se compara períodos menores de tempo. Em outras palavras, há correlações maiores na forma geral das curvas e perde-se correlação quando se vai para o detalhe das curvas.

Quando se impõe algum atraso ou avanço temporal entre as séries de parâmetros a correlação aumenta, na maioria dos casos. Em relação ao Semidiâmetro Solar as correlações com outros índices avaliados aumentaram, com uma única exceção. Nestes casos, os valores máximos de correlação ocorreram para uma defasagem em torno de cem dias entre as séries comparadas. A exceção foi com o Índice de Flares cuja correlação máxima ocorreu para as séries sem defasagem.

Algumas duplas de parâmetros apresentam dois picos de correlações máximas, ou um máximo secundário de correlações para defasagens diferentes. Entre Semidiâmetro e Irradiância há um segundo pico de máximas correlações para além de um ano de defasagem. Entre o Semidiâmetro Solar e a Contagem de Manchas há um pico secundário para uma defasagem de 200 dias.

Quando examinamos as séries de dados do Semidiâmetro Solar com as da Irradiância, incluindo ou não as atividades de pico, encontramos que para as séries completas a moda dos valores de defasagem onde ocorrem as maiores correlações se dá em torno de zero enquanto que para a série restrita, sem as atividades de pico, esta moda se dá em torno de 100 dias de defasagem. No primeiro caso a variação do Semidiâmetro é simultânea com a variação da Irradiância, enquanto que no caso restrito a variação do Semidiâmetro ocorre 100 dias antes daquela da Irradiância. Algumas críticas aos valores variação de Semidiâmetro Solar encontrados se dão ao fato de que alguns autores encontram correlações positivas com as variações de Irradiância enquanto que outros encontram anticorrelações. Os resultados aqui encontrados quando examinamos as séries de dados incluindo ou não atividades de pico permitem explicar estas aparentes controvérsias.

**Figura do Sol**

A forma do Sol é complexa, há uma tendência geral descendente em torno da qual os valores flutuam ora acima ora abaixo. Há cinco pontos com uma depressão profunda de 0,1 segundos de arco em relação a seus vizinhos, eles estão em torno de zero graus, 5 graus, 12 graus, 29 graus e 72 graus. Acima da reta se destacam três pontos, a 9 graus, a 38 graus e a 50 graus.

De maneira geral, as faixas de latitudes se mantêm iguais de 1998 a 1999, crescem de 1999 a 2000 se mantêm iguais entre 2000 e 2001, tornam a crescer entre 2001 e 2002 e depois caem de 2002 para 2003. Cada faixa de latitude tem, entretanto, sua evolução temporal independente. Há faixas de latitude solar onde as variações do Semidiâmetro são mais suaves como entre 22,5 e 30,0 graus e entre 45,0 e 52,5 graus. Há faixas de latitude onde estas são mais fortes como entre 7,5 e 22,5 graus, entre 30,0 e 45,0 graus e acima de 67,5 graus.

Embora nossa coleção de dados seja bem grande ainda não é suficiente para um estudo mais conclusivo da variação da figura do Sol, pois, o Astrolábio Solar impõem uma severa restrição nas latitudes solares observadas em cada época do ano. Neste sentido é muito desejável que se implemente o Heliômetro Solar que está em vias de ser operacionalizado no Observatório Nacional (Reis Neto, 2005)

**PERSPECTIVAS FUTURAS**.

Está em andamento no ON o projeto do Heliômetro Solar que deverá entrar em operação em 2006. Este instrumento utiliza uma técnica desenvolvida por Bessel para observar paralaxe de estrelas que consiste em cortar a lente objetiva ao meio segundo um plano que contém o eixo ótico e tornar a colar as partes com um pequeno deslocamento entre elas na direção do corte. Com este procedimento obtêm-se duas imagens deslocadas de uma distância angular. No nosso caso, faremos deslocar uma imagem da outra de uma distância pouco maior que diâmetro solar e no CCD detector teremos duas bordas opostas do Sol. Medindo a distância entre estas bordas e conhecendo o deslocamento das imagens, podemos obter, por diferença, o diâmetro solar na latitude solar observada. Este instrumento nos dá duas vantagens sobre o Astrolábio Solar: as observações são instantâneas, não há que se esperar o Sol se deslocar ao longo de uma linha durante alguns minutos, assim, o número de observações pode se multiplicar por um valor considerável. Fazendo o eixo do instrumento girar observa-se qualquer latitude solar em qualquer hora ou data, isto permite o monitoramento da figura do Sol diariamente. Com o Heliômetro observa-se, da mesma forma que com Astrolábio, os bordos do Sol, podemos então utilizar diversas técnicas já desenvolvidas para redução de dados do Astrolábio.

Esperamos brevemente poder contar com este instrumento que irá multiplicar nossos dados e nossas possibilidades para estabelecer com maior precisão as variações do Semidiâmetro Solar em função do tempo e em função de suas latitudes.